\documentclass[11pt]{article}
\usepackage{epsfig}
\usepackage{latexsym}
\usepackage{amsmath}
\usepackage{verbatim}
\usepackage{mathrsfs}

\usepackage{amssymb,amsmath}
\usepackage{amsthm}
\usepackage{amscd}
\usepackage{amssymb}
\usepackage{amsfonts}
\usepackage{url}
\usepackage{slashed}
\usepackage[latin1]{inputenc}
\usepackage[english]{babel}

 \csname
@addtoreset\endcsname{equation}{section}

\def\gz0{\gamma^{0}}

\def\scs#1{\section{\sc #1}}
\def\scss#1{\subsection{\sc #1}}

\def\a{\alpha}
\def\b{\beta}
\def\g{\gamma}

\def\d{\delta}

\def\e{\epsilon}

\def\l{\lambda}
\def\L{\Lambda}
\def\m{\mu}
\def\n{\nu}

\def\r{\rho}

\def\s{\sigma}

\def\O{\Omega}

\def\cA{{\cal A}}
\def\cB{{\cal B}}
\def\cC{{\cal C}}
\def\cD{{\cal D}}

\def\cF{{\cal F}}
\def\cG{{\cal G}}
\def\cH{{\cal H}}
\def\cI{{\cal I}}

\def\cK{{\cal K}}
\def\cL{{\cal L}}

\def\cP{{\cal P}}

\def\cR{{\cal R}}

\def\cV{{\cal V}}

\def\be{\begin{equation}}
\def\ee{\end{equation}}
\def\bs{\begin{split}}
\def\es{\end{split}}
\def\bea{\begin{eqnarray}}
\def\eea{\end{eqnarray}}
\def\ba{\begin{array}}
\def\ea{\end{array}}
\def\bec{\begin{center}}
\def\ec{\end{center}}
\def\ba{\begin{align}}
\def\ena{\end{align}}

\def\non{\nonumber}

\def\12{\frac{1}{2}}

\def\ra{\rightarrow}

\begin{document}

\begin{flushright}
{\today}
\end{flushright}

\vspace{25pt}

\begin{center}
%%%%%%%%%%%%%%%%%%%%%%%%%%%%%%%%%%%%%%%%%%%%%%%%%%%%%%%%%%%%%%%%%%%%%%%%
{\Large\sc Higher-Spin Interactions: four-point functions and beyond}\\
%%%%%%%%%%%%%%%%%%%%%%%%%%%%%%%%%%%%%%%%%%%%%%%%%%%%%%%%%%%%%%%%%%%%%%%%
\vspace{25pt}
{\sc M.~Taronna}\\[15pt]
{\sl\small
Scuola Normale Superiore and INFN\\
Piazza dei Cavalieri, 7\\I-56126 Pisa \ ITALY \\
e-mail: {\small \it m.taronna@sns.it}}\vspace{10pt}
%%%%%%%%%%%%%%%%%%%%%%%%%%%%%%%%%%%%%%%%%%%%%%%
\vspace{35pt}

{\sc\large Abstract}

\end{center}

%%%%%%%%%%%%%%%%%%%%%%%%%%%%%%%%%%%%%%%%%%%%%%%%%

{In this work we construct an infinite class of four-point functions for massless higher-spin fields in flat space that are consistent with the gauge symmetry. In the Lagrangian picture, these reflect themselves in a peculiar non-local nature of the corresponding non-abelian higher-spin couplings implied by the Noether procedure that starts from the fourth order. We also comment on the nature of the colored spin-2 excitation present both in the open string spectrum and in the Vasiliev system, highlighting how some aspects of String Theory appear to reflect key properties of Field Theory that go beyond its low energy limit. A generalization of these results to n-point functions, fermions and mixed-symmetry fields is also addressed.}

%%%%%%%%%%%%%%%%%%%%%%%%%%%%%%%%%%%%%%%%%%%%%%%%

\setcounter{page}{1}
\pagebreak
{\linespread{0.75}\tableofcontents}
\newpage

%%%%%%%%%%%%%%%%%%%%%%%%%%%%%%%%%%%%%%%%%%%%%%%

%%%%%%%%%%%%%%%%%%%%%%%%%%%%%%%%%%%%%%%%%%%%%%%

\scs{Introduction}

%%%%%%%%%%%%%%%%%%%%%%%%%%%%%%%%%%%%%%%%%%%%%%%

Higher-Spin (HS) theories have been over the last decades an intense field of research that has attracted an increasing attention, starting from the works of the late 80's by Fradkin and Vasiliev \cite{Fradkin:1986qy} and Vasiliev \cite{Vasiliev:1988sa,Unfolded} that opened the way to the first classically consistent examples of non-abelian interactions of this type. However, we are still far from a satisfactory understanding of the problem, so much so that only in the last few years a reasonable understanding of the \emph{free} HS theory has been attained (see e.g. \cite{solvay}). Two distinct approaches have indeed come to terms with a problem that, in some respects, dates back to the early days of Quantum Field Theory (QFT) (see, e.g.~\cite{Majorana:1932rj}). The first is a ``metric-like'' approach, initiated in the works of Singh and Hagen \cite{SH}, Fronsdal \cite{Fronsdal} and de Wit and Freedman \cite{de Wit:1979pe}, and reconsidered more recently by Francia and Sagnotti \cite{FranciaSagnotti}. In their works the authors of \cite{FranciaSagnotti} proposed a \emph{geometric} reinterpretation of the free-field equations that can be expressed as
\be
\frac{1}{\square^{\,n}}\ \partial\cdot{\cR^{\,[n]}}_{;\m_1\ldots\m_{2n+1}}\,=\,0\ ,
\ee
for odd spins $s\,=\,2n\,+\,1$, and
\be
\frac{1}{\square^{\,n-1}}\ {\cR^{\,[n]}}_{;\m_1\ldots\m_{2n}}\,=\,0\ ,
\ee
for even spins $s\,=\,2n$, together with the related \emph{minimal} Lagrangian formulation \cite{minimal,exchange}, that rests for any $s$ on at most two additional fields and simplifies the previous BRST (Becchi, Rouet, Stora, Tyutin) constructions of \cite{OldBRST}. This form of the free equations includes the three familiar lower-spin examples, given by the linearized Einstein equations for $s\,=\,2$, by the Maxwell equations for $s\,=\,1$ and formally also by the Klein-Gordon equation for $s\,=\,0$, together with \emph{non-local} equations for spin larger than two. One can thus have, somehow, an intuition, even if restricted to a single spin at a time, of possible generalizations of the geometric framework of Maxwell theory and Einstein gravity to HS, pointing also to a possible key role of non-localities, that may be more and more fundamental at the interacting level, together with an eventual reconsideration of QFT from a more general perspective. More recently, these results were also generalized to reducible HS \emph{free} fields, starting from the ``triplet'' system and recovering similar interesting \emph{non-local} structures \cite{francia10}. The constructions for symmetric (spinor)-tensors that we have just outlined afford also interesting generalizations to the case of mixed-symmetry fields of the type $\phi_{\,\m_1\ldots\m_{s_1};\n_1\ldots\n_{s_2}\ldots}$, whose non-local geometric equations were first proposed in \cite{Bekaert:2002dt}, while the Lagrangian formulation was initiated with the pioneering work of Labastida \cite{labastida} and was completed only a few years ago in \cite{mixed}. A second kind of approach is the ``frame-like'' one, was developed mostly by Vasiliev and collaborators \cite{Vasiliev:1988sa,Unfolded}, generalizing the Cartan-Weyl framework to HS, and led to the Vasiliev system. Despite the remarkable success of Vasiliev's approach, only recently has it been possible to arrive at a covariant description of all bosonic flat space cubic interactions in \cite{cubicFT} by purely field theoretical methods. At the same time, starting from a String Theory vantage point all consistent cubic interactions involving bosonic and fermionic fields were obtained in \cite{cubicstring}\footnote{Further off-shell completions were presented in \cite{cubicoff-shell}.}. This extended previous results, including the works of the 80's by Bengtsson, Bengtsson and Brink \cite{lightcone} and the important works of Metsaev \cite{Metsaev}, in the light-cone formulation, and the works of Berends, Burgers and van Dam \cite{bbvd} in a covariant formulation that were then reconsidered and extended by Boulanger and others in \cite{BoulangerCubic,ZinovievCubic}, and were recently exploited in the interesting work of Bekaert, Joung and Mourad \cite{Euihun}.

This paper is aimed at improving our understanding of HS interactions, extending and generalizing the results of \cite{cubicstring}. Actually, the string results and their structure may give new insights on field theory properties that manifest themselves when looking at HS fields, pointing out and motivating a closer relation between ST and HS gauge theories and resonating with the long-held feeling that ST draws its origin from a \emph{generalized} Higgs effect responsible for its massive excitations (see e.g. \cite{Gross,AmatiCiafaloniVeneziano,Higgs,cubicstring}). The crux of the matter has long been to construct a consistent deformation of the free system at the quartic order. It is indeed this the case in which the HS program has encountered along the years an insurmountable barrier, both in the metric-like formulation and in the frame-like one, in which Vasiliev's system unfortunately does not provide a transparent answer. Indeed, only recently in \cite{Seed} the chain of higher-derivative terms found in the work of Fradkin and Vasiliev \cite{Fradkin:1986qy} and weighted by inverse powers of the cosmological constant $\L$ was recognized to be related, in the case of the gravitational coupling, to a higher-derivative \emph{seed} that is nicely associated to the simpler flat space cubic vertex, whose exact structure can be recovered, in a suitable scaling limit, wiping out the lower members of the tail. Moreover, even the origin of the spin-$2$ excitation present in the Vasiliev system is still unclear from a field theory perspective, since it can be dressed with Chan-Paton factors like any excitation belonging to the open bosonic string. This would make the ``graviton\footnote{We call it \emph{graviton} here with a little abuse of language since it admits colors.}'' colored, in contrast with standard field theory results pointing out inconsistencies of this kind of option \cite{coloredspin2} (strictly speaking with finitely many fields). To reiterate, at the quartic order a number of difficulties have piled up along the years, starting from the no-go results \cite{NoGo} and the inconsistencies found by Weinberg in \cite{WeinbergNoGo}, up to the inconsistency pointed out in \cite{Spin3Obstruction} for the Berends-Burgers-van Dam cubic coupling of spin-$3$ fields (for a recent review see for instance \cite{NogoRev} and references therein). Four-point functions for HS fields have been and still are somehow the most intriguing source of difficulties. Here we shall try to address these questions studying a class of solutions to the Noether procedure and discussing the role of Lagrangian \emph{non-localities} along lines that are actually in the spirit of the previous work \cite{NonLocalNoether}. As we shall see, some non-localities turn out to naturally arise at the Lagrangian level as soon as massless HS particles are considered. This reflects very peculiar and subtle aspects of the corresponding tree-level S-matrix amplitudes that clash with the factorization property usually assumed in the framework of the S-matrix theory.

It is instructive to introduce our arguments starting from the usual field theory perspective on the problem of HS interactions given by the Noether procedure, that played a key role in the construction of supergravity \cite{SUGRA}, and in its various incarnations has played a crucial role in order to solve for cubic HS couplings in explicit cases \cite{bbvd,cubicFT}. From this point of view the HS problem can be reformulated as equivalent to finding, order by order in the number of fields, a deformation of the free system of the form
\be
S[\phi]\,=\,\sum_s S^{\,(2)}[\phi_{\m_1\ldots\m_s}]\,+\,\e\, S^{\,(3)}[\phi_{\m_1\ldots\m_s}]\,+\,\e^{\,2}\,S^{\,(4)}[\phi_{\m_1\ldots\m_s}]\,+\,O(\e^{\,3})\ ,\label{Deform}
\ee
including at least one field of spin $s\,>\,2$ and where the contribution $S^{\,(3)}$ is cubic, $S^{\,(4)}$ is quartic, and so on. Consistency of the deformation \eqref{Deform} translates into an equivalence class of deformations of the linearized gauge symmetries of the type
\be
\d_{\L}\,\phi_{\,\m_1\ldots\m_s}\,=\,\d^{\,(0)}_{\L}\,\phi_{\,\m_1\ldots\m_s}\,+\,\e\,\d^{\,(1)}_{\L}\,\phi_{\,\m_1\ldots\m_s}\,+\, \e^{\,2}\, \d^{\,(2)}_{\L}\,\phi_{\,\m_1\ldots\m_s}\,+\,O(\e^{\,3})\ ,
\ee
leaving invariant $S[\phi]$ order by order, and defined modulo local redefinitions of fields and gauge parameters of the form
\be
\begin{split}
&\phi_{\,\m_1\ldots\m_s}\,\ra\,\phi_{\,\m_1\ldots\m_s}\,+\,\e\,f(\phi)_{\,\m_1\ldots\m_s}\,+\,O(\e^{\,2})\ ,\\
&\L_{\,\m_1\ldots\m_{s-1}}\,\ra\,\L_{\,\m_1\ldots\m_{s-1}}\,+\,\e\,\zeta(\phi,\L)_{\,\m_1\ldots\m_{s-1}}\,+\,O(\e^{\,2})\ .
\end{split}
\ee
Interestingly enough, there is a translation of this procedure that is closely related to a BRST formulation of HS fields \cite{OldBRST,NonLocalNoether,triplets,FormalBRST,BRSTrev} and in general to String Theory \cite{stringtheory} and to String Field Theory (SFT), both open \cite{OSFT} and closed \cite{CSFT}. Indeed, given \emph{any} nilpotent anti-derivation one can retrace the well known theory of Free Differential Algebras (FDA) \cite{FDA} in order to build fully non-linear \emph{covariant} HS equations of motion\footnote{We could think of them as zero curvature equations taking the freedom to use the word \emph{curvature} in this slightly different framework as a generalization of the results presented in \cite{francia10} where it was shown how the non local curvature equations of \cite{FranciaSagnotti}, as well as a new kind of non-local equations related to the triplet system, can be obtained integrating out the various auxiliary fields. The idea that we have in mind is an extension of those arguments at the full non-linear level that should produce in principle a consistent non-linear deformation of the usual linearized HS curvatures.}. The choice of the framework of FDA in order to describe the dynamics of a physical system is not a new idea, since it lies at the heart of the Vasiliev system and of some supergravity models as well as being a particular incarnation of a very general construction that appeared also in a number of different avatars like $L_{\infty}$ algebras \cite{homotopy}, of which the BV formalism \cite{BV} is one example, or again homological perturbation theory, $Q$-manifolds \cite{Qmanifold} and so on. For instance, the Vasiliev system is indeed formulated in terms of a FDA based on the nilpotent de Rham differential and can be presented in the remarkably compact form
\be
\widehat{F}\,=\,\frac{i}{2}\ d Z^{\,i}\wedge d Z_{\,i}\ \widehat{\Phi}\star\kappa\ ,\qquad \widehat{D}\widehat{\Phi}\,=\,0\ .
\ee
where in these equations the $\star$ operation implements the higher-spin algebra in the frame-like approach.
Leaving aside a detailed discussion of these equations (see e.g.~\cite{Vasiliev:1988sa,Unfolded,solvay} for more details), let us stress that the HS \emph{curvature} is here defined as
\be
\widehat{F}\,=\,d\widehat{A}\,+\,\widehat{A}\star\wedge\ \widehat{A}\ .
\ee
However, the de Rham differential does not define any local propagating degrees of freedom within its cohomology classes, and one is forced to introduce infinitely many auxiliary fields recovering a system of equations in the so called \emph{unfolded formulation}. The end result can be regarded as a Hamiltonian-like first-order rewriting of some linear, at the free level, or fully non-linear and non-local equations, at the interacting level. It is conceivable in this sense that non-localities, even those of the type $\tfrac1{\square}$, need to be expanded in terms of perturbatively local contributions in order to recover this \emph{unfolded} form of the equations, in which any higher-derivative term is redefined as a new field. As a result, the theory will be related to some massive parameter whose role is played here by the cosmological constant $\L$.
The starting point of our construction, in analogy with and generalizing what has been done in constructing SFT \cite{OSFT,CSFT}, for example, is to consider a FDA built from a nilpotent anti-derivation $Q$ capable of identifying the degrees of freedom of free massless HS fields \emph{directly} in its cohomology classes, ending up with non-linear, and eventually manifestly \emph{non-local}, equations of the form\footnote{See e.g. \cite{ParentForm} for interesting discussions about the relations between first order systems based on the de Rham differential and systems based on a BRST differential.}
\be
\cR(\Phi)\,=\,Q\Phi\,+\,G(\Phi)\,=\,0\ .
\ee
From this point of view we are going to extend what was previously considered in the framework of the triplet system in \cite{cubic}, stressing however the role played by the simpler gauge-fixed system\footnote{From now on, with a little abuse of notation, we shall refer to these systems involving transverse and possibly traceless fields as \emph{on-shell} systems or \emph{on-shell gauge-fixed} systems or similar circumlocutions.}
\be
\cL\,=\,\frac{1}{2}\ \phi(-p)\,\star\,p^{\,2}\,\phi(p)\ ,\label{on-shell}
\ee
for bosons, and
\be
\cL\,=\,\frac{1}{2}\ \bar{\psi}(-p)\,\star\,\slashed{p}\ \psi(p)\ ,\label{on-shellf}
\ee
for fermions, where here and throughout this paper we shall refer to the $\star$ as an inner-product in the metric-like formalism whose role is to contract the corresponding indices of the various polarization tensors. As we shall see, these provide the simplest settings in order to search for consistent deformations of the HS free theory and actually can be regarded as a starting point for any possible off-shell completion whose structure, at the end, can make some geometric features of the theory more manifest. Restricting the attention for brevity to the bosonic case of eq.~\eqref{on-shell}, the \emph{master} field $\phi(p\,,\xi)$ is initially subject to the \emph{transversality} constraint
\be
p\cdot\partial_{\xi}\,\phi(p\,,\xi)\,=\,0\ ,
\ee
and is a generating function of the form
\be
\phi(p\,,\xi)\,=\,\sum_{s\,=\,0}^{\infty}\,\frac{1}{s!}\ \phi_{\,\m_1\ldots\m_s}(p)\,\xi^{\m_1}\ldots\,\xi^{\m_s}\ .
\ee
whose components carry an arbitrary bosonic, in general also reducible, representation of the Lorentz group\footnote{In principle we can consider also generating functions of mixed-symmetry fields.}. This system is gauge invariant under the transformation
\be
\d\phi(p\,,\xi)\,=\,p\cdot\xi\,\L(p\,,\xi)\ ,
\ee
with
\be
p\cdot\partial_{\xi}\,\L(p\,,\xi)\,=\,0\ ,\qquad p^{\,2}\L(p\,,\xi)\,=\,0\ ,
\ee
where the components of the generating function of gauge parameters satisfy algebraic constraints similar to those satisfied by $\phi(p\,,\xi)$.

To reiterate, our aim here is to combine the simplifications coming from these gauge fixed systems with the off-shell structure encoded by the FDA. We are then able to recognize a simpler incarnation of the Noether procedure at the level of the $n$-point functions that realize both the \emph{linearized} gauge symmetries and the \emph{global} symmetries of the free system, to be contrasted with the $n$-point Lagrangian couplings, that nonetheless can be directly extracted from these data. Hence, defining by $\tilde{\cK}_{12\ldots n}(p_{\,i}\,,\xi_i\,,\ldots)$ the generating functions of color-ordered HS $n$-point functions, that by definition we consider as formal series whose relative coefficients are in general \emph{functions} of the Mandelstam-like invariants, one is led in the FDA framework to the off-shell homogeneous equations
\be
\left(Q_1\,+\,Q_2\,+\,\ldots\,+\,Q_n\right)\,\tilde{\cK}^{\,\text{FDA}}_{12\ldots n}\,=\,0\ .\label{FDANoether}
\ee
These equations can be solved at the level of the gauge-fixed theory \eqref{on-shell} where they simplify into
\be
p_{\,i}\cdot\partial_{\xi_i}\,\tilde{\cK}^{\,\textbf{on-shell}}_{12\ldots n}(p_{\,i}\,,\xi_i\,,\ldots)\,\approx 0\ ,\qquad i\,=\,1\ ,\ldots\ ,n\ ,\label{onshellN}
\ee
in which the approximate equality means on-shell and modulo divergences. The strategy is then to extend this result adding traces and/or divergences in order to recover the \emph{same} equality
\be
p_{\,i}\cdot\partial_{\xi_i}\,\tilde{\cK}^{\,\textbf{off-shell}}_{12\ldots n}(p_{\,i}\,,\xi_i\,,\ldots)\,\approx 0\ ,\qquad i\,=\,1\ ,\ldots\ ,n\ ,
\ee
but now in any off-shell framework and where the equality is modulo the full Lagrangian equations of motion (EoM), without using any divergence constraint. Equivalently, the same strategy is to complete the on-shell gauge-fixed solution to a full solution of \eqref{FDANoether}.
\begin{figure}[htbp]
\begin{center}
\resizebox{10cm}{!}{\psfig{figure=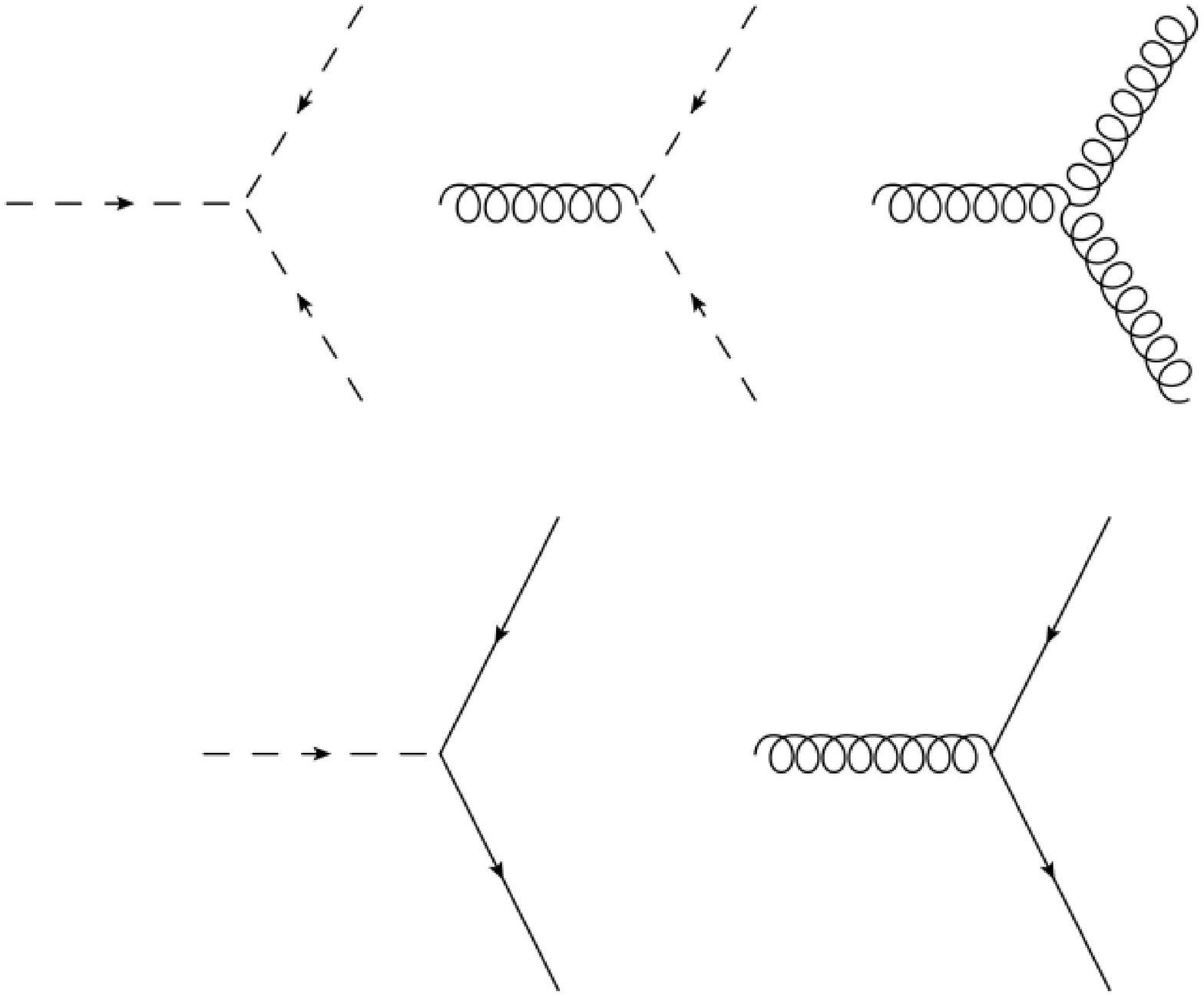,width=15cm}}
\caption{Building blocks for the $\cG^{\,(i)}_{12\ldots n}$'s.} \label{fig:intro}
\end{center}
\end{figure}
The solution of this problem can be expressed in terms of powers of the standard color-ordered $n$-point functions $\cG_{12\ldots n}^{\,(i)}$ of the most general theory built from a gauge boson, a scalar field and a spin-$1/2$ fermion, that are uniquely specified by the cubic couplings in fig.~\ref{fig:intro}. One should add, in principle, also the usual Yang-Mills quartic couplings, that however can be reconstructed here interpreting them as counterterms restoring the \emph{linearized} gauge invariance of the tree-level amplitudes, thus making clear the fundamental role of the latter with respect to the former. Notice that eqs.~\eqref{FDANoether} and \eqref{onshellN} fix basically the dependence of the kernel $\tilde{\cK}_{12\ldots n}$ on the symbols, but leave open the possibility of multiplying each independent gauge invariant tensor structure with functions of Mandelstam-like invariants. Moreover, let us underline a kind of \emph{correspondence} between tree-level $n$-point functions and $n$-point Lagrangian couplings. The former are associated to a linearized gauge symmetry related to the free system of eq.~\eqref{on-shell} together with the corresponding global symmetries, while the latter encode somehow \emph{geometrical} principles, together with a fully non-linear deformation of the original gauge symmetries that could be captured, in principle, by a resummation of the full tower of non-linear couplings. For instance, in the case of spin-$2$ this resummation rebuilds the usual Riemaniann geometry, as observed by Deser in \cite{Deser}, and references therein, and in the case of HS fields should encode, in a similar fashion, the HS geometry.
%This picture points to a kind of interplay between linearized gauge symmetries and fully non-linear ones, or amplitudes and Lagrangian couplings, or again global symmetries and \emph{geometry}, and as we shall see this can be encoded, on one side of this correspondence in the linearized free system \eqref{on-shell} that specifies the most general form of the tree-level $n$-point functions.
One should also gain a deeper understanding of the \emph{non-linear} symmetries and of the global symmetries of the system, that need to be related to some HS algebra and whose analysis we leave for the future.

In this work we take the aforementioned perspective in order to study how field theory can in principle overcome some difficulties encountered over the years. The main result can be roughly summarized by the generating function
\be
\tilde{\cK}_{1234}\left(\Xi_i\right)\,=\,-\,\frac{1}{su}\ \left[1\,-\,su\sum_{c\in\cF}\cG^{(c)\,\text{fermi}}_{1234}\left(\Xi_i\right)\right]\,\exp\left[-\,su\sum_{d\in\cB}\cG^{(d)\,\text{bose}}_{1234} \left(\Xi_i\right)\right]\ ,
\ee
of four-point S-matrix amplitudes involving massless HS fields where the $\cG^{(d)\,\text{bose}}_{1234}$'s and the $\cG^{(c)\,\text{fermi}}_{1234}$'s are generically gauge boson four-point functions as discussed in Section~\ref{sec:YMfour} and in Appendix~\ref{sec:fermionic}. They have been multiplied by Mandelstam invariants in order to get only single poles while their current exchange parts reconstruct HS exchanges after combining properly their power expansion. Moreover, the generating function should be considered modulo functions of the Mandelstam variables that do not introduce poles of order grater than one.
Appendix~\ref{app:A} contains a preliminary classification of $n$-point functions involving HS fields so that looking back at the three-point case, that is the only one in which the kernel $\tilde{\cK}$ coincides with the Lagrangian coupling generating function, we are able to recover the results of \cite{cubicFT} in the bosonic case and of \cite{cubicstring} in the bosonic and fermionic case, that in the gauge-fixed theory \eqref{on-shell} take the form
\begin{multline}
\tilde{\cK}_{123}\,=\,\left(1\,+\,\d\,+\,\slashed{\xi}_{\,1}\,+\,\slashed{\xi}_{\,2}\,+\,\slashed{\xi}_{\,3}\right)\\
\times\,\exp\left\{\sqrt{\frac{\a^{\,\prime}\!\!}{2}}\ \Big[(\xi_1\cdot\xi_2\,+\,1)\,\xi_3\cdot p_{\,12}\,+\,(\xi_2\cdot\xi_3\,+\,1)\,\xi_1\cdot p_{\,23}\,+\,(\xi_3\cdot\xi_1\,+\,1)\,\xi_2\cdot p_{\,31}\Big]\right\}\ .
\end{multline}
Let us remark again that this result should be considered as a generating function, whose relative coefficients between different terms can be chosen arbitrarily at the cubic level while they ought to be fixed by consistency with the quartic results or any other consistency condition. In this sense the exponential used by String Theory could leave the way to other functions. Here $\d$ contracts indices belonging to two spinor-tensors and the $1$ is the purely bosonic part of the coupling.
This structure of the four-point functions (and of their $n$-point counterparts) arise as the solution of the Noether procedure for a general field theory with massless particles of any spin. Moreover, while embodying an infinite class of local quartic couplings, it also contains the seeds for the difficulties that have been faced along the years and also, hopefully, the way in which field theory can in principle bypass all no-go results of \cite{NoGo,WeinbergNoGo}. However, the physical interpretation and the need for the corresponding Lagrangian non-localities is still a subtle issue and the peculiar form of the amplitudes one arise at clashes in general with commonly accepted ideas about S-matrix structure that reflect some difficulties in defining an S-matrix for massless particles (see e.g. the discussion in the introduction of \cite{WeinbergNoGo} or \cite{SmatrixBook}). The full amplitude generating function, containing also Chan-Paton factors \cite{Paton:1969je}, is finally recovered summing over all non-cyclic permutations of the external legs, as
\begin{multline}
\cA(\Phi_1,\Phi_2,\ldots,\Phi_n)\,=\,\sum_\s\text{Tr}\Big[\Phi_1(\xi_1)\,\Phi_{\s(2)}(\xi_{\s(2)})\, \ldots\, \Phi_{\s(n)}(\xi_{\s(n)})\Big]\\\star_{\,12\ldots n} \tilde{\cK}_{1\s(2)\ldots\s(n)}(p_{\,i},\xi_{\,i})\ ,\label{IntroAmplitude}
\end{multline}
where the trace is over the color indices carried by the polarization generating functions $\Phi_{i}$, recovering in this way a kind of generalized \emph{open-string}-like form. Interestingly, from four points onwards, there are more possibilities, since some permutations with respect to the labels $\{1\,,\ldots\,,n\}$ of the $\cG_{12\ldots n}^{(i)}$'s are independent for $n\,\geq\,4$. This means that one can combine two or more totally cyclic independent kernels, eliminating the Chan-Paton factors and recovering in this way a kind of \emph{closed-string}-like amplitudes from which the usual gravitational four-point functions emerge, together with HS generalizations. In the four-point case, for instance, for each $\cG^{(i)}_{1234}$ there are two independent options and one can recover in this fashion the \emph{closed-string}-like kernel
\be
\tilde{\cK}(\xi_i\,,\xi_i^{\,\prime})\,=\, \left(\sum_\s\ \tilde{\cK}_{1\s(2)\s(3)\s(4)}(p_{\,i},\xi_{\,i})\ \tilde{\cK}_{1\s(2)\s(4)\s(3)}(p_{\,i},\xi^{\,\prime}_{\,i})\right)\ ,\label{Shs2}
\ee
together with analogous generalizations to higher orders described in Appendix~\ref{app:A}. These results are analyzed in a number of examples pointing out some differences between the graviton and the colored spin-$2$ fields but leaving for the future a detailed analysis of the generalized \emph{closed-string}-like amplitudes together with possible generalizations of the Bern-Carrasco-Johansson (BCJ) construction of \cite{BCJ} to HS.

A general lesson to be drown from the results that we have summarized is that HS $n$-point functions appear to go in hand with a peculiar feature: while they are still built from current exchanges and local terms, they generally factorize only on (infinite) \emph{subsets} of the available exchanges that lack finite numbers of lower-spin contributions clashing with the usual factorization property of the S-matrix. This unusual feature is tantamount to a \emph{non-local} nature of HS Lagrangian couplings, in particular when expanded around flat-space. These anyway result from conventional amplitudes built from Feynman propagators and one can expect them to be still compatible with the notion of causality and with the cluster property. Nonetheless, potential clashes with tree-level unitarity may be possibly related to the fact that infinitely many degrees of freedom ought to contribute to the same residue. In this respect we can only anticipate that within what we shall call \emph{minimal scheme} even admitting non-localities no pole arising in the amplitude can lack an interpretation as an external particle participating in the physical process. This can give some hope to arrive to an understanding of background independent HS interactions, even though we are not able at present to give a definite answer about the consistency of the proposed scheme in flat space. We choose to take this point of view in this paper having also in mind a deformation of the flat-space results to constant curvature backgrounds or to massive fields.

Among other things, we shall discuss from the same perspective the role of the spin-$2$ excitation present in the Vasiliev system that admits in principle Chan-Paton factors, trying to give an answer to a puzzle pointed out in \cite{exchange} together with a very interesting open question about its true nature. Indeed, at the massless level a mixing between the singlet part of colored spin-$2$ components and a combination that is strictly uncolored and plays the role of a graviton, anticipated in \cite{exchange}, is here justified by the existence of two different kinds of amplitudes, the first of the \emph{open-string}-type and the second of the \emph{closed-string}-type. Here, however, we shall see that a non-abelian colored spin-$2$ field brings about non-localities, and from this point of view it is naturally related to the massive excitations present in the Open String spectrum while only the spin-$2$ components interacting with a \emph{closed-string}-like amplitude can be directly related to the usual graviton. The non-local nature of HS Lagrangian couplings puts our discussion of QFT on more general grounds, that ought to be better understood, as we anticipated. From this point of view important questions arise concerning both the nature and the geometrical meaning, if any, of the non-localities that we find in this work.

Our explicit construction leads us to propose a milder constraint that we name as \emph{minimal scheme}. As we shall explain in more detail in the following:
\begin{itemize}
\item it is equivalent to locality in all standard frameworks,
\item it allows for the presence of a very restricted set of non-localities whenever HS fields are considered recovering insofar as possible the commonly expected structure of the S-matrix,
\end{itemize}
At the same time this discussion, and in particular our generalization of \emph{open-string}-like and \emph{closed-string}-like amplitudes, reinforces the feeling that ST hides within its structure a number of potentially profound lessons for Field Theory.

The plan of the paper is the following. In Section~$2$, restricting the attention to the case of symmetric fields, we briefly review the symbol calculus of \cite{cubicstring}, with special emphasis on the extension of the formalism to a full BRST framework that entails also anticommuting coordinates related to ghosts and auxiliary fields. In Section~$3$ we recast the usual Noether procedure in the form of a FDA, describing briefly the case of cubic and quartic interactions and pointing out how the information on the non-linear deformations of gauge symmetries are nicely encoded in the free system \eqref{on-shell}. In Section~$4$ we turn to describe four-point amplitudes, first in the simpler case of Yang-Mills theory and then in the general setting of HS theories, pointing out the differences between \emph{open-string}-like amplitudes and \emph{closed-string}-like amplitudes and how the usual no-go theorems can be overcome, in principle at least, by the solution proposed. In Section~$5$ we describe the new features of HS couplings, with special emphasis on \emph{non-localities}. Our conclusions are summarized in Section~$6$. Finally, the Appendices address the extension to fermionic four-point functions, to $n$-point functions and to theories containing mixed-symmetry (spinor-)tensors together with some results on the relation between the gauge-fixed theory of eqs.~\eqref{on-shell} and \eqref{on-shellf} and the off-shell completions.

%%%%%%%%%%%%%%%%%%%%%%%%%%%%%%%%%%%%%%%%%%%%%%%%%%%%%%

\scs{Symbol calculus}\label{simbol}

%%%%%%%%%%%%%%%%%%%%%%%%%%%%%%%%%%%%%%%%%%%%%%%%%%%%%%

In this section we review a few basic techniques that will be used in this paper, among which symbol calculus is possibly the most relevant. Following \cite{cubicstring}, a simplification relies on the introduction of auxiliary variables $\xi_{\,i}^{\,\m}$ that make it possible to define \emph{generating functions} of HS tensor fields. By convention, we shall denote such generating functions as
\be
\phi_{\,i}(p_{\,i},\xi_i)\,=\,\sum_{s=0}^{\infty}\,\frac{1}{s!}\ \phi_{\,i\,\m_1\ldots\m_s}(p_{\,i})\ \xi_i^{\,\m_1}\ldots\xi_i^{\,\m_s}\ .
\ee
The auxiliary coordinates $\xi_i^{\m}$ can be considered, with a little abuse of notation, as ``phase space'' coordinates together with the $x^{\,\m}$'s, in close analogy to the Weyl-Wigner set up, and can be extended to a similar kind of super phase-space $\cH$ adding corresponding anti-commuting ``coordinates'' and ``momenta'' that play the role of \emph{ghosts} for the $x$ and $\xi$ variables in a standard BRST treatment. The construction is very simple and proceeds via the introduction of the anti-commuting coordinates
\be
\begin{pmatrix}
  \theta^{0}& &\theta&&\partial_{\,\bar{\theta}}\\
  \partial_{\theta^{0}}& &\bar{\theta}&&\partial_{\theta}
\end{pmatrix}\ ,
\ee
following techniques and ideas similar to those used in \cite{BRSTrev,cubic}. Here, for convenience, we have arranged the $\theta$'s taking into account the ghost number, so that the first line contains ghost-number $+1$ ``coordinates'' and ``momenta'' while the second line contains their ghost-number $-1$ counterparts. To make a link with the previous constructions let us stress that our $\theta$-variables are nothing but the ``symbols'' of the string-like operators
\be
\begin{pmatrix}
  c_{0}& &c_{-1}& &c_{1}\\
  b_{0}& &b_{-1}&&b_{1}
  \end{pmatrix}\ ,
\ee
exactly as $\xi$ is the ``symbol'' of the oscillator $\a_{-1}$. While any formula that we will write hereafter can be translated in the operator setting by using the aforementioned dictionary, we choose to use and extend the symbol notation of \cite{cubicstring} working with functions on some kind of superspace and dealing with generating functions instead of Hilbert-space states. As usual, one can put some structure on $\cH$, defining an adjoint operation by
\begin{align}
{x}^{\dagger}\,=\,x\ ,\quad {p}^{\dagger}\,=\,p\ ,\quad \xi^{\,\dagger}\,=\,\partial_{\xi}\ ,\quad {\theta^0}^{\dagger}\,=\,\theta^0\ ,\quad {\partial_{\theta^0}}^{\dagger}\,=\,\partial_{\theta^0}\ ,\quad \theta^{\dagger}\,=\,\partial_{\bar{\theta}}\ ,\quad\bar{\theta}^{\,\dagger}\,=\,\partial_{\theta}\ ,
\end{align}
so that one is naturally led to consider the graded space of functions over $\cH$, that we call $\O(\cH)$, defining over it a nilpotent derivation given by a BRST charge of ghost number $1$. The BRST charge should encode the free propagating degrees of freedom of the system in its cohomology classes. The simplest choice is related to reducible HS gauge fields and is solely built from the mass-shell and transversality constraints without taking into account trace constraints. Following the usual BRST setting (see e.g. \cite{OldBRST,triplets,BRSTrev}), one thus obtains
\be
Q\,=\,\theta^0\,p^{\,2}\,+\,\theta\,p\cdot\partial_\xi\,+\,\partial_{\bar{\theta}} \,p\cdot\xi\,-\,\theta\,\partial_{\bar{\theta}}\,\partial_{\theta_0}\ ,\label{BRST}
\ee
that is one possible gauge-invariant off-shell completion of $p^{\,2}$. Other consistent completions are related to the Fronsdal setting \cite{Fronsdal} or to the related compensator setting of \cite{FranciaSagnotti} as well as to other frameworks, that we shall describe briefly for completeness in Appendix~\ref{app:quadratic}, remarking their links with the gauge-fixed system described at the quadratic level by \eqref{on-shell}, to which all can be reduced. The major advantage of performing off-shell computations with this BRST setting will be clear in the following. For instance, one can have some kind of intuition related to the super phase-space $\cH$ making also manifest the link with the tensionless limit of ST, which is naturally formulated in this language \cite{triplets}. Moreover, one can now move between gauge-fixed and off-shell theory projecting onto the purely bosonic symbols $\xi$ and imposing or relaxing the transversality constraint on the generating functions.

Applying $Q$ to the most general ghost-number $0$ HS \emph{super-field} $\Phi\in\O(\cH)$
\be
\Phi\,(p,\theta^0,\xi,\theta,\bar{\theta})\,=\,\phi(p,\xi)\,+\,\theta^0\,\bar{\theta}\,C(p,\xi)\,+ \,\theta\,\bar{\theta}\,D(p,\xi)\ ,\label{superfield}
\ee
that from now on we consider in full generality as a matrix valued field
\be
\Phi\,=\,\Phi^a \, T^a\ ,
\ee
where the $T^a$'s are matrices associated to the Chan-Paton factors \cite{Paton:1969je}, one gets the equation of motion
\begin{multline}
Q\,\Phi\,=\,\theta^0\Big(p^{\,2}\phi(\xi)\,-\,p\cdot\xi\,C(\xi)\Big)\,+ \,\theta\Big(p\cdot\partial_\xi\,\phi(\xi)\,+\,C(\xi)\,-\,p\cdot\xi D(\xi)\Big)\\+\,\theta^0\,\theta\,\bar{\theta}\Big(p^{\,2}D(\xi)\,-\,p\cdot\partial_\xi C(\xi)\Big)\,=\,0\ .\label{tripleteqns}
\end{multline}
As usual, one can easily check the \emph{invariance} of this system under the gauge transformation
\be
\d\Phi\,=\,Q\,\L\ ,
\ee
where $\L$ is a ghost-number $-1$ parameter of the form
\be
\L(p,\theta^0,\xi,\theta,\bar{\theta})\,=\,\bar{\theta}\,\L(p,\xi)\ ,
\ee
and hence
\be
\d\Phi\,=\,p\cdot\xi\,\L(p,\xi)\,+\,\theta^0\bar{\theta}\,(-p^{\,2})\,\L(p,\xi)\,+ \,\theta\bar{\theta}\,p\cdot\partial_{\xi}\,\L(p,\xi)\ .
\ee
The nontrivial cohomology classes at ghost number $0$ contain exactly the physical degrees of freedom of a spin $s$, $s-2$ down to spin $1$ or $0$, respectively for odd or even $s$.
The system so far described is actually the \emph{triplet system} \cite{triplets} that can be recovered from the tensionless limit of the free part of open String Field Theory, so that the supercoordinates provide an off-shell completion of $p^{\,2}$. The next ingredient that we are going to introduce is a pairing that is important to construct Lagrangians, and more generally singlets. A pairing or contraction can be defined generalizing the $\star$-inner-product introduced in \cite{cubicstring} and given by
\begin{equation}
\star\,:\ \Big(\Phi_{\,1}(p_1\,,\xi_1)\,,\,\Phi_{\,2}(p_2\,,\xi_2)\Big) \ \ra \ \Phi_{\,1}\star\,\Phi_{\,2}\,=\,\exp\Big(\partial_{ \xi_1}\cdot\,\partial_{\xi_2}\Big)\,\Phi_{\,1}(p_1\,,\xi_1)\ \Phi_{\,2}(p_2\,,\xi_2)\Big|_{\xi_i\,=\,0}\ ,\label{contra}
\end{equation}
where $\xi_1$ and $\xi_2$ are only the bosonic symbols. The point here is to extend the contraction operator in the exponent of \eqref{contra}, letting it act over the whole HS super phase-space $\cH$. Actually, it is possible to define two extensions having this property, that we choose to label respectively as $\star$ and $\tilde{\star}$. They are associated to different projections of the superfields $\Phi\in\O(\cH)$ with respect to the $\theta^0$ coordinate, and can be defined as
\begin{multline}
\tilde{\star}\,:\ \Big(\Phi_{\,1}(p_1,\theta^0_1,\xi_1,\theta_1,\bar{\theta}_1)\,,\,\Phi_{\,2}(p_2,\theta^0_2,\xi_2,\theta_2, \bar{\theta}_2)\Big) \ \ra \ \Phi_{\,1}\,\tilde{\star}\ \Phi_{\,2}\\=\,\exp\Big(\partial_{ \xi_1}\!\!\cdot\partial_{\xi_2}\,-\,\partial_{\theta_1}\,\partial_{\bar{\theta}_2}\,+ \,\partial_{\bar{\theta}_1}\,\partial_{{\theta}_2}\Big)\,\Phi_{\,1}(p_1,\theta^0,\xi_1,\theta_1,\bar{\theta}_1)\ \Phi_{\,2}(p_2,\theta^0,\xi_2,\theta_2,\bar{\theta}_2)\,\Big|_{\xi_i\,=\,\theta_i\,=\,\bar{\theta}_i\,=\,0}\ ,\label{supercontra2}
\end{multline}
and
\be
\star\,:\ \Big(\Phi_{\,1}(p_1,\theta^0_1,\xi_1,\theta_1,\bar{\theta}_1)\,,\,\Phi_{\,2}(p_2,\theta^0_2,\xi_2,\theta_2, \bar{\theta}_2)\Big) \ \ra \ \Phi_{\,1}\star\,\Phi_{\,2}\,=\,\int d\theta^0\,\left[\Phi_{\,1}\,\tilde{\star}\ \Phi_{\,2}\right]\ \theta^0\ ,\label{supercontra}
\ee
where by convention
\be
p_{\,1}\,+\,p_{\,2}\,=\,0\ ,
\ee
and both reduce to eq.~\eqref{contra} if one restricts the attention to the bosonic coordinates and ghost-number zero super-fields $\Phi_{\,i}$. Let us mention for completeness that the aforementioned $\star$-operations encode the usual Hermitian products on the Fock spaces for the corresponding oscillators. To conclude this section we anticipate that the inner-product $\star$ will play a role in the off-shell framework while the inner-product $\tilde{\star}$ will play a role in the computation of scattering amplitudes, and more generally of $n$-point functions. To reiterate, this formalism provides a natural framework to move back and forth between these two ways of formulating the problem, exploiting both the FDA structure that appears off-shell and the simplifications that are present in the gauge-fixed theory. Nonetheless, the latter represents by itself a well defined system in which the problem of finding consistent deformations of the free theory is well posed. In the next sections we are going to consider, from a purely field theoretical point of view, the problem of constructing consistent gauge-invariant Lagrangians and covariant equations of motion for HS fields, studying in detail the Noether procedure and reinterpreting it in terms of FDA.

%%%%%%%%%%%%%%%%%%%%%%%%%%%%%%%%%%%%%%%%%%%%%%%%%%%%%%

\scs{Structure of the interactions}

%%%%%%%%%%%%%%%%%%%%%%%%%%%%%%%%%%%%%%%%%%%%%%%%%%%%%%

In this section, following and extending various ideas that have appeared in the literature on the subject including \cite{CSFT,cubic}, we are going to analyze and reformulate the Noether procedure, translating it in terms of a Free Differential Algebra (FDA) or in general $L_\infty$-algebra \cite{FDA,homotopy}. This kind of formalism has been somehow a source of inspiration for this work but we want to stress that the results that we are going to present can be in principle recast in other off-shell frameworks.

%%%%%%%%%%%%%%%%%%%%%%%%%%%%%%%%%%%%%%%%%%%%%%%%%%%%%%

\scss{HS covariant equations and gauge transformations}

%%%%%%%%%%%%%%%%%%%%%%%%%%%%%%%%%%%%%%%%%%%%%%%%%%%%%%

We begin here describing the formal setup along the lines of \cite{CSFT}, then considering in the next subsections its explicit realization. In order to construct a consistent HS gauge field theory we are going to work within the HS super phase-space $\cH$ introduced in Section~\ref{simbol}. There is, indeed, a very general construction that can be realized in principle whenever one is able to define an identically nilpotent operator $Q$. Following ideas and techniques that are involved in the construction of Free Differential Algebras (FDA) \cite{FDA}, one can try and define HS equations of motion of the general form
\be
\cR(\Phi)\,=\,Q\,\Phi\,+\,G(\Phi)\,=\,0\ ,\label{Rphi}
\ee
with $G(\Phi)$ a ghost-number one function of the type
\be
G(\Phi)\,=\,\frac{1}{2!\!}\,[\Phi,\Phi]\,+\,\frac{1}{3!\!}\,[\Phi,\Phi,\Phi]\,+\,\ldots \ .\label{Gphi}
\ee
Here one could also add, in principle, a linear term that could be regarded as a deformation of the BRST charge $Q$, while the ellipsis in eq.~\eqref{Gphi} indicate higher-order contributions in the HS superfield $\Phi\,$ given by $n$-ary totally symmetric multilinear products
\be
[\,\underbrace{\ \cdot\ ,\ldots,\ \cdot\ }_n\,]\ :\ \ \underbrace{\O(\cH)\,\otimes\,\ldots\,\otimes\,\O(\cH)}_{n}\,\ra\,\O(\cH)\label{brakets}
\ee
that generalize the standard structure constants ${f^a}_{b_1\ldots b_n}$ of FDA's. In order to streamline the notation, we define two other different operations whose role is to encode the two options that show up in multiplying brackets together. These are indicated respectively by $\diamond$ and $\circ$ and can be defined by\footnote{Notice that in principle the $\diamond$-operation can be expressed totally in terms of $\circ$, making possible to express formally all $n$-ary products starting from the identity:
\be
[\Phi_1,\ldots,\Phi_n]\,=\,(\ldots(1\circ\Phi_1)\circ\ldots)\circ\Phi_n\ ,
\ee
while we will use the convention
\begin{align}
([\Psi])\circ[\Phi_1,\ldots,\Phi_n]\,&=\,[[\Psi],\Phi_1,\ldots,\Phi_n]\ ,\\
[\Psi]\circ([\Phi_1,\ldots,\Phi_n])\,&=\,[\Psi,[\Phi_1,\ldots,\Phi_n]]\ ,
\end{align}
in order to distinguish which of the terms plays the role of $\Psi$ in eqs.~\eqref{rules2} and \eqref{rules3}.
}
{\allowdisplaybreaks
\begin{align}
[\Phi_{\,1},\ldots,\Phi_{\,n}]\,\diamond\,[\Psi_{\,1},\ldots,\Psi_{\,q}]&\,\equiv\, [\Phi_{\,1},\ldots,\Phi_{\,n},\Psi_{\,1},\ldots,\Psi_{\,q}]\ ,\label{rules1}\\
\Psi\,\circ\,[\Phi_{\,1},\ldots,\Phi_{\,n}]&\,\equiv\,[\Psi,\Phi_{\,1},\ldots,\Phi_{\,n}]\ ,\label{rules2}\\
[\Phi_{\,1},\ldots,\Phi_{\,n}]\,\circ\,\Psi&\,\equiv\,[\Phi_{\,1},\ldots,\Phi_{\,n},\Psi]\ ,\label{rules3}\\
1\,\circ\,\Psi\,\equiv\,\Psi\,\circ\,1&\,\equiv\,[\Psi]\ ,\label{rules4}\\
[\Phi]\diamond 1\,=\,1\diamond [\Phi]&\,\equiv\,[\Phi]\ ,\label{rules5}
\end{align}}
\!\!\!while we do not require these definitions to extend to other cases that are not of the form displayed in the latter equations.
This kind of notation is used only in this section to encode the formal structures here introduced as a notational device in order to keep track of the \emph{bracket algebra}. In terms of these operations eq.~\eqref{Rphi} takes the suggestive form
\be
\cR(\Phi)\,=\,e_{\,\diamond}^{\,[\Phi]}\,-\,1\,=\,0\ ,\label{curv}
\ee
where the exponential sum the whole series in eq.~\eqref{Rphi} through the operation $\diamond$ and where a single element bracket has been defined by
\be
[\Phi]\,\equiv\,Q\,\Phi\ .
\ee
It is then possible to write down a non-linear gauge transformation given by
\begin{multline}
\d\Phi\,=\,\left(\L\star\frac{\d}{\d\Phi}\right)\cR(\Phi)\,=\,\L\,\circ\,e_{\,\diamond}^{\,[\Phi]}\\\equiv\, Q\,\L\,+\,[\L,\Phi]\,+\,\frac{1}{2!\!}\,[\L,\Phi,\Phi]\,+\,\ldots\,
+\,\frac{1}{n!\!}\,[\L,\underbrace{\Phi,\ldots,\Phi}_n\ ]\,+\,\ldots\ ,\label{delta}
\end{multline}
under which eq.~\eqref{curv} transforms covariantly provided the condition
\be
\left(\cR(\Phi)\star\frac{\d}{\d\Phi}\right)\,\cR(\Phi)\,=\, \left(e_{\,\diamond}^{\,[\Phi]}\,-\,1\right)\,\circ\,e_{\,\diamond}^{\,[\Phi]}=\,0\label{constr}
\ee
holds, where the notation is meant to stress that the object within the parenthesis should be interpreted as the $\Psi$ in eq.~\eqref{rules2}. We should emphasize that eq.~\eqref{constr} ought to be valid identically, independently of the superfield $\Phi$ as a property of the brackets themselves. Moreover, we have introduced a derivative with respect to the HS super-field $\Phi$, given by
\begin{multline}
\frac{\d}{\d\Phi(x_1,\theta_1^0,\xi_{\,1},\theta_{\,1},\bar{\theta}_{\,1})}\ \Phi(x_2,\theta_2^0,\xi_{\,2},\theta_{\,2},\bar{\theta}_{\,2})\\=\,\d(x_1-x_2)\,\d(\theta^0_1-\theta^0_2)\,\exp\Big(\xi_1\cdot\xi_2 -\theta_1\bar{\theta}_2\,+\,\bar{\theta}_1\theta_2\Big)\ .
\end{multline}
so that one is led, by convention, to
\begin{align}
\left(\L\star\frac{\d}{\d\Phi}\right)\,[\Phi]\,&=\,[\L]\ ,\label{rules6}\\
\left(\L\star\frac{\d}{\d\Phi}\right)\,[\,\underbrace{\Phi,\ldots,\Phi}_n\,]\,& =\,n\,[\L,\underbrace{\Phi,\ldots,\Phi}_{n-1}\,]\label{rules7}\ .
\end{align}
No distinction is needed between left and right derivatives, since the field $\Phi$ has ghost number zero. At this point we have not said much about the brackets, aside from the condition \eqref{constr}. Interestingly, it turns out that a nice algebraic structure, called in the mathematical literature \emph{homotopy associativity} \cite{homotopy}, is encoded in \eqref{constr}. In order to make this structure manifest, it is convenient to expand \eqref{constr} order by order in $\Phi$, obtaining the conditions
\begin{gather}
Q^{\,2}\Phi\,=\,0\ ,\vphantom{\Big|}\\
Q\,[\Phi,\Phi]\,+\,2\,[Q\,\Phi,\Phi]\,=\,0\ ,\vphantom{\Big|}\label{binary}\\
Q\,[\Phi,\Phi,\Phi]\,+\,3[Q\,\Phi,\Phi,\Phi]\,+\,3[[\Phi,\Phi],\Phi]\,=\,0\ ,\vphantom{\Big|}\label{ternary}\\
\ldots\vphantom{\Big|}\nonumber
\end{gather}
The first is automatically satisfied given the BRST operator \eqref{BRST} while the second, once the BRST charge has been chosen, is an equation for the bilinear product $[\ \cdot\ ,\ \cdot\ ]$, that should be constructed so as to make the BRST charge $Q$ a derivation. The third equation relates the cubic term to the quadratic one by the condition that the failure of the BRST charge to be a derivation at this order should be compensated by the possible failure of the Jacobi identity for the quadratic product. One can keep going along these lines, studying all other relations that, if solved consistently, give rise to full gauge covariant HS equations of motion. Exploiting the consistency condition \eqref{constr}, one can also make a further step, verifying the on-shell closure of the gauge algebra to all orders and computing, formally at least, the HS algebra of gauge transformations that takes the form
\be
[\d_{\,\L_1},\d_{\,\L_2}]\,\Phi\,\approx\,\d_{F(\L_1,\L_2,\Phi)}\,\Phi\ ,\label{gaugealgebra}
\ee
where $F(\L_1,\L_2,\Phi)$ is some function encoding the structure constants of the HS, or more generically, Field Theory algebra. Computing the commutator of two gauge transformation one indeed recovers
\begin{multline}
[\d_{\,\L_1},\d_{\,\L_2}]\,\Phi\,= \,\L_{\,[2}\circ\left(\L_{\,1]}\,\circ\,e_{\,\diamond}^{\,[\Phi]}\right)\,\circ\,e_{\,\diamond}^{\,[\Phi]}\\=\, \left(\L_{\,[1}\,\circ\,e_{\,\diamond}^{\,[\Phi]}\right)\,\circ\,\L_{\,2]}\,\circ\,e_{\,\diamond}^{\,[\Phi]}\,=\, \left([\L_{[1}]\,\diamond\,e_{\,\diamond}^{\,[\Phi]}\right)\,\circ\,[\L_{2]}]\,\diamond\, e_{\,\diamond}^{\,[\Phi]}\ ,\label{it}
\end{multline}
where we repeatedly used eq.~\eqref{delta} together with the fact that $\L_{\,1}\,\circ\, e_{\,\diamond}^{\,[\Phi]}$ has vanishing ghost-number and where we have used the notation $\left(\L_{\,1}\,\circ\, e_{\,\diamond}^{\,[\Phi]}\right)$ in order to make clear that the object within the brackets  has to be interpreted as $\Psi$ in eq.~\eqref{rules2}. Eq.~\eqref{it} can be further simplified using \eqref{constr}. Hence, making use of the relation
\begin{multline}
0\,=\,\left(\L_{1}\star\frac{\d}{\d\Phi}\right)\left(\L_{2}\star\frac{\d}{\d\Phi}\right) \left\{\vphantom{\Bigg|}\left(e_{\,\diamond}^{\,[\Phi]}\,-\,1\right)\,\circ\,e_{\,\diamond}^{\,[\Phi]}\right\}\,= \,\left([\L_{1},\L_{2}]\,\diamond\,e_{\,\diamond}^{\,[\Phi]}\right) \,\circ\,e_{\,\diamond}^{\,[\Phi]}\\+\,\left([\L_{[2}]\,\diamond\,e_{\,\diamond}^{\,[\Phi]}\right)\,\circ\,[\L_{1]}]\,\diamond\, e_{\,\diamond}^{\,[\Phi]}\,+ \,\left(e_{\,\diamond}^{\,[\Phi]}\,-\,1\right)\,\circ\,[\L_1,\L_2]\,\diamond\,e_{\,\diamond}^{\,[\Phi]}\ ,
\end{multline}
that can be recovered by taking into account the odd ghost number of $\L_i$ together with the rules \eqref{rules6} and \eqref{rules7}, one obtains
\be
[\d_{\,\L_1},\d_{\,\L_2}]\,\Phi\,\approx\,\left([\L_1,\L_2]\,\diamond\,e_{\,\diamond}^{\,[\Phi]}\right)\,\circ\,e_{\,\diamond}^{\,[\Phi]}\ ,
\ee
where we have used the on-shell EoM's, so that $\left(e_{\,\diamond}^{\,[\Phi]}\,-\,1\right)\,\approx\,0$\ \footnote{Notice that we can go on-shell after having computed the derivatives. Hence in our conventions on-shell \be \Psi\,\circ\,\left(e_{\,\diamond}^{\,[\Phi]}\,-\,1\right)\,\neq\,0\ ,\ee while \be \left(e_{\,\diamond}^{\,[\Phi]}\,-\,1\right)\,=\,0\ .\ee}.
This implies that
\be
F(\L_1,\L_2,\Phi)\,=\,[\L_1,\L_2]\,\diamond\,e_{\,\diamond}^{\,[\Phi]}\,=\,[\L_1\,,\L_2]\,+\,O(\Phi)\ .
\ee
To reiterate what we have done so far, before going ahead and giving a more detailed description of the actual n-ary products, we want to stress that this construction, together with a choice of the BRST charge \eqref{BRST}, is a very general setting from which field theories originate. In particular, the FDA is a formal translation of the Noether procedure and can deal both with local and non-local field theories. We leave a further analysis on the important issue of HS symmetries and HS \emph{geometry}, especially in the light of the present discussion, for future works.

%%%%%%%%%%%%%%%%%%%%%%%%%%%%%%%%%%%%%%%%%%%%%%%%%%

\scss{Lagrangian Form}

%%%%%%%%%%%%%%%%%%%%%%%%%%%%%%%%%%%%%%%%%%%%%%%%%%

Starting from the formal structure given by the FDA, let us now study a possible Lagrangian formulation that is clearly important if one wants to quantize the theory, compute scattering amplitudes or define a starting point to pursue the Batalin-Vilkovisky program \cite{BV}. This step can be actually addressed using the formalism developed so far, and indeed in this setting\footnote{Analogous results can be obtained in other settings that are different from the triplet system, if the corresponding pairing is constructed.} it is straightforward to integrate eq.~\eqref{curv} to a full Lagrangian given by
\be
\cL\,=\,\int d\theta^0 \left\{\,e^{\,\Phi}\right\}\ ,\label{lagr}
\ee
where by definition
\be
\left\{\Phi^{\,n+1}\right\}\,\equiv\,\{\underbrace{\Phi,\ldots,\Phi}_{n+1}\,\}\,=\,\Phi\ \tilde{\star}\ [\,\underbrace{\Phi,\ldots,\Phi}_n\,]\ ,
\ee
with $\tilde{\star}$ defined in \eqref{supercontra2} and hence
\be
\{1\}\,=\,0\ ,\qquad\{\Phi\}\,=\,0\ .
\ee
At this point, having in mind this incarnation of the Noether procedure, it is worthwhile for the ensuing discussion to give an explicit definition of the $n$-ary products \eqref{brakets} in terms of some \emph{color-ordered} kernels $\cK_{12\ldots n}$ playing the role of generalized \emph{structure constants}. Here, we define
\be
\cK_{12\ldots n}\,=\,\cK_{12\ldots n}\left(p_{\,i},\partial_{\theta^0_i}\,;\xi_i,\theta_i,\bar{\theta}_i\right)\ ,\quad 1\leq i\leq n\ .
\ee
to be any Lorentz invariant function defined over $\cH$ with ghost number equal to zero. To be explicit, we have added $\partial_{\theta^0_i}$ as the ghost coordinate related to $p_{\,i}$, or $\theta_i$ and $\bar{\theta}_i$ as the ghost coordinates related to $\xi_i$. In particular, the building blocks of the bosonic part of $\cK_{12\ldots n}$ are\footnote{In this paper we choose to not consider terms proportional to the Levi-Civita tensor $\e_{\m_1\ldots\m_D}$ that, for any fixed number of legs in $\cK_{12\ldots n}$, can give rise to non-trivial contributions only for some lower dimensions. It would be interesting to analyze this kind of options for $D\,=\,3$.}
\be
\xi_{i}\cdot\xi_j\ ,\qquad p_{\,i}\cdot\xi_j\ ,\qquad p_{\,i}\cdot p_{\,j}\ ,
\ee
while the building blocks of the complete kernel defined over the whole super phase-space $\cH$ are the ghost-number zero terms (see also \cite{cubic})
\be
\xi_{i}\cdot\xi_j\ ,\qquad p_{\,i}\cdot\xi_j\ ,\qquad p_{\,i}\cdot p_{\,j}\ ,\qquad \theta_i\, \partial_{\theta^0_j}\ ,\qquad \theta_i\,\bar{\theta}_j\ .
\ee
One can then express the $n$-ary products in terms of the $\cK_{12\ldots n}$'s, writing
\begin{multline}
[\Phi_1,\ldots,\Phi_n](\Xi)\,=\, \sum_\s\ \left[\Phi_{\s(1)}\left(\,\Xi_{\s(1)}\right)\ldots\Phi_{\s(n)}\left(\,\Xi_{\s(n)}\right)\right]\\\star_{\,1,\ldots,n}\ \cK_{\,0\,\s(1)\s(2)\ldots \s(n)}\left(\hat{\Xi},\hat{\Xi}_1,\ldots,\hat{\Xi}_n\right)\ \theta^0\ ,
\end{multline}
where the sum is over all permutations $\s$ of the external legs and we have used the short-hand notation
\be
\Xi_i\,=\,\left(p_{\,i},\theta^0_i,\xi_i,\theta_i,\bar{\theta}_i\right)\ ,\qquad \hat{\Xi}_i\,=\,\left(p_{\,i},\partial_{\theta^0_i},\xi_i,\theta_i,\bar{\theta}_i\right)\ ,
\ee
for these collections of relevant variables. Let us stress that we have defined the $n$-ary products in terms of \emph{color-ordered} Kernels that, as such, are intimately related to \emph{open-string}-like couplings. In the following we will also discuss other types of Kernels that one can call similarly of the \emph{closed-string}-type in order to underline how string results actually reflect interesting field theory properties that are still to be completely understood. As one can expect, \emph{closed-string}-like kernels also contribute, along similar lines, to the definition of $n$-ary products, thus leading to distinguish two different kinds of couplings. We begin in the next sections by addressing the problem of finding the structure of the $\cK_{123}$'s and $\cK_{1234}$'s that is consistent with \eqref{constr}, discussing also some implications at the level of the S-matrix. We shall see in practice the advantages of starting with the \emph{on-shell} results in a gauge-fixed system in order to recover eventually a consistent \emph{off-shell} deformation. Afterwards, describing in this field theory setting the results of \cite{cubicstring}, we shall turn to analyze the problem of the trilinear product and its relation to a tree-level four-point S-matrix, recognizing along lines that are actually in the spirit of the previous work \cite{NonLocalNoether} the role of non-localities whenever HS fields are taken into account and analyzing the nature of the difficulties that are to be faced when addressing this problem.

%%%%%%%%%%%%%%%%%%%%%%%%%%%%%%%%%%%%%%%%%%

\scss{Binary product}\label{sec:binary}

%%%%%%%%%%%%%%%%%%%%%%%%%%%%%%%%%%%%%%%%%%

As we have seen in the previous section, a fully consistent binary product is bound to satisfy the condition
\be
Q\,[\Phi_1(\Xi_1),\Phi_2(\Xi_2)](\Xi)\,+\,[Q_1\Phi_1(\Xi_1),\Phi_2(\Xi_2)](\Xi)\, +\,[\Phi_1(\Xi_1),Q_2\Phi_2(\Xi_2)](\Xi)\,=\,0\ .
\ee
In terms of the kernel $\cK_{123}\left(\hat{\Xi}_1,\hat{\Xi}_2,\hat{\Xi}_3\right)$ this equation translates into
\be
(Q_1\,+\,Q_2\,+\,Q_3)\,\ \cK_{123}\left(\hat{\Xi}_1,\hat{\Xi}_2,\hat{\Xi}_3\right)\theta^0_1\,\theta^0_2\,\theta^0_3\,=\,0\ ,\label{Consist1}
\ee
and, in order to find a solution to \eqref{Consist1} one can exploit the results obtained in \cite{cubicFT,cubicstring} where a kernel $\cK_{123}$ was presented in the on-shell gauge-fixed form as a consistent deformation of the \emph{universal} quadratic action proportional to $p^{\,2}$ of eq.~\eqref{on-shell} and then extended off-shell\footnote{For a brief review of this results see Appendix~\ref{app:cubic}.}. Hence, in the on-shell gauge-fixed form of \cite{cubicFT,cubicstring}, the binary product so far considered takes the form
\be
[\phi_{\,2}\,,\phi_{\,3}\,](p_{\,1},\xi_1)\,=\,\sum_\s\ \Big[\phi_{\,\s(2)}\left(p_{\,\s(2)},\xi_{\s(2)}\right)\,\phi_{\,\s(3)}\left(p_{\,\s(3)},\xi_{\s(3)}\right)\Big]\star_{\,2,3} \exp\left(\cG_{1\s(2)\s(3)}\right)\ ,
\ee
where $\phi_{\,2}$ and $\phi_{\,3}$ are matrix valued, the sum is over the two permutations of the set $\{2,3\}$,
\be
\cG_{123}(p_{\,i},\xi_i)\,=\,\sqrt{\frac{\a^{\,\prime}\!\!}{2}}\ \Big[(\xi_{\,1}\cdot\xi_{\,2}\,+\,1)\,\xi_{\,3}\cdot p_{\,12}\,+\,(\xi_{\,2}\cdot\xi_{\,3}\,+\,1)\,\xi_{\,1}\cdot p_{\,23}\,+\,(\xi_{\,3}\cdot\xi_{\,1}\,+\,1)\,\xi_{\,2}\cdot p_{\,31}\Big]\ ,\label{G}
\ee
and by definition
\be
p_{\,ij}\,=\,p_{\,i}\,-\,p_{\,j}\ .
\ee
The complete off-shell result can then be recovered also in the formalism presented here, completing \eqref{G} on the whole $\cH$. This step can be afforded with standard techniques \cite{BRSTrev}, and the result is of the form
\begin{multline}
\cG^{\,\text{off-shell}}_{123}\left(\hat{\Xi}_{\,i}\right)\,=\,g^{(1)}_{ijkl}\ \xi_i\cdot\xi_j\,\xi_k\cdot p_{\,l}\,+\,g^{(2)}_{ijkl}\ \theta_i\bar{\theta}_j\,\xi_k\cdot p_{\,l}\,+\,g^{(3)}_{ijkl}\ \xi_i\cdot\xi_j\,\theta_k \partial_{\,\theta^0_l}\,+\,g^{(4)}_{ijkl}\ \theta_i\bar{\theta}_j\,\theta_k\partial_{\,\theta^0_l}\\+\,g^{(1)}_{ij}\xi_i\cdot p_{\,j}\,+\,g^{(2)}_{ij}\theta_i\partial_{\,\theta^0_j}\,+\,g^{(3)}_{ij}\theta_i \bar{\theta}_{j}\,+\,g^{(4)}_{ij}\xi_i\cdot \xi_{j}\ ,\label{offG}
\end{multline}
where all terms of $\cG$ in eq.~\eqref{G} have been supplemented with corresponding partners in the full $\cH$, so that the coefficients satisfy the symmetry relations
\be
g^{(1)}_{ijkl}\,=\,g^{(1)}_{jikl}\ ,\quad g^{(3)}_{ijkl}\,=\,g^{(3)}_{jikl}\ ,\quad g^{(4)}_{ijkl}\,=\,g^{(4)}_{kjil}\ ,
\ee
together with a cyclicity property with respect to the external legs. The condition of eq.~\eqref{Consist1} fixes all the remaining coefficients of the expansion \eqref{offG}, so that after some algebra, one is led to the general solution
\begin{align}
\cG^{\,\text{off-shell}}_{123}\left(\hat{\Xi}_{\,i}\right)\,&=\,\cA\left(g^{(1)}_{ijkl}\ \xi_i\cdot\xi_j\,\xi_k\cdot p_{\,l}\,+\,g^{(2)}_{ijkl}\ \theta_i\bar{\theta}_j\,\xi_k\cdot p_{\,l}\,+\,g^{(3)}_{ijkl}\ \xi_i\cdot\xi_j\,\theta_k \partial_{\,\theta^0_l}\,+\,g^{(4)}_{ijkl}\ \theta_i\bar{\theta}_j\,\theta_k\partial_{\,\theta^0_l}\vphantom{\frac{1}{2}}\right)\nonumber\\&+\,\cB\left(\xi_1\cdot p_{\,23}\,+\,\xi_2\cdot p_{\,31}\,+\,\xi_3\cdot p_{\,12}\,-\,\theta_1\partial_{\,\theta^0_{23}}\,-\,\theta_2\partial_{\,\theta^0_{31}} \,-\,\theta_3\partial_{\,\theta^0_{12}}\vphantom{\frac{1}{2}}\right)\nonumber\\&+\,\cC\left(\theta_1 \bar{\theta}_{1}\,+\,\theta_2 \bar{\theta}_{2}\,+\,\theta_3 \bar{\theta}_{3}\,+\,\frac{1}{2}\left(\xi_1\cdot \xi_{1}\,+\,\xi_2\cdot \xi_{2}\,+\,\xi_3\cdot \xi_{3}\right)\right)\ ,\label{offG2}
\end{align}
where $\cA$, $\cB$ and $\cC$ are free relative coefficients that cannot be fixed simply from \eqref{Consist1}, while the coefficients $g^{(l)}_{ijkl}$ can be read in table~\ref{table:coeff}.
\begin{table}[ht]
\centering % used for centering table
\begin{tabular}{c c c c c}
\hline\hline %inserts double horizontal lines
Index combination & $g^{(1)}$ & $g^{(2)}$ & $g^{(3)}$ & $g^{(4)}$ \\ [0.5ex] % inserts table
%heading
\hline % inserts single horizontal line
1231 & 1 & -2 & -1 & 1 \\ % inserting body of the table
1232 & -1 & 0 & 1 & -1\\
1233 & 0 & 0 & 0 & 1\\
1211 & 0 & 0 & 1 & 0\\
1212 & -1 & 2 & 0 & 0\\
1213 & -1 & 2 & 0 & 0\\
1221 & 1 & -2 & 0 & -4\\
1222 & 0 & 0 & -1 & 1\\
1223 & 1 & 0 & 0 & -1\\
1111 & 0 & 0 & 0 & 0\\
1112 & -1 & -2 & 1 & 0\\
1113 & 1 & 2 & -1 & 0\\
1121 & -2 & -4 & 1 & 1\\
1122 & -6 & -12 & -5 & -4\\
1123 & 0 & -2 & -1 & -1\\
1131 & 2 & 4 & -1 & -1\\
1132 & 0 & 2 & 1 & 1\\
1133 & 6 & 12 & 5 & 4\\
2131 & & 0 & &\\
2132 & & 2 & &\\
2133 & & 0 & &\\
2111 & & 0 & &\\
2112 & & 2 & &\\
2113 & & 0 & &\\
2121 & & -2 & &\\
2122 & & 0 & &\\
2123 & & -2 & &\\
 [1ex] % [1ex] adds vertical space
\hline %inserts single line
\end{tabular}
\caption{\emph{Coefficients entering the solution of \eqref{Consist1}. The empty entries can be recovered from the ones given using cyclicity property of the indices.}} % title of Table
\label{table:coeff} % is used to refer this table in the text
\end{table}
Here, for generality, we have also considered the additional gauge-invariant term proportional to $\cC$ that on-shell is non-zero only for reducible fields, and in this framework can be considered as a further mixing term among the components of the various triplets. We stress that one needs to consider the \emph{whole} expression for $\cG^{\,\text{off-shell}}$ as given in \eqref{offG2}, choosing\footnote{The coefficient $\cC$ does not play any role on-shell for reducible fields.}
\be
\cA\,=\,\cB\ ,
\ee
if one wants to reproduce after gauge fixing the same $\cG$ of eq.~\eqref{G}. Moreover, at least in principle, it should be possible to decompose the vertex related to \eqref{offG2} into the Fronsdal vertices given in \cite{cubicFT,cubicstring} for irreducible fields, although ST is naturally formulated in this picture that, as we can see, bears also a simpler relation to a FDA. Anyway, we have to say that any other off-shell completion would be at this stage equally meaningful, as we discuss in Appendix~\ref{app:quadratic} and \ref{app:cubic}. Both the three-point off-shell vertices and the on-shell amplitudes thus take the form
\be
\cA_{\,3}\,=\,\sum_\s\ \text{Tr}\Big[\phi_{\,1}\left(p_{\,1},\xi_{1}\right)\, \phi_{\,\s(2)}\left(p_{\,\s(2)},\xi_{\s(2)}\right)\,\phi_{\,\s(3)}\left(p_{\,\s(3)},\xi_{\s(3)}\right)\Big]\star_{\,1,2,3} \exp\left(\cG_{1\s(2)\s(3)}\right)\ ,\label{3coupling}
\ee
with $\cG$ given, respectively, by eq.~\eqref{G} or by eq.~\eqref{offG}. We emphasize here that the operator $\cG_{123}$ encodes the tensor structures that one could recover at the cubic level in any HS theory and indeed, apart from any specific case, the \emph{most} general kernel $\cK_{123}$ that is consistent with gauge invariance and does not contain traces, is\footnote{For a on-shell proof of this statement see Appendix~\ref{app:mostgeneral}.}
\be
\cK_{123}(p_{\,i},\xi_i)\,=\,a\left(\cG^{(0,1)}_{123}\,,\,\cG^{(0,2)}_{123}\,,\,\cG^{(0,3)}_{123}\,,\,\cG^{(1)}_{123}\right)\ ,
\ee
where $a(z_1,z_2,z_3,w)$ is an arbitrary function symmetric under exchange of the $z_i$'s, that encodes the overall relative coefficients between the various couplings, while both the $\cG^{(0,i)}_{123}$'s and $\cG^{(1)}_{123}$, after a complete on-shell gauge fixing, reduce to
\begin{align}
\cG^{(0,1)}_{123}\,&=\,\sqrt{\frac{\a^{\,\prime}\!\!}{2}}\ \ \xi_{\,1}\cdot p_{\,23}\ ,\quad
\cG^{(0,2)}_{123}\,=\,\sqrt{\frac{\a^{\,\prime}\!\!}{2}}\ \ \xi_{\,2}\cdot p_{\,31}\ ,\quad
\cG^{(0,3)}_{123}\,=\,\sqrt{\frac{\a^{\,\prime}\!\!}{2}}\ \ \xi_{\,3}\cdot p_{\,12}\ ,\label{G1}\\
\cG^{(1)}_{123}\,&=\,\sqrt{\frac{\a^{\,\prime}\!\!}{2}}\ \Big[\,\xi_{\,1}\cdot\xi_{\,2}\,\xi_{\,3}\cdot p_{\,12}\,+\,\xi_{\,2}\cdot\xi_{\,3}\,\xi_{\,1}\cdot p_{\,23}\,+\,\xi_{\,3}\cdot\xi_{\,1}\,\xi_{\,2}\cdot p_{\,31}\Big]\label{G2}\ ,
\end{align}
and satisfy individually the conditions
\be
p_{\,i}\cdot\partial_{\xi_i}\ \cG_{123}^{(i, k)}\,\approx\,0\ ,
\ee
where $\approx$ means on-shell and modulo divergence terms. We stress that these equations implement, at the level of the $3$-point functions, the linearized part of the gauge transformations of the fields related to the free system and to the physical requirement of decoupling unphysical degrees of freedom. In principle, as we have done in \eqref{offG2}, one could also consider more complicated functions of three additional totally cyclic arguments depending on
\be
\theta_i \bar{\theta}_{i}\,+\,\frac{1}{2}\,\xi_i\cdot \xi_{i}\ ,\qquad\ i\,=\,1\ ,2\ ,3\ ,
\ee
that, being proportional to traces, vanish identically on-shell for \emph{irreducible} fields. Apart from terms proportional to traces, the four terms in which $\cG^{\,\text{off-shell}}_{123}$ splits are of order one and three in the $\xi_i$'s, and their complete expressions can be easily read off from eq.~\eqref{offG2}. In the following, in order to simplify the discussion we shall concentrate on a particular exponential generating function in terms of the single operator \eqref{offG2}, keeping in mind however, the possibility of playing with relative coefficients. The important issue of constraining the coupling function at this order and the spectrum by purely field theoretic arguments is left for future work.

\begin{figure}[htbp]
\begin{center}
\resizebox{10cm}{!}{\psfig{figure=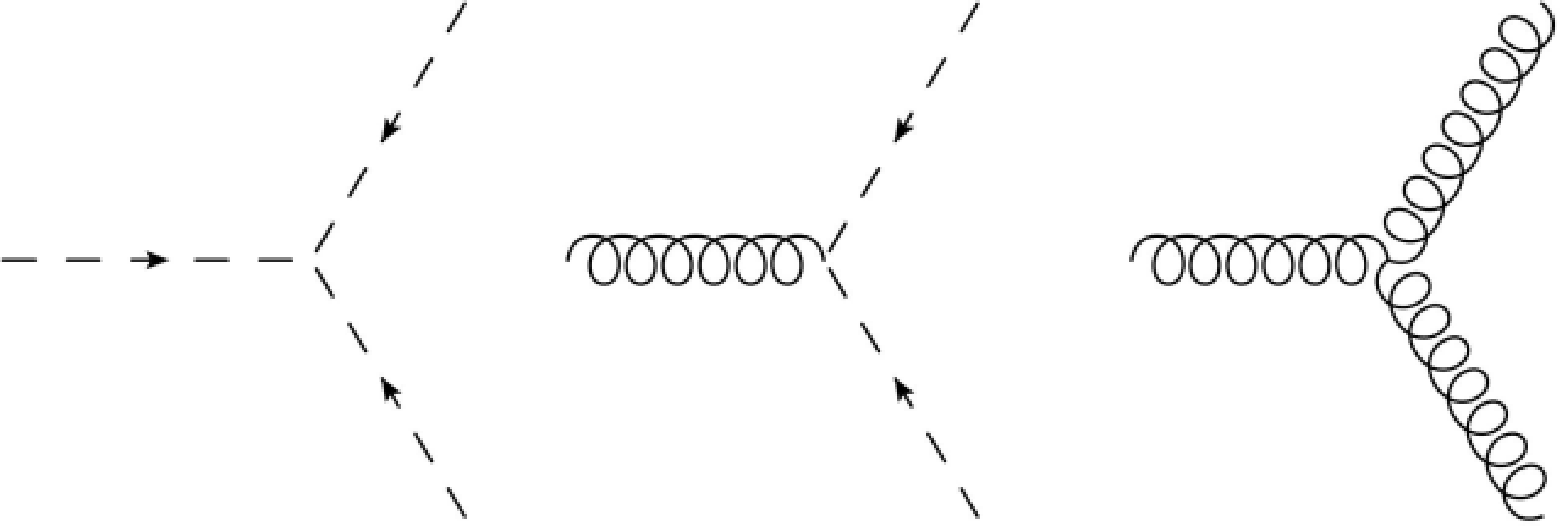,width=15cm}}
\caption{Cubic couplings for scalars and gauge boson corresponding to the pieces in which $\cG_{123}$ splits.} \label{fig:cubicboson}
\end{center}
\end{figure}

Finally, we would like to emphasize the simple form of \eqref{G1} and \eqref{G2} and their link with the cubic tree-level Feynman rules of a theory containing a gauge boson and a scalar field. It is indeed true that the expressions \eqref{G1} and \eqref{G2} coincide with the corresponding tree-level expression for the color-stripped three-point amplitudes of fig.~\ref{fig:cubicboson}, where we have also considered, for generality, a three-scalar amplitude that is just a constant overall factor and does not affect gauge invariance by definition.
Here, $\xi_1$, $\xi_2$ and $\xi_3$ are the polarization tensors of the gauge boson while the scalar wavefunction is conventionally set to one. This observation will be important in the following, when we shall generalize the construction to higher orders. Moreover, since $\cG_{123}$ is totally antisymmetric under permutations of the external legs
\be
\cG_{123}\,=\,-\,\cG_{132}\ ,
\ee
the same is true for any odd power of them. As a result, in this case the sum in \eqref{3coupling} will generate the group theory factors that are appropriate, at odd levels, for the classical groups for which the $T^{a}$'s belong to the fundamental representation. In this construction, of course, the levels reflect the spins of the states, rather than their masses as in String Theory. On the other hand, even powers of $\cG_{123}$ are totally symmetric under permutations of the external legs, and this suffices to generate the group theory factors that are appropriate at even levels.

%%%%%%%%%%%%%%%%%%%%%%%%%%%%%%%%%%%%%%%%%%%%%%%

\scss{Ternary product}\label{sec:ternary}

%%%%%%%%%%%%%%%%%%%%%%%%%%%%%%%%%%%%%%%%%%%%%%%

In this section we analyze eq.~\eqref{ternary} in order to recover the general form of a consistent ternary product. As in the previous case, it can be of use to exploit the full super phase-space $\cH$ in which the picture provided by the FDA is available. As we have anticipated, we consider a kernel representation of the ternary product given by some function $\cK_{1234}\left(\hat{\Xi}_1,\hat{\Xi}_2,\hat{\Xi}_3,\hat{\Xi}_4\right)$, whose labels $\{1\,,2\,,3\,,4\}$ are meant to emphasize that we are restricting the attention to color-ordered kernels, thus splitting eq.~\eqref{ternary} into independent color-ordered contributions. Restricting the attention to one of these, eq.~\eqref{ternary} takes the form
\begin{multline}
\left[(Q_1\,+\,Q_2\,+\,Q_3\,+\,Q_4)\ \cK_{1234}\left(\hat{\Xi}_1,\hat{\Xi}_2,\hat{\Xi}_3,\hat{\Xi}_4\right)\right]\ \theta^0_1\,\theta^0_2\,\theta^0_3\,\theta^0_4\\=\,-\,
\left[\cK_{(12|a}\left(\hat{\Xi}_1,\hat{\Xi}_2,\hat{\Xi}_a\right)\,\theta^0_a\ \star_a\ \cK_{\tilde{a}|34)}\left(\hat{\Xi}_{\tilde{a}},\hat{\Xi}_3,\hat{\Xi}_4\right)\right]\ \theta^0_1\,\theta^0_2\,\theta^0_3\,\theta^0_4\ ,\label{Consist2}
\end{multline}
where we have adopted the notation
\be
\cK_{(12|a}\cK_{a|34)}\,=\,\cK_{12a}\cK_{a34}\,+\,\cK_{23a}\cK_{a14}\ ,
\ee
in order to recover a cyclic quantity under permutations of the external legs. It is now relatively simple to work out a particular solution of this equation as for instance
\begin{multline}
\cK_{1234}\left(\hat{\Xi}_1,\hat{\Xi}_2,\hat{\Xi}_3,\hat{\Xi}_4\right)\\=\,-\,\left(\int d\theta^0_a\,\cK_{(12|a}\left(\hat{\Xi}_1,\hat{\Xi}_2,\hat{\Xi}_a\right)\,\theta^0_a\right) \frac{\tilde{\star}_a}{p_{\,a}^{\,2}}\left(\int d\theta^0_{\tilde{a}}\,\cK_{\tilde{a}|34)}\left(\hat{\Xi}_{\tilde{a}},\hat{\Xi}_3,\hat{\Xi}_4\right)\, \theta^0_{\tilde{a}}\right)\ .\label{soluzfour}
\end{multline}
This statement can be proved observing that, acting with $Q_1\,+\,Q_2$ on $\cK_{12a}$, one can exploit \eqref{Consist1} obtaining
\be
Q_1\,+\,Q_2\,\ra\,-\,Q_a\ ,
\ee
while acting with $Q_3\,+\,Q_4$ on $\cK_{a34}$ one can similarly obtain
\be
Q_3\,+\,Q_4\,\ra\,-\,Q_{\tilde{a}}\ .
\ee
Now, since by definition
\be
p_{\,a}\,+\,p_{\,\tilde{a}}\,=\,0\ ,
\ee
one is finally led to
\be
Q_1\,+\,Q_2\,+\,Q_3\,+\,Q_4\,\ra\,-\,\left(\theta^0_a\,+\,\theta^0_{\tilde{a}}\right)\,p_{\,a}^{\,2}\ ,
\ee
which cancels the non-locality and yields after few manipulations the right-hand side of \eqref{Consist2}. The solution so far obtained to the Noether procedure was recognized already in \cite{cubicstring} by standard on-shell techniques, but it is only a particular solution of the \emph{non-homogeneous} equation \eqref{Consist2} and still leaves some freedom that, as we shall see, is closely related to the possibility of recovering a non-vanishing four-point S-matrix amplitude. Indeed, one can always add to this solution another contribution that solves identically the \emph{homogeneous} equation
\be
(Q_1\,+\,Q_2\,+\,Q_3\,+\,Q_4)\,\tilde{\cK}_{1234}\,=\,0\ .\label{Consist3}
\ee
Summarizing, the most general form of the solution to \eqref{Consist2}, and hence to the Noether procedure at this order, is
\begin{multline}
\cK_{1234}\left(\hat{\Xi}_1,\hat{\Xi}_2,\hat{\Xi}_3,\hat{\Xi}_4\right)\\=\,-\,\left(\int d\theta^0_a\,\cK_{(12|a}\left(\hat{\Xi}_1,\hat{\Xi}_2,\hat{\Xi}_a\right)\,\theta^0_a\right) \frac{\tilde{\star}_a}{p_{\,a}^{\,2}}\left(\int d\theta^0_{\tilde{a}}\,\cK_{\tilde{a}|34)}\left(\hat{\Xi}_{\tilde{a}},\hat{\Xi}_3,\hat{\Xi}_4\right)\, \theta^0_{\tilde{a}}\right)\\+\,\tilde{\cK}_{1234}\left(\hat{\Xi}_1,\hat{\Xi}_2,\hat{\Xi}_3,\hat{\Xi}_4\right)\ ,\label{soluz}
\end{multline}
This form manifests the relation between $\cK_{1234}$, the current exchange part and simpler quantities like $\tilde{\cK}_{1234}$ that solve \emph{homogeneous} equations and entail the linearized gauge symmetries (tantamount to Ward identities) of the free system \eqref{on-shell}. In this form the Lagrangian coupling can be, in principle, both local and non-local and in the next sections we shall analyze its properties. Before going ahead, let us concentrate on four-point S-matrix amplitudes, that have been over the years a key obstruction to building a consistent theory of massless HS particles in flat space. The following section is meant to clarify the meaning of $\tilde{\cK}_{1234}$ and its relation to an S-matrix amplitude.

%%%%%%%%%%%%%%%%%%%%%%%%%%%%%%%%%%%%%%%%%%%%%%%%%%%%%%%%%%%%%%%

\scs{Four-point scattering amplitudes}\label{sec:four-point}

%%%%%%%%%%%%%%%%%%%%%%%%%%%%%%%%%%%%%%%%%%%%%%%%%%%%%%%%%%%%%%%

In this section we recover the Feynman rules from the Lagrangian \eqref{lagr}, computing the four-point S-matrix amplitudes and clarifying the role of the kernel $\tilde{\cK}_{1234}$ introduced in the previous section.
In order to recover the Feynman rules for the theory \eqref{lagr}, one should begin by choosing an appropriate gauge. In analogy with String Field Theory it is natural to choose the Feynman-Siegel gauge for the superfields $\Phi$ in \eqref{superfield}, that in our notation reads
\be
\partial_{\theta^0}\,\Phi(\Xi)\,=\,0\ ,
\ee
and sets to zero all components of $\Phi$ proportional to $\theta^0$ $(C\,=\,0)$. The propagator then simplifies, since the kinetic term takes the diagonal form
\be
\cL^{(2)}\,=\,\Phi\,\tilde{\star}\,(-p^{\,2})\Phi\ ,
\ee
and one obtains the propagator
\be
\cP\,=\,\frac{\tilde{\star}}{p^{\,2}\!\!}\ ,
\ee
that is nicely expressed in terms of the contraction $\tilde{\star}$ in \eqref{supercontra2}. The color-ordered Feynman rules for the cubic and quartic interactions associated to the color-ordered kernels so far considered can be recovered integrating the Lagrangian vertices over the $\theta^0$ coordinates, and read
\begin{align}
\cV^{(3)}_{123}\,&=\,\int d\theta^0_1 d\theta^0_2 d\theta^0_3\ \cK_{123}\left(\hat{\Xi}_1,\hat{\Xi}_2,\hat{\Xi}_3\right)\ \theta^0_1\,\theta^0_2\,\theta^0_3\ ,\\
\cV^{(4)}_{1234}\,&=\,-\,\int \prod_{i=1}^4 d\theta^0_i\nonumber\\&\qquad\times\left(\int d\theta^0_a\,\cK_{(12|a}\left(\hat{\Xi}_1,\hat{\Xi}_2,\hat{\Xi}_a\right)\,\theta^0_a\right)\frac{\tilde{\star}_a}{p_{\,a}^{\,2}}\left(\int d\theta^0_{\tilde{a}}\,\cK_{\tilde{a}|34)}\left(\hat{\Xi}_{\tilde{a}},\hat{\Xi}_3,\hat{\Xi}_4\right)\, \theta^0_{\tilde{a}}\right)\theta^0_1\,\theta^0_2\,\theta^0_3\,\theta^0_4\nonumber\\&+\,\int \prod_{i=1}^4 d\theta^0_i\tilde{\cK}_{1234}\left(\hat{\Xi}_1,\hat{\Xi}_2,\hat{\Xi}_3,\hat{\Xi}_4\right)\,\theta^0_1\,\theta^0_2\,\theta^0_3\,\theta^0_4\ .
\end{align}
One can now compute the HS four-point amplitudes, ending up with
\begin{multline}
\cA_4\,=\,\sum_{\s}\text{Tr}\left(\Phi_{1}(\Xi_1)\,\Phi_{\s(2)}(\Xi_{\s(2)})\,\Phi_{\s(3)}(\Xi_{\s(3)})\, \Phi_{4}(\Xi_{\s(4)})\right)\\\star_{\,1,2,3,4}\int \prod_{i=1}^4 d\theta^0_i\ \tilde{\cK}_{1\s(2)\s(3)\s(4)}\left(\hat{\Xi}_1,\hat{\Xi}_2,\hat{\Xi}_3,\hat{\Xi}_4\right)\, \theta^0_1\,\theta^0_2\,\theta^0_3\,\theta^0_4\ ,\label{A}
\end{multline}
where the trace is over the color indices and where only the contribution coming from $\tilde{\cK}_{1234}$ is still present, while the contribution given by the current exchanges is completely canceled by the first contribution to the quartic coupling. In some sense what we have recovered here is the equivalence between the Noether procedure at the Lagrangian level together with its links to a non-abelian deformation of the gauge symmetries, and the decoupling of unphysical states at the S-matrix level that is intrinsically related to the linearized part of the gauge symmetry. Moreover, this construction clarifies the role of $\tilde{\cK}_{1234}$ that, indeed, coincides on-shell, after a complete gauge fixing, with the four-point amplitude generating function. In the following we shall provide the general \emph{tree-level} form for $\tilde{\cK}_{1234}$, both in the framework of the on-shell gauge-fixed system \eqref{on-shell} and off-shell, extracting the \emph{Lagrangian} couplings and commenting on the issue of non-localities, also in relation to the content of Weinberg's theorem of \cite{NoGo}. Let us emphasize that we have split the quartic Lagrangian couplings into portions that are generically non-local, putting on more general grounds their form along lines that are actually in the spirit of \cite{NonLocalNoether}. From this point of view the quartic couplings are to be regarded as \emph{counterterms} canceling the non-vanishing linearized gauge variation of the \emph{current exchange amplitude} and can be in principle non-local as well, canceling in this case the portion of the current exchange whose gauge variation \emph{cannot} be compensated by local terms. $\cK_{1234}$ is indeed explicitly non-local if $\tilde{\cK}_{1234}$ does not contain \emph{all} current exchanges with the correct overall coefficients.

Having relaxed the standard \emph{locality} hypothesis as we have done so far we need to find some physical alternative that has to give a rationale for the possibly non-local answer. This is what we call \emph{minimal scheme}. It is defined via a number of constraints on the couplings, and hence on the cubic coupling function together with the spectrum of the theory, that have to satisfy altogether the following prescriptions:
\begin{itemize}
\item any particle propagating in some exchange gives rise to non-vanishing four-point amplitudes where it plays the role of an external state\footnote{With this constraint we ensure that the S-matrix be non trivial and that all current exchanges be built from states present in the spectrum constraining both the latter and the coupling functions of the theory.},
\item any quartic coupling does not contain portions that are identically gauge invariant under the linearized gauge variation\footnote{With this requirement we avoid for simplicity local quartic couplings that are proportional to the amplitude itself multiplied with Mandelstam variables in order to get a local object. This requirement is not so strict and can be eliminated in the most general case.},
\item all exchanges generated that admit a gauge invariant completion (possibly non-local) whose form \emph{does not} spoil the initial exchange, have to admit a \emph{local} gauge invariant completion to a \emph{non-vanishing} four-point scattering amplitude\footnote{This is equivalent to considering the maximal set of gauge invariant amplitudes enforcing a correspondence between residues and the propagating particles present in the spectrum. More in detail if a residue is present, its coefficient cannot be different from that of the corresponding current exchange.},
\item only those non-vanishing current exchanges that \emph{do not} admit a gauge invariant completion preserving the initial exchange have to be removed by \emph{non-local} quartic couplings\footnote{Non-localities should never touch the correspondence between residues and exchanges.}.
\end{itemize}
At the end only the current exchanges of the latter type, if any, will not contribute to the amplitudes while any residue present in a given scattering amplitude will be associated with one of the propagating degrees of freedom present in the spectrum. This iterative procedure fixes the non-vanishing entries of the cubic coupling function and in principle also some the relative coefficients between different cubic couplings and enforces constraints on the possible spectra playing the same job of locality in the lower-spin cases. Other possibilities are related to quartic local couplings that are gauge invariant under the linearized gauge variation. This types of couplings are proportional to $\tilde{\cK}_{1234}$ but contain a sufficient number of Mandelstam variables in the numerator in order to give rise to a local object. Any coupling of this type can give in principle a consistent local field theory with no cubic coupling but since the tensorial structure is the same of $\tilde{\cK}_{1234}$, these kinds of options are clearly encoded into choices of the relative functions of the Mandelstam variables that weight each different contribution to $\tilde{\cK}_{1234}$. In the following, we shall restrict our attention to the minimal scheme and pursuing this kind of program, that entails the usual Noether procedure, we are going to explore in which sense the usual notion of \emph{local field theory} may be overcome leaving though a systematic analysis of those constraints and of their solutions for the future\footnote{It can be useful to stress that for the lower-spin cases the solutions to the minimal scheme are all local.}. Moreover, we want to mention that some solutions that comply to the minimal scheme explicitly clash with many commonly accepted ideas about the structure of the S-matrix for massless particles. However, given the enormous difficulties present on this subject related to various assumptions whose origin cannot be proved in a rigorous sense starting from the usual causality and unitarity hypothesis \cite{SmatrixBook}, we make the choice of exploring the minimal setting that can give rise to non-trivial HS interactions trying to understand in which sense their non-triviality makes them different from their lower-spin counterparts and having always in mind their deformation to non-zero constant curvature backgrounds and to massive theories. We also leave for the future the important question of clarifying wheatear the minimal scheme here proposed is sufficient to imply global symmetries of higher spin, as it does for their lower-spin counterparts.

%%%%%%%%%%%%%%%%%%%%%%%%%%%%%%%%%%%%%%%%%%%%%%%

\scss{The Yang-Mills example}\label{sec:YMfour}

%%%%%%%%%%%%%%%%%%%%%%%%%%%%%%%%%%%%%%%%%%%%%%%

In the previous sections we have obtained the general form of the quartic coupling working in the framework of FDA and stressing the role of linearized gauge symmetries related to the free system. In order to make the construction explicit let us apply these techniques to the familiar case of Yang-Mills theory coupled to scalar fields in the adjoint representation. In order to study this problem, it is instructive to restrict momentarily  the attention to the on-shell gauge-fixed part of the kernel and to begin by finding a solution for $\tilde{\cK}_{1234}$ up to divergences and mass-shell conditions for the external states. In this framework the gauge-fixed part of the kernel $\tilde{\cK}_{1234}$ can be recovered solving the equations
\be
p_{\,i}\cdot\partial_{\xi_{\,i}}\ \tilde{\cK}_{1234}(p_{\,j},\xi_{\,j})\,\approx\,0\ ,\label{quarticS}
\ee
where $\approx$ means that the result is to hold on-shell for the external states and up to terms proportional to divergences. Actually, the physical meaning of \eqref{quarticS}, and more generally of \eqref{Consist3}, amounts to the decoupling condition for unphysical polarizations (Ward identities) at the level of the $S$-matrix, that has the form of the linearized gauge symmetry of the free system. We emphasize here the analogy between these constraints and those satisfied by the Kernels \eqref{G1} and \eqref{G2}. The only difference is that the solution of the decoupling condition is no more, in general, a Lagrangian coupling, simply because from the quartic order Lagrangian couplings do not coincide with S-matrix amplitudes but rather differ from them in the current exchange parts. What happens in the three-point case is actually an accident from this point of view, and it is actually simpler to solve homogeneous equations like \eqref{Consist3} rather than \eqref{Consist2}.
Following our strategy, one can start from the current exchange amplitude that can be constructed from the cubic Feynman rules encoded in \eqref{G}, and restricting the attention to a single color-ordered contribution, one is left with
\be
\cA^{(exch.)}_{1234}\,=\,-\,\frac{1}{s}\ \cG_{12a}\,\star_{\,a} \cG_{a34}\,-\,\frac{1}{u}\ \cG_{41a}\,\star_{\,a} \cG_{a23}\ ,\label{exch}
\ee
where only the $s$ and $u$ channels contribute, we use the bosonic $\star$ of eq.~\eqref{contra}, and the Mandelstam variables are
\be
s\,=\,-\,(p_{\,1}\,+\,p_{\,2})^2\ ,\quad t\,=\,-\,(p_{\,1}\,+\,p_{\,3})^2\ ,\quad u\,=\,-\,(p_{\,1}\,+\,p_{\,4})^2\ ,
\ee
with all ingoing momenta. This current exchange \eqref{exch} is not gauge invariant, and its linearized gauge variation reads
\begin{multline}
\d_4 \cA^{(c.e.)}_{1234}\,=\,p_{\,4}\cdot\partial_{\xi_4}\ \cA^{(c.e.)}_{1234}\\=\,-\,2\,\xi_2\cdot p_{\,4}\,+\,\xi_1\cdot p_{\,4}\,+\,\xi_3\cdot p_{\,4}\,-\,2\,\xi_1\cdot\xi_3\,\xi_2\cdot p_{\,4}\,+\,\xi_1\cdot\xi_2\,\xi_3\cdot p_{\,4}\,+\,\xi_2\cdot\xi_3\,\xi_1\cdot p_{\,4}\ ,\label{gaugevar}
\end{multline}
so that the whole point of the Noether procedure, as we have stressed, is to produce a \emph{local}, or possibly \emph{non-local}, counterterm whose linearized gauge variation cancels this contribution. The totally cyclic counterterm can be worked out relatively easily in this case: it is \emph{local}, and is given by
\begin{multline}
\cV^{\text{YM}}_{\,1234}\,=\,2\,\xi_1\cdot\xi_3\,\xi_2\cdot\xi_4\,- \,\xi_1\cdot\xi_4\,\xi_2\cdot\xi_3\,-\,\xi_1\cdot\xi_2\,\xi_3\cdot\xi_4\\+\,2\,\left(\xi_1\cdot\xi_3\,+\,\xi_2\cdot\xi_4\right)\,- \,\xi_1\cdot\xi_2\,-\,\xi_2\cdot\xi_3\,-\,\xi_3\cdot\xi_4\,-\,\xi_4\cdot\xi_1\ ,\label{cV}
\end{multline}
so that it coincides precisely with the corresponding color-ordered contribution to the Yang-Mills quartic coupling generating function that can be deduced from the Yang-Mills Lagrangian coupled to scalar fields.

Returning to our discussion, we would like to reiterate that we have recovered the Yang-Mills quartic Lagrangian generating function $\cV^{\,\text{YM}}_{1234}$ following the strategy outlined in the previous section and imposing the decoupling condition of unphysical degrees of freedom at the level of the amplitude, that coincides with the linearized gauge invariance of the free system. One then recovers the full generating function of Yang-Mills four-point amplitudes in color-ordered form, given by
\be
\cG_{1234}(p_{\,i},\xi_{\,i})\,=\,-\,\frac{1}{s}\ \cG_{12a}\star_{\,a} \cG_{a34}\,-\,\frac{1}{u}\ \cG_{41a}\star_{\,a} \cG_{a23}\,+\,\cV^{\,\text{YM}}_{1234}\ ,\label{G1234}
\ee
that is actually a solution for $\tilde{\cK}_{1234}(p_{\,i},\xi_{\,i})$. In a similar fashion, any Lagrangian vertex can be recovered as a counterterm that guarantees the gauge invariance property of the corresponding amplitude while, as pointed out in the previous section, the key physical content of the Noether procedure is to produce amplitudes that decouple unphysical degrees of freedom and  encode in principle consistent non-abelian deformations of the gauge symmetry. Moreover, we want to underline, as also stressed in \cite{NonLocalNoether}, that from Noether procedure alone there is no general argument forcing to choose local counterterms for $\cV_{1234}$. It is then interesting at least in principle to analyze the most general quartic coupling, studying also non-local solutions in this simple toy model. In this case, for instance, we can conceive to define the kernel as
\be
\tilde{\cK}_{1234}\,=\,\l\,\cG_{1234}\ ,\label{ampl}
\ee
with $\l$ an overall coefficient that does not affect the defining property of eq.~\eqref{quarticS}. This choice would led to a \emph{non-local} quartic Lagrangian coupling of the form
\be
\cV^{\,(4)}_{1234}(p_{\,i},\xi_{\,i})\,=\,-\,\frac{\l\,-\,1}{s}\ \cG_{12a}\star_{\,a} \cG_{a34}\,-\,\frac{\l\,-\,1}{u}\ \cG_{41a}\star_{\,a} \cG_{a23}\,+\,\l\,\cV^{\,\text{YM}}_{1234}\ ,\label{G1234non}
\ee
where we have subtracted the current exchange contribution in eq.~\eqref{exch}. Similarly, one could have also started from a more general current exchange of the form
\be
\cA^{(exch.)}_{1234}\,=\,-\,\frac{\a}{s}\ \cG_{12a}\,\star_{\,a} \cG_{a34}\,-\,\frac{\b}{u}\ \cG_{41a}\,\star_{\,a} \cG_{a23}\ ,\label{exch}
\ee
weighted by different constants $\a$ and $\b$, that can be interpreted as parameterizing a violation of the Jacobi identity, and yielding to the \emph{non-local} Lagrangian quartic coupling
\be
\cV^{\,(4)}_{1234}(p_{\,i},\xi_{\,i})\,=\,-\,\frac{\l\,-\,\a}{s}\ \cG_{12a}\star_{\,a} \cG_{a34}\,-\,\frac{\l\,-\,\b}{u}\ \cG_{41a}\star_{\,a} \cG_{a23}\,+\,\l\,\cV^{\,\text{YM}}_{1234}\ .\label{G1234non2}
\ee
The meaning, if any, of these class of solutions, that manifest themselves in this setting creates a sort of \emph{ambiguity} related to the various choices for the parameters $\l$, $\a$ and $\b$, so much so that if one wants to relax the \emph{locality} constraint one clearly needs to replace it with something else. Our observations move from the fact that whatever the choice for these coefficients the amplitude that one recovers at the end is always given by eq.~\eqref{ampl}, whose residues have a fixed form matching the current exchange part only if $\l\,=\,\a\,=\,\b$. Hence, we are led again to the conclusion that the only choice leading to a physically meaningful setting is exactly $\a\,=\,\b\,=\,\l$, in which the current exchange contributions extracted from \eqref{ampl} can be \emph{entirely} related to the cubic part of the theory via the current exchanges, as the \emph{minimal scheme} requires, while all other choices violate it\footnote{Whenever a finite number of degrees of freedom were to contribute to the residue, arbitrary choices of $\a$, $\b$ and $\l$ would clearly result in violations of \emph{tree-level} unitarity, since the residue would not match the current exchange contribution of eq.~\eqref{exch} that takes into account the correct degrees of freedom that ought to be propagating. The situation may well be different if an infinite number of degrees of freedom contributes to the same residue.}.
An opposite situation presents itself when the theory possesses cubic couplings leading to current exchanges whose violation of gauge invariance leads to amplitudes that \emph{cannot} factorize on the initial exchanges. In such cases the quartic coupling becomes \emph{intrinsically} non-local, as we shall see in the next sections\footnote{Notice that in the YM case just presented as well as in all standard lower-spin examples, including classical gravity, the minimal scheme implies locality.}, and either the cubic couplings cannot be pursued to a consistent theory because non-localities create further inconsistencies or one needs to resort to the minimal scheme or to other similar frameworks.

The full amplitude generating function is then obtained summing over all color orderings as
\begin{multline}
\cA(\Phi_1,\Phi_2,\Phi_3,\Phi_4)\,=\,\sum_\s\text{Tr}\Big[\Phi_1(\xi_1)\,\Phi_{\s(2)}(\xi_{\s(2)})\, \Phi_{\s(3)}(\xi_{\s(3)})\, \Phi_{\s(4)}(\xi_{\s(4)})\Big]\\\star_{\,1234} \cG_{1\s(2)\s(3)\s(4)}(p_{\,i},\xi_{\,i})\ ,\label{SYangMills}
\end{multline}
where we consider a matrix valued generating function
\be
\Phi(p,\xi)\,=\,\phi(p)\,+\,A(p)\cdot\xi\ ,
\ee
encoding both the scalar wave-function and the polarization tensor of gauge bosons. The sum is over all permutations of three elements, in order to recover the usual group theoretical factors
\be
f^{abe}f^{cde}\,\sim\,\text{Tr}\left([T^a, T^b][T^c,T^d]\right)\ ,
\ee
together with a sum over the non-cyclic permutations of the external legs. Moreover, we have also expressed the amplitude in terms of the bosonic contraction \eqref{contra} acting on the $\xi_{i}$'s. It is interesting to observe that the kernel $\cG_{1234}$ satisfies some simple relations like
\be
\cG_{1234}\,+\,\cG_{2134}\,+\,\cG_{2314}\,=\,0\ ,\qquad\cG_{1234}\,=\,\cG_{4321}\ ,\label{ide}
\ee
that, together with the cyclicity in the external legs, leave two independent objects, say for instance
\be
\cG_{1234}\ ,\qquad\cG_{1243}\ .
\ee
This is to be confronted with the case of three-point functions, where a single independent ordering of the external legs
\be
\cG_{123}\ ,
\ee
was available. Analyzing more in detail what we have gained, as for three-point amplitudes, one can look more closely at the various contributions to $\cG_{1234}$, distinguishing them by their order in the symbols $\xi_i$'s. In this case, for each of the independent terms $\cG_{1234}$ and $\cG_{1243}$, one can extract three different contributions of order $0$, $2$ and $4$ in the symbols, given by
\begin{align}
a_{-1}(s,t,u)\,\cG_{1234}^{(-1)}(p_{\,i})\,=\,-\,&\left(2\,t\,-\,2\,\frac{s\,u}{t}\right)\,\cG_{1234}^{(-1)}(p_{\,i})\,=\,\cG_{1234}(p_{\,i},\l\,\xi_i)\Big|_{\l\,=\,0}\,=\,\frac{t\,-\,u}{s} \,+\,\frac{t\,-\,s}{u}\ ,\\
\cG^{(1)}_{1234}(p_{\,i},\xi_i)\,&=\,\left(\frac{d}{d\l}\right)^{\,2}\ \left[ \cG_{1234}(p_{\,i},\l\,\xi_i)\right]\Big|_{\l\,=\,0}\ ,\\
\cG^{(2)}_{1234}(p_{\,i},\xi_i)\,&=\,\left(\frac{d}{d\l}\right)^{\,4}\ \left[ \cG_{1234}(p_{\,i},\l\,\xi_i)\right]\Big|_{\l\,=\,0}\ .
\end{align}
The first is a function of the Mandelstam variables that is related to the four-scalar amplitude. Here, by convention we have defined $\cG^{(-1)}_{1234}$ as the factorized contribution with a scalar exchange, encoding in the function
\be
a_{-1}(s,t,u)\,=\,-\,\left(2\,t\,-\,2\,\frac{s\,u}{t}\right)\ ,
\ee
the residue of the spin-$1$ exchange. On the other hand, the other two $\cG_{1234}^{(i)}$'s are related, respectively, to the two scalar -- two gauge boson amplitude and to the four gauge boson amplitude, whose structure has been recovered here enforcing \emph{linearized} gauge invariance.

To summarize, we have recovered the analogs of the three-point $\cG_{123}^{(0)}$ and $\cG_{123}^{(1)}$. They are:
\be
\cG^{(1)}_{1234}(p_{\,i},\xi_i)\ ,\quad \cG^{(2)}_{1234}(p_{\,i},\xi_i)\ ,\quad \cG^{(1)}_{1243}(p_{\,i},\xi_i)\ ,\quad \cG^{(2)}_{1243}(p_{\,i},\xi_i)\ ,
\ee
and can be related to the tree-level S-matrix amplitudes
\begin{align}
\cA(\phi_{\,1},\phi_{\,2},A_3,A_4)\,&=\,\sum_\s \text{Tr}\Big[\phi_{\,1}\,\phi_{\,\s(2)}\,A_{\s(3)}\cdot\xi_{\s(3)}\, A_{\s(4)}\cdot\xi_{\s(4)}\Big]\,\star_{\,1234} \cG^{(1)}_{1\s(2)\s(3)\s(4)}(p_{\,i},\xi_{\,i})\ ,\\
\cA(A_1,A_2,A_3,A_4)\,&=\,\sum_\s \text{Tr}\Big[A_1\cdot\xi_1\,A_{\s(2)}\cdot\xi_{\s(2)}\,A_{\s(3)}\cdot\xi_{\s(3)}\, A_{\s(4)}\cdot\xi_{\s(4)}\Big]\nonumber\\&\qquad\qquad\qquad\qquad\qquad\qquad\qquad\qquad\star_{\,1234} \cG^{(2)}_{1\s(2)\s(3)\s(4)}(p_{\,i},\xi_{\,i})\ .
\end{align}
The role of the overall constant factor, that at the cubic level corresponds to the three-scalar coupling, is played here by the basic building block of a four-scalar amplitude given by
\be
\cG^{(-1)}_{1234}(p_{\,i})\,=\,-\,\frac{1}{s}\,-\,\frac{1}{u}\,=\,\frac{t}{s\,u}\ .\label{cG-1}
\ee
The other $\cG^{(i)}_{1234}$'s are on the contrary color-ordered amplitudes for the processes involving two or four gauge bosons. Explicitly
\begin{multline}
\cG^{(1)}_{1234}(p_{\,i},\xi_{i})\,=\,-\,\left[\frac{1}{s}\,\left( \cG^{(0)}_{12a}\,\star_{a}\cG^{(1)}_{a34}\,+\, \cG^{(1)}_{12a}\,\star_{a}\cG^{(0)}_{a34}\,+\,\cG^{(0)}_{12a}\,\cG^{(0)}_{a34}\Big|_{\xi_a\,=\,0}\right)\right.\\\left.
+\,\frac{1}{u}\,\left( \cG^{(0)}_{41a}\,\star_{a}\cG^{(1)}_{23a}\,+\, \cG^{(1)}_{41a}\,\star_{a}\cG^{(0)}_{23a}\,+\,\cG^{(0)}_{12a}\,\cG^{(0)}_{a34}\Big|_{\xi_a\,=\,0}\right)\right]\\ +\,2\,\left(\xi_1\cdot\xi_3\,+\,\xi_2\cdot\xi_4\right)\,- \,\xi_1\cdot\xi_2\,-\,\xi_2\cdot\xi_3\,-\,\xi_3\cdot\xi_4\,-\,\xi_4\cdot\xi_1\ ,\label{G01}
\end{multline}
and
\begin{multline}
\cG^{(2)}_{1234}(p_{\,i},\xi_{i})\,=\,-\,\left[\frac{1}{s}\  \cG^{(1)}_{12a}\,\star_{a}\cG^{(1)}_{a34}\,+\,\frac{1}{u}\ \cG^{(1)}_{41a}\,\star_{a}\cG^{(1)}_{23a}\right]\\\,+\,2\,\xi_1\cdot\xi_3\,\xi_2\cdot\xi_4\,- \,\xi_1\cdot\xi_4\,\xi_2\cdot\xi_3\,-\,\xi_1\cdot\xi_2\,\xi_3\cdot\xi_4\ .\label{G02}
\end{multline}
The Kernels $\cG_{1234}^{(i)}$ that we have found here can be used to find the most general solution for a kernel $\tilde{\cK}_{1234}$ satisfying \eqref{quarticS} in a theory with scalars and gauge bosons that is compatible with the cubic couplings in fig.~\ref{fig:cubicboson}. The corresponding solution reads
\be
\tilde{\cK}_{1234}\,=\,\ a_{-1}(s,t,u)\,\cG^{(-1)}_{1234}\,+\,a_0(s,t,u)\,\cG^{(0)}_{1234}\,+\,a_1(s,t,u)\,\cG^{(1)}_{1234}\,+\,a_2(s,t,u)\,\cG^{(2)}_{1234} ,
\ee
where the $a_i(s,t,u)$ are functions of the Mandelstam variables that do not introduce further physical pole with respect to the $\cG_{1234}^{(i)}$'s, laying the freedom left by Noether procedure in building a consistent theory. They encode the residues of the various processes as well as further local quartic couplings that are gauge invariant under the linearized gauge symmetry. In this case, by consistency, $\tilde{\cK}_{1234}$ does not contain exchanges with spin greater than one, while the $\cG^{(i)}_{1234}$'s are defined in eqs.~\eqref{cG-1}, \eqref{G01} and \eqref{G02} and we can see the reason why we have left a free slot for $\cG_{1234}^{(0)}$, since we can consider a further contribution linear in the symbols $\xi_i$ and defined as
\begin{multline}
\cG^{(0)}_{1234}\,=\,-\,\frac{1}{s}\,\left(\xi_{1}\cdot p_{\,2a}\,+\,\xi_2\cdot p_{\,a1}\,+\,\xi_3\cdot p_{\,4\tilde{a}}\,+\,\xi_4\cdot p_{\,\tilde{a}3}\right)\\-\,\frac{1}{u}\,\left(\xi_{1}\cdot p_{\,b4}\,+\,\xi_2\cdot p_{\,3\tilde{b}}\,+\,\xi_3\cdot p_{\,\tilde{b}2}\,+\,\xi_4\cdot p_{\,1b}\right)\ ,
\end{multline}
where by convention
\be
p_{\,a}\,=\,-\,p_{\,1}\,-\,p_{\,2}\ ,\qquad p_{\,\tilde{a}}\,=\,-\,p_{\,a}\ ,\qquad p_{\,b}\,=\,-\,p_{\,1}\,-\,p_{\,4}\ ,\qquad p_{\,\tilde{b}}\,=\,-\,p_{\,b}\ ,
\ee
so that the Yang-Mills plus scalar example is finally recovered with the choice
\be
\tilde{\cK}^{\text{YM}}_{1234}\,=\,-\,\left(2\,t\,-\,\frac{s\,u}{t}\right)\,\cG^{(-1)}_{1234}\,+\,\cG^{(1)}_{1234}\,+\,\cG^{(2)}_{1234}\ .
\ee
Other non-standard examples related to a theory with gauge bosons and scalars can arise whenever one takes into account the corresponding quartic amplitude that can be extracted from the kernel
\be
\left[\cG_{1234}^{\,(1)}\right]^{\,2}\ .
\ee
In this case one can recover the local quartic couplings that are linked to the highest-derivative cubic coupling involving two or three gauge bosons\footnote{The analogous term of the form $\left[\cG_{1234}^{\,(0)}\right]^{\,4}$ cannot be considered here, since it gives rise to amplitudes that propagate HS fields.} that in generating function form read
\be
\xi_1\cdot p_{\,23}\,\xi_2\cdot p_{31}\,+\,\text{cyclic}\ ,\qquad \xi_1\cdot p_{\,23}\,\xi_2\cdot p_{31}\,\xi_3\cdot p_{\,12}\ .\label{higher}
\ee
In the following we shall keep always in mind that these amplitudes arise as particular combination of powers of the building blocks so far considered. In general one could also choose, admitting some redundancy in the description, to consider generating functions of these derived amplitudes together with the previous ones.

To conclude this section, we want to emphasize the role of the scattering amplitudes in comparison with the Lagrangian couplings that we have recovered in \eqref{soluz}. As we have seen, it is possible to work directly at the amplitude level, where the gauge symmetry is realized linearly, extracting the quartic couplings as counterterms needed in order to guarantee the linearized gauge invariance. We want to emphasize here that, although the decoupling condition for the unphysical polarizations does not fix the relative functional coefficients between $G^{\,(0)}_{1234}$, $G^{\,(1)}_{1234}$ and $G^{\,(2)}_{1234}$, these can be completely fixed requiring either locality or that all current exchanges contained in $\tilde{\cK}_{1234}$ match the corresponding ones built from the cubic couplings and viceversa whenever possible\footnote{We are referring here, in particular, to our minimal scheme and to the portion of the correlation function related to the current exchanges, leaving aside any local quartic coupling that is gauge invariant under the linearized gauge transformations and is hence proportional to the amplitude itself.}. In the following we will push forward these observations, generalizing the results to HS gauge fields, with special attention to the nature of the four-point Lagrangian couplings that for gauge bosons can be local, but in this approach result explicitly from subtractions between different non-local objects and can be, in general, non-local as well.

%%%%%%%%%%%%%%%%%%%%%%%%%%%%%%%%%%%%%%%%%%%%%%%

\scss{The HS case}

%%%%%%%%%%%%%%%%%%%%%%%%%%%%%%%%%%%%%%%%%%%%%%%

In this section we are going to consider the general case of HS four-point couplings, extending the ideas of the previous section. In order to arrive at a systematic description of HS four-point amplitudes, we proceed as before, considering the gauge-fixed theory of eq.~\eqref{on-shell} in which all fields are transverse and carry, for simplicity, an \emph{irreducible} representation of the Lorentz group. At the end, we shall comment on the Lagrangian couplings that arise after subtracting the current exchange portions. In order to achieve this goal we need to exhibit the general kernel $\tilde{\cK}_{1234}$ satisfying
\be
(Q_1\,+\,Q_2\,+\,Q_3\,+\,Q_4)\,\tilde{\cK}_{1234}\,=\,0\ ,
\ee
that is equivalent on-shell to the simpler condition
\be
p_{\,i}\cdot\partial_{\xi_{\,i}}\ \tilde{\cK}_{1234}(p_{j},\xi_{j})\,\approx\,0\ .\label{decoupling}
\ee
Actually, one can construct a general ansatz for a solution to eq.~\eqref{decoupling} starting from the results obtained in the previous section and generalizing what happens in the three-point case where, as we have discussed, the HS couplings are related to the gauge-boson ones. Focusing on a \emph{generating function}, it suffices to exponentiate the kernels obtained in the previous section, so that a solution to eq.~\eqref{decoupling} can be given by\footnote{Notice that taking into account the total number of Lorentz invariant quantities available at this order and counting the number of constraints put by gauge invariance one can see that the solution in eq.~\eqref{Opengenfunc} is constructed in terms of a sufficient number of color ordered independent building blocks. In general, considering all possible four-point gauge-boson amplitudes one would end up with a redundant but possibly more transparent description.}
\be
\tilde{\cK}_{1234}(p_{\,j},\xi_{\,j})\,=\,-\,\frac{1}{su}\ \exp\left[-\,su\,\left(\cG^{(0)}_{1234}\,+\,\cG^{(1)}_{1234}\,+\,\cG^{(2)}_{1234}\right)\right]\ .\label{Opengenfunc}
\ee
In general, admitting some redundancy as anticipated in the previous section, we could also add to the exponent all other planar four-point amplitudes $\cG^{(n>2)}_{1234}$ that can be built from the higher derivative cubic couplings involving two or three gauge bosons in eq.~\eqref{higher}.
We could call this result with a little abuse of language, of the \emph{open-string}-type since it is \emph{planar} and hence is naturally associated to Chan-Paton factors \cite{Paton:1969je}. Moreover, $\tilde{\cK}_{1234}$ should be considered modulo arbitrary relative functions of the Mandelstam variables, that are not constrained by eq.~\eqref{decoupling} and play the role of relative weights between the various totally cyclic terms in the expansion of \eqref{Opengenfunc}, while we have considered a fixed ordering $1234$ of the external legs so that only the $\cG^{(i)}_{1234}$'s enter and contribute to the correct channels reproducing the HS exchanges. The full amplitude, that we can call again of the \emph{open-string}-type, is recovered as usual by
\begin{multline}
\cA(\Phi_1,\Phi_2,\Phi_3,\Phi_4)\,=\,\sum_\s\text{Tr}\Big[\Phi_1(\xi_1)\,\Phi_{\s(2)}(\xi_{\s(2)})\, \Phi_{\s(3)}(\xi_{\s(3)})\, \Phi_{\s(4)}(\xi_{\s(4)})\Big]\\\star_{\,1234} \tilde{\cK}_{1\s(2)\s(3)\s(4)}(p_{\,i},\xi_{\,i})\ ,\label{Shs}
\end{multline}
where now $\Phi_i(\xi_i)$ is an arbitrary matrix valued generating function containing all totally symmetric HS polarization tensors, while the trace is over the color indices. We clearly recover the results of the previous section as soon as we restrict the attention to the linear part in the $\cG^{(i)}_{1234}$'s. Going ahead, we have chosen a dependence as
\be
su\,\cG^{(i)}_{1234}(p_{\,i},\xi_{\,i})\ ,
\ee
multiplying with $(su)$ the kernels, in order to avoid higher-order poles as soon as one considers HS fields. For instance, the form of a four-point scattering amplitude of the open-string-type in the case of four spin-$2$ fields becomes here
\be
\tilde{\cK}_{1234}\,\sim\,-\,\frac{1}{su}\ \sum_{\a\,+\,2\b\,+\,4\g\,=\,8}\ a_{\a,\b,\g}(s,t,u)\left[-\,su\,\cG^{(0)}_{1234}\right]^\a\left[-\,su\,\cG^{(1)}_{1234}\right]^\b \left[-\,su\,\cG^{(2)}_{1234}\right]^\g \ ,\label{OFourspin2}
\ee
where the $a_{\a,\b,\g}(s,t,u)$'s are some functions that do not introduce additional \emph{physical} poles in the Mandelstam variables and that are to be fixed, in our minimal scheme, confronting them with the corresponding current exchange amplitudes, and hence relating them to the cubic coupling function, whose arbitrariness is in turn constrained by the minimal scheme itself.
The on-shell form of the quartic coupling generating function, or on-shell trilinear product kernel, can be now extracted exploiting the on-shell version of eq.~\eqref{soluz} that reads here
\be
\cK_{1234}\,=\,-\,\cK_{(12|a} \star_a\,\frac{\cP(\xi_a,\xi_{\tilde{a}})}{p_{\,a}^{\,2}}\,\star_{\tilde{a}} \,\cK_{\tilde{a}|34)}\,+\, \tilde{\cK}_{1234}\ ,\label{V4onshell}
\ee
where $\cP(\xi_a,\xi_{\tilde{a}})$ is the propagator numerator, all kernels are considered to be on-shell and, in order to be explicit,
\be
\tilde{\cK}_{1234}\,=\,-\,\frac{1}{su}\ \sum_{\a,\,\b,\,\g}\ a_{\a,\b,\g}(s,t,u)\left[-\,su\,\cG^{(0)}_{1234}\right]^\a\left[-\,su\,\cG^{(1)}_{1234}\right]^\b \left[-\,su\,\cG^{(2)}_{1234}\right]^\g \ .\label{OFourspingenfunc}
\ee
This form, as in the spin-$1$ case, encodes in principle also non-minimal choices, which reflect some freedom left by Noether procedure, related to local quartic couplings that are gauge invariant under the linearized gauge variation and whose tensorial structure is exactly as in the amplitude, but multiplied by a sufficient number of Mandelstam variables that suffice to eliminate all poles. Moreover, although \eqref{OFourspin2} is a generic planar color-ordered expression consistent with gauge invariance that one can write for four spin-$2$ fields it does not exhaust all the possibilities, as it was the case for the cubic couplings. Indeed, we have at our disposal two independent kernels $\cG_{1234}$ and $\cG_{1243}$ and another available option is to combine them together using $\cG_{1243}$ in place of the color factor. One ends up, in this way, with the following type of \emph{derived} kernel
\be
\cK(\xi_{\,i}\,,\xi^{\,\prime}_{\,i})\,=\,\left(\sum_\s\ \tilde{\cK}_{1\s(2)\s(3)\s(4)}(p_{\,i},\xi_{\,i})\ \tilde{\cK}_{1\s(2)\s(4)\s(3)}(p_{\,i},\xi^{\,\prime}_{\,i})\right)\ ,\label{Shs3}
\ee
whose contributions of the form
\be
\left[\cG_{1243}\right]^{\,\a}\,\left[\cG_{1243}\right]^{\,\b}\ ,
\ee
with neither $\a\,=\,0$ nor $\b\,=\,0$, could be called, with a little abuse of language, \emph{closed-string}-like amplitudes\footnote{To be precise, we can call in this way \emph{only} the contributions with $\a\,=\,\b$. The other contributions, that show up starting from spin-$3$, do not satisfy the analog of \emph{level matching} but are not ruled out here by gauge invariance. We cannot exclude at this stage that they are not ruled out by other arguments, but we leave a more detailed analysis of these potentially interesting options for the future.}. Here we have literally replaced the Chan-Paton contribution in eq.~\eqref{Shs} with the kernel $\tilde{\cK}_{1243}$, again defined in eq.~\eqref{Opengenfunc}, so that for tree-level scattering amplitudes involving totally symmetric fields, one recovers the generating function
\begin{multline}
\cA(\xi_1,\xi_2,\xi_3,\xi_4)\,=\,\frac{1}{stu}\ \sum_\s\ e^{-\,s_{1\s(2)}s_{1\s(3)}\,\left(\cG^{(0)}_{1\s(2)\s(4)\s(3)}(\xi_{\,i})\,+\,\cG^{(1)}_{1\s(2)\s(4)\s(3)}(\xi_{\,i}) \,+\,\cG^{(2)}_{1\s(2)\s(4)\s(3)}(\xi_{\,i})\right)} \\ \times e^{-\,s_{1\s(2)}s_{1\s(4)}\,\left(\cG^{(0)}_{1\s(2)\s(3)\s(4)}(\xi_{\,i})\,+\,\cG^{(1)}_{1\s(2)\s(3)\s(4)}(\xi_{\,i}) \,+\,\cG^{(2)}_{1\s(2)\s(3)\s(4)}(\xi_{\,i})\right)}\ .\label{Closedgenfunc}
\end{multline}
Here by definition
\be
s_{ij}\,=\,-\,(p_{\,i}\,+\,p_{\,j})^2\ ,
\ee
the sum is over all permutation of the three elements $\{234\}$ and one has, again, the freedom to multiply each totally cyclic gauge-invariant term in the expansion of \eqref{Closedgenfunc} with arbitrary relative functions $a_i(s,t,u)$ of the Mandelstam variables that give rise to amplitude with single poles at most (see e.g. eq.~\eqref{OFourspingenfunc}). Moreover:
\begin{itemize}
 \item one can in principle constrain these functions relating any current exchange contribution belonging to $\tilde{\cK}$ to the corresponding contribution obtained from the cubic couplings of the theory via the minimal scheme,
 \item the only \emph{non-local} contributions to the quartic Lagrangian coupling are those related to the exchanged particles that cannot be made gauge invariant with the addition of a local counterterms and that, for this reason, can never belong to $\tilde{\cK}$.
\end{itemize}
We leave a more detailed analysis of these issues related to \emph{non-local} field theories for the future, trying to understand their eventual \emph{geometric} rationale in which sense, if any, they can be consistent with unitarity, even though they clash with commonly accepted ideas about the structure of S-matrix poles like factorization.

Before going on with our discussion, it can be of interest to comment more in details on the nature of the couplings that we have obtained for spin-$2$ external particles as a toy model of more general cases, extracting the current exchange part and identifying the cubic couplings involved. It is also important to discuss the difference between the \emph{open-string}-like couplings of eq.~\eqref{Opengenfunc} and the \emph{closed-string}-like ones of eq.~\eqref{Closedgenfunc}. Let us begin considering the coupling in eq.~\eqref{Closedgenfunc} associated to
\be
\cA (\xi_1,\xi_2,\xi_3,\xi_4)\,=\,\ldots\,+\,\frac{1}{stu}\ \sum_\s \,s^{\,2}_{1\s(2)}s_{1\s(3)}s_{1\s(4)}\,\cG^{(2)}_{1\s(2)\s(4)\s(3)}\,\cG^{(2)}_{1\s(2)\s(3)\s(4)}\,+\,\ldots\ ,
\ee
and contributing to the four spin-$2$ scattering amplitude. Explicitly this contribution is given by
\begin{multline}
\cA(\xi_1,\xi_2,\xi_3,\xi_4)\,=\,-\,\sum_\s\left[s\left(\frac{1}{s}\ \cG^{\,(1)}_{12a}\star_{\,a} \cG^{\,(1)}_{a34}\,+\,\frac{1}{u}\ \cG_{41a}^{\,(1)}\star_{\,a} \cG_{a23}^{\,(1)}\,-\,\cV^{\,YM}_{1234}\right)\right.\\\left.\times\,\left(\frac{1}{s}\ \cG_{12a}^{\,(1)}\star_{\,a} \cG_{a34}^{\,(1)}\,+\,\frac{1}{t}\ \cG_{13a}^{\,(1)}\star_{\,a} \cG_{a24}^{\,(1)}\,+\,\cV^{\,YM}_{1243}\right)\right]\ ,\label{Grav}
\end{multline}
so that one recovers, as expected, a \emph{non-planar} amplitude and the various contributions conspire after some algebra to yield
\be
\cA_{1234}\,=\,-\frac{1}{s}\ \left(\cG_{12a}^{\,(1)}\right)^{\,2}\star_{\,a} \left(\cG_{a34}^{\,(1)}\right)^{2}\,-\frac{1}{t}\ \left(\cG_{13a}^{\,(1)}\right)^{\,2}\star_{\,a} \left(\cG_{a42}^{\,(1)}\right)^{2}\,-\frac{1}{u}\ \left(\cG_{14a}^{\,(1)}\right)^{\,2}\star_{\,a} \left(\cG_{a23}^{\,(1)}\right)^{2}\,+\,\ldots\ ,
\ee
where the ellipses represent local terms and where the current exchange amplitude have been completely reconstructed so that one can manifestly observe a four-point function involving the minimal coupling of two spin-$2$ fields with a propagating spin-$2$, since the number of derivatives entering the current exchange is precisely $4$. This result resonates with the fact that this particular four-point function \eqref{Grav} is exactly the standard ``four-graviton'' four-point function, written in a form analogous to that obtained in the field theory limit in \cite{KLT}. On the contrary, let us now consider the \emph{open-string}-like amplitude
\be
\cA_{1234}\,=\,-\,\frac{1}{su}\,\left(-\,s\,u\,G^{\,(2)}_{1234}\right)^2\ ,
\ee
that can be recovered from eq.~\eqref{OFourspin2}. In this case we see a different structure that is given explicitly by
\be
\cA_{1234}\,=\,-\,s\,u\left(\frac{1}{s}\ \cG_{12a}^{\,(1)}\star_{\,a} \cG_{a34}^{\,(1)}\,+\,\frac{1}{u}\ \cG_{41a}^{\,(1)}\star_{\,a} \cG_{a23}^{\,(1)}\,-\,\cV^{\,YM}_{1234}\right)^2\ ,
\ee
so that, extracting the pole part in order to read off the current exchange contribution, one recovers
\be
\cA_{1234}\,=\,-\,\frac{u}{s}\ \left(\cG_{12a}^{\,(1)}\right)^{\,2}\star_{\,a} \left(\cG_{a34}^{\,(1)}\right)^{\,2}\,-\,\frac{s}{u}\ \left(\cG_{41a}^{\,(1)}\right)^{\,2}\star_{\,a} \left(\cG_{a23}^{\,(1)}\right)^{\,2}\,+\,\ldots\ .
\ee
Here as before the ellipsis represent \emph{local} terms while, using functions $a_i(s,t,u)$ with no unphysical pole, one can only increase the power of the additional Mandelstam variables in the numerator that are actually necessary in order to guarantee both the absence of unphysical poles and the decoupling of transverse \emph{unphysical} polarizations. This translates into the fact that the coupling in which the current exchange factorizes this time involves the exchange of a spin-$3$ excitation and is of the form
\be
\cV_3\,\sim\,\left[\,\cG^{\,(1)}_{ijk}\right]^{\,2}\left[\,\cG^{\,(0)}_{ijk}\right]\ ,
\ee
as one can see looking at the residue, that is of order six in the momenta. In principle, one could go ahead, considering higher powers of the Mandelstam variables that are associated with HS exchanges building a full overall function $a(s,t,u)$ that does not introduce additional \emph{physical poles}. For instance, one possibility could be the following gauge-invariant amplitude
\be
\cA_{1234}\,=\,-\,\frac{1}{s\,u}\,e^{\,-t}\left(\vphantom{\frac{1}{2}}u\ \cG_{12a}\star_{\,a} \cG_{a34}\,+\,s\ \cG_{41a}\star_{\,a} \cG_{a23}\,-\,s\,u\,\cV^{\,\text{YM}}_{1234}\right)^2\,+\,\ldots\ ,\label{fullampl}
\ee
where the exponential of $t$ accounts for an infinite number of exchanges, or more complicated examples related to the results in \cite{Euihun,cubicstring}, while the ellipsis stand for terms containing also powers of $\cG^{(1)}_{1234}$, as in eqs.~\eqref{Opengenfunc} and \eqref{OFourspin2}, whose relative functions can be fixed employing the minimal scheme. In principle, this kind of structure is needed if one wants to construct a consistent quartic amplitude that factorizes into an infinite number of exchanges. However, let us stress that here only spins higher than two can propagate, even if a coupling to lower-spin fields does in principle exist and in contrast to the previous case where the exchange of a spin-$2$ field was present. The counterpart of this peculiar aspect turns out to be non-localities at the Lagrangian level, as expected from the result of \cite{coloredspin2}, since some of the corresponding lower-spin exchanges \emph{do not} admit any local gauge-invariant completion. Moreover, this means that:
\begin{itemize}
\item a massless colored spin-$2$ field can have a charge of spin strictly higher than $2$,
\item a full theory producing such four-point functions has to contain an infinite tower of HS fields.
\end{itemize}
The latter can be inferred since, by requiring that any propagating HS particle brings non-trivial interactions, as soon as a spin-$3$, say, propagates one can look at processes with also spin-$3$ external particles recovering again a propagating particle of higher spin and so on. Other possibilities are then available, since one can in principle consider also powers of $\cG^{(0)}_{1234}$ recovering exchanges where no minimal coupling is present but at the price of increasing the exchanged particle minimum spin or even other options related to the other derived kernels $\cG^{(n>2)}_{1234}$ but recovering in the four spin-$2$ case consistent abelian self-interactions. It is important to point out that, from this perspective, potential clashes with \emph{tree-level unitarity} need to be analyzed taking into account that an infinite number of exchanges is present. For instance, it is no more clear in this case how to disentangle all contributing residues outside the radius of convergence of the series of exchanges. Things would have been clearly different if only a finite number of exchanges had contributed to any given pole of the tree-level amplitude. Certainly, a deeper understanding of non-local field theories is needed in order to clarify such peculiar features that actually may be considered as the counterpart of an infinite number of higher and higher-derivative cubic couplings contributing to the same residue and might well led to an inconsistent answer in a Minkowski background. The latter implications could be appreciated deforming these results to (A)dS backgrounds or to massive fields where concrete examples of this kind are available or studying the tensionless limit of ST at the quartic order, from which one can expect to recover similar types of results. More information can also come solving for the most general theory that is consistent with the minimal scheme, and we plan to address these problems in the future.

The planar spin-$2$ example may clarify the role of the spin-$2$ excitation present in the Vasiliev system, that in principle can be dressed with Chan-Paton factors making its interpretation debatable. For some time the relation of such spin-$2$ excitation with gravity and/or with the \emph{massive} spin-$2$ excitation present in \emph{open} string theory was somehow unclear, as pointed out in \cite{exchange}. Indeed, it was argued that although the cubic coupling of two massive open-string spin-$2$ excitations with a graviton is forbidden by momentum conservation, this is not true in the tensionless limit, whenever one reaches a regime where the massive open string spin-$2$ becomes massless. This observation implies a potential mixing that can be already appreciated from the results of \cite{cubicstring}. In fact, among the limiting cubic couplings of the massive spin-$2$, the lowest derivative one is exactly the same as that of the graviton. Our present discussion puts on clearer grounds these issues, since at the quartic order two different possibilities show up distinguishing two options. One of these, given in eq.~\eqref{OFourspin2}, is naturally endowed with Chan-Paton factors while the other, eq.~\eqref{Closedgenfunc}, is closely related to gravity. They coexist in the amplitude of four spin-$2$ excitations and hence one is led to conclude that in the massless case the two options for spin-$2$ can give rise indeed to a \emph{Cabibbo-like} mixing between the combination of spin-$2$ fields that interacts as gravity and the singlet component interacting with \emph{open-string}-like four-point amplitudes, as anticipated in \cite{exchange}. From this perspective, we can imagine that the quartic couplings of the Vasiliev system with trivial $O(1)$ Chan-Paton factors, is of the form in eq.~\eqref{Shs3}, up to the usual overall functions $a_i(s,t,u)$ that are needed for consistency to match an infinite number of exchanges. We see, indeed, that expanding eq.~\eqref{Shs3} we can explicitly recognize the presence of both \emph{open-string}-like and \emph{closed-string}-like couplings both in the spin-$2$ case and for HS fields. Obviously, the mixing so far considered is expected to disappear whenever the theory breaks the HS-symmetry. In this case, the colored spin-$2$ field, that brings about non-localities, becomes massive, while a combination of the massless spin-$2$ fields remains massless, playing the role of the graviton. Again, it is tempting to believe that behind the string structure of the interactions there are some field theory properties that have to be understood and that may be intimately related to HS theories.

%%%%%%%%%%%%%%%%%%%%%%%%%%%%%%%%%%%%%%%%%%%%%%%

\scss{Off-shell ternary product}

%%%%%%%%%%%%%%%%%%%%%%%%%%%%%%%%%%%%%%%%%%%%%%%

Although the on-shell results contain the essential physical information about both HS four-point functions and the corresponding Lagrangian couplings, for completeness it is interesting to discuss also the off-shell extension. In order to solve this problem one can use various off-shell frameworks and in the following we concentrate on the super phase-space $\cH$ while other off-shell completions in Fronsdal's and other settings will be briefly considered in Appendix~\ref{app:quadratic} and in Appendix~\ref{app:cubic}. Here, in analogy with Section~\ref{sec:binary}, we can simply write the extension of \eqref{cV} over $\cH$, recovering the off-shell structure
\be
\cV^{\,\text{YM\ off-shell}}_{1234}\,=\,f^{\,(1)}_{ijkl}\ \xi_i\cdot\xi_j\,\xi_k\cdot\xi_l\,+\,f^{\,(2)}_{ijkl}\ \xi_i\cdot\xi_j\,\theta_k\bar{\theta}_l\,+\,f^{\,(3)}_{ijkl}\
\theta_i\bar{\theta}_j\,\theta_k\bar{\theta}_l\,+\,f^{\,(1)}_{ij}\ \xi_i\cdot\xi_j\,+\,f^{\,(2)}_{ij}\ \theta_i\bar{\theta}_j\ .\label{offcV}
\ee
where we can see the two contributions that are quartic and quadratic in the symbols $\xi$'s and $\theta$'s. The coefficients can be uniquely fixed by eq.~\eqref{Consist2}, that in this case reads
\be
\left\{(Q_1\,+\,Q_2\,+\,Q_3\,+\,Q_4)\,\cV^{\,\text{off-shell}}_{1234}\,+\,\left[\cG^{\,\text{off-shell}}_{(12|a}\,\theta_a^0\,\star_a\, \cG^{\,\text{off-shell}}_{a|34)}\right]\right\}\,\theta^0_1\theta^0_3\theta^0_3\theta^0_4\,=\,0\ ,
\ee
together with the useful boundary condition for $\cV_{1234}$ that is given up to traces and divergences by eq.~\eqref{cV}. Hence, eq.~\eqref{offcV} differs from eq.~\eqref{cV} only by trace terms and one can straightforwardly relate the coefficients $f^{\,(n)}_{ijkl}$ and $f^{\,(n)}_{ij}$ to those in the table~\ref{table:coeff}. In terms of eq.~\eqref{offcV} it is relatively easy to write down the kernel $\cK_{1234}$ associated to the ternary product of the theory as
\begin{multline}
\cK_{1234}\,=\,\left\{-\,\left(\int d\theta^0_a\,\cK^{\,\text{off-shell}}_{(12|a}\left(\hat{\Xi}_1,\hat{\Xi}_2,\hat{\Xi}_a\right)\,\theta^0_a\right) \frac{\tilde{\star}_a}{p_{\,a}^{\,2}}\left(\int d\theta^0_{\tilde{a}}\,\cK^{\,\text{off-shell}}_{\tilde{a}|34)}\left(\hat{\Xi}_{\tilde{a}},\hat{\Xi}_3,\hat{\Xi}_4\right)\, \theta^0_{\tilde{a}}\right)\right.\\\left.+\ \tilde{\cK}^{\,\text{off-shell}}_{1234}\left(\hat{\Xi}_1,\hat{\Xi}_2,\hat{\Xi}_3,\hat{\Xi}_4\right)\vphantom{\int}\right\}
\,\theta^0_1\,\theta^0_2\,\theta^0_3\,\theta^0_4\ ,\label{V4}
\end{multline}
where $\cK_{123}^{\,\text{off-shell}}$ is the off-shell cubic coupling of Section~\ref{sec:binary}, while $\tilde{\cK}_{1234}^{\,\text{off-shell}}$ takes the form
\begin{multline}
\tilde{\cK}^{\,\text{off-shell}}_{1234}\,=\,-\,\frac{1}{su}\ \sum_{n_i=0}^\infty \frac{1}{n_0!\,n_1!\,n_2!}\ \ a_{n_0,n_1,n_2}(s,t,u)\,\left[-\,s\,u\,\cG^{\,(0)\,\text{off-shell}}_{1234}\right]^{\,n_0} \\\times\, \left[-\,s\,u\,\cG^{\,(1)\,\text{off-shell}}_{1234}\right]^{\,n_1} \left[-\,s\,u\,\cG^{\,(2)\,\text{off-shell}}_{1234}\right]^{\,n_2}\ .
\end{multline}
Here, the $\cG^{\,(i)\,\text{off-shell}}_{1234}$'s are defined to be the corresponding off-shell completions of the $\cG_{1234}^{\,(i)}$'s in the previous section and can be extracted from the following kernel
\be
\cG^{\,\text{off-shell}}_{1234}\,=\,-\,\frac{1}{s}\ \cG^{\,\text{off-shell}}_{12a}\star_{\,a} \cG^{\,\text{off-shell}}_{a34}\,-\,\frac{1}{u}\ \cG^{\,\text{off-shell}}_{41a}\star_{\,a} \cG^{\,\text{off-shell}}_{a23}\,+\,\cV^{\,\text{YM\ off-shell}}_{1234}\ ,
\ee
distinguishing quadratic and quartic terms in the symbols. Instead $\cG^{\,\text{off-shell}}_{123}$ is given in eq.~\eqref{offG2} and finally,
\begin{align}
-\,s\,u\,\cG_{1234}^{(0)\,\text{off-shell}}\,&=\,u\,\left(\xi_{1}\cdot p_{\,2a}\,+\,\xi_2\cdot p_{\,a1}\,+\,\xi_3\cdot p_{\,4\tilde{a}}\,+\,\xi_4\cdot p_{\,\tilde{a}3}\right)\nonumber\vphantom{\frac{1}{2}}\\
&-\,u\left(\theta_1\partial_{\theta^0_2}\,-\,\theta_2\partial_{\theta^0_1}\, +\,\theta_3\partial_{\theta^0_4}\,-\,\theta_4\partial_{\theta^0_3}\right)\non\vphantom{\frac{1}{2}}\\
&+\,s\,\left(\xi_{1}\cdot p_{\,b4}\,+\,\xi_2\cdot p_{\,3\tilde{b}}\,+\,\xi_3\cdot p_{\,\tilde{b}2}\,+\,\xi_4\cdot p_{\,1b}\right)\non\vphantom{\frac{1}{2}}\\
&-\,s\left(-\theta_1\partial_{\theta^0_4}\,+\,\theta_4\partial_{\theta^0_1}\,-\,\theta_3\partial_{\theta^0_2}\,+\,\theta_2\partial_{\theta^0_3}\right)\ ,\vphantom{\frac{1}{2}}\label{G0off}
\end{align}
where by definition
\be
p_{\,a}\,=\,-\,p_{\,1}\,-\,p_{\,2}\ ,\qquad p_{\,\tilde{a}}\,=\,-\,p_{\,a}\ ,\qquad p_{\,b}\,=\,-\,p_{\,1}\,-\,p_{\,4}\ ,\qquad p_{\,\tilde{b}}\,=\,-\,p_{\,b}\ .
\ee
%
%To reiterate, here the functions $a_n(s,t,u)$ as well as the coupling function at the cubic level are to be chosen in our minimal scheme so that on-shell the non-local portions in $\tilde{\cK}_{1234}$ are canceled without introducing additional unphysical poles of the form
%
%\be
%\frac{1}{s}\ ,\qquad \frac{1}{t}\ ,\qquad \frac{1}{u}\ ,
%\ee
%
%that would introduce seemingly violations of the usual notion of \emph{causality} at the level of the amplitude. This implies that some \emph{non-local} four-point vertices are left out in \eqref{V4} giving the amplitude an intrinsic \emph{non-local} nature as soon as we increase the spin of the external particles. This points out a kind of clash for HS interactions between the analyticity and the factorization properties of the corresponding S-matrix amplitudes.
Analogous equations can be recovered also for closed-string-like couplings along the lines of the previous discussion so that the definition\footnote{We write $\cK$ without the label $(1234)$ in order to stress that a \emph{closed-string}-like kernel is not any more color-ordered and planar.} of $\tilde{\cK}$ is slightly changed and the color-ordering prescription is not used, while summing over all channels in the current exchange part.

%%%%%%%%%%%%%%%%%%%%%%%%%%%%%%%%%%%%%%%%%%%%%%%

\scss{Weinberg's theorem revisited}\label{sec:Weinberg}

%%%%%%%%%%%%%%%%%%%%%%%%%%%%%%%%%%%%%%%%%%%%%%%

In this section we take a closer look, in light of the previous discussion, at a key no-go theorem on the subject, in order to understand as much as possible its assumptions and in which sense one can go beyond them, clarifying hopefully the meaning of the results proposed so far. Indeed, one of the strongest arguments that has been presented over the years is Weinberg's 1964 Theorem of \cite{WeinbergNoGo} (see e.g. \cite{NogoRev} for a review and also for an interesting discussion of its interpretation). It is an S-matrix argument based on the analysis of a would be S-matrix element with $N$ external particles with momenta $p_{\,i}$, $i\,=\,1,\ldots,N$ and one massless spin-$s$ particle of momentum $q$ and polarization tensor $\phi_{\,\m_1\ldots\m_s}(q)$. In the following we shall review this argument explicitly in the case with arbitrary \emph{massless} particles entering the process and restricting the attention to the consistent cubic vertices studied in Section~\ref{sec:binary}. The idea is to analyze the case in which the momentum $q$ of one of the particles participating in the scattering process tends to zero, called also \emph{soft limit}. This limit encodes the long distance behavior, if any, of the interactions, which is dominated by the pole part, and it is very interesting since it gives constraints coming from very general and model independent IR properties. The dominant pole generates in this limit a resonance, so that one can factorize the amplitude, eliminating any \emph{local} contact interaction and leaving \emph{only} the contribution associated to the current exchange\footnote{We depart here from the original Weinberg proof that has been given in the S-matrix language assuming some commonly accepted ideas about the pole structure of the S-matrix. In this respect, the usual factorization property translates here into perturbative locality of the corresponding Lagrangian theory in its on-shell form.}. Actually, this is the contribution on which Weinberg concentrated in order to develop his argument, and in the following we shall study precisely the same contribution in our explicit setting recognizing what vertices contribute to long-distances and what vertices give instead a vanishing contribution in the same limit and reinterpreting Weinberg's conclusions. The explicit form of the S-matrix amplitude becomes in this limit
\begin{multline}
S(p_{\,1},\phi_{\,1};\ldots;p_{\,N},\phi_{\,N};q,\phi)\,\approx\,\sum_{i=1}^N S(p_{\,1},\phi_{\,1};\ldots;p_{\,i}+q,\tilde{\xi}_i;\ldots;p_{\,N},\phi_N) \\\star_{\,i}\frac{\hat{\cP}(\tilde{\xi}_i,\zeta_1)}{2p_{\,i}\cdot q}\star_{\,1}\left[\exp\Big(G_{123}(\zeta_1,\zeta_2,\zeta_3)\Big)\star_{\,2,3} \phi_{\,i}(-p_{\,i},\zeta_2)\,\phi(-q,\zeta_3)\right]\ ,\label{softlimit}
\end{multline}
where $\cP(\tilde{\xi}_i,\zeta_1)$ is the propagator numerator and, apart from the pole factor, the dependence on $q$ has been completely factorized solely into $\cG_{123}$. Using for the momenta of the particles participating to the factorized scattering process the parametrization
\be
p_{\,1}\,=\,p_{\,i}\,+\,q\ ,\qquad p_{\,2}\,=\,-p_{\,i}\ ,\qquad p_{\,3}\,=\,-q\ ,
\ee
one then recovers
\be
\cG_{123}\,=\,2\,\left(1\,+\,\zeta_1\cdot\zeta_2\right)\,\zeta_3\cdot p_{\,i}\,+\,\left(1\,+\,\zeta_2\cdot\zeta_3\right)\,\zeta_1\cdot (q\,-\,p_{\,i})\,-\,2\,\left(1\,+\,\zeta_3\cdot\zeta_1\right)\,\zeta_2\cdot q\ ,\label{limiting}
\ee
where we have made use of momentum conservation together with the transversality constraint
\be
p_{\,i}\cdot\zeta_i\,=\,0\ .
\ee
First of all, from this form one can immediately conclude that for $s\,>\,3$ the relevant tensor structure contributing at long distances, whenever present, is always given by
\be
\frac{(\zeta_3\cdot p_i)^{s}}{2p_i\cdot q}\ .
\ee
Second, we are now in a position to see whether or not the amplitude that we are recovering in this limit decouples the unphysical degrees of freedom and what are the cubic couplings that contribute. This physical requirement, as we have shown in Section~\ref{sec:ternary}, is precisely the content of the Noether procedure from a Lagrangian point of view. Hence, let us perform a linearized gauge transformation for the HS particle $\phi$ whose momentum $q$ goes to zero. The unphysical polarizations are given by
\be
\d\phi(-q,\zeta_3)\, =\,-q\cdot\zeta_3\,\L(-q,\zeta_3)\ ,
\ee
and performing this substitution in \eqref{softlimit} one finally ends up with
\begin{multline}
\d\,S(p_{\,1},\phi_{\,1};\ldots;p_{\,N},\phi_{\,N};q,\L)\,\approx\,\sum_{i=1}^N S(p_{\,1},\phi_{\,1};\ldots;p_{\,i}+q,\tilde{\xi}_i;\ldots;p_{\,N},\xi_N) \\\star_{\,i}\ \hat{\cP}(\tilde{\xi}_i,\zeta_1)\ \star_{\,1}\left[\left(1\,+\,\zeta_1\cdot\zeta_2\right) \exp\Big(\cG_{123}(\zeta_1,\zeta_2,\zeta_3)\Big)\star_{\,2,3} \phi_{\,i}(-p_{\,i},\zeta_2)\,\L(-q,\zeta_3)\right]\ ,\label{softlimit2}
\end{multline}
where the offending pole has been canceled by the terms proportional to momentum squared produced by the $\star$-contraction of $q\cdot\zeta_3$ with $\cG_{123}$. We can recognize here the most dangerous contribution\footnote{It is important to stress that in the limit $q\ra 0$ we recover only the order zero contribution in $q$ while, by consistency, \emph{all} contributions have to cancel identically. This underlines the \emph{no-go} character of this argument from which one can only extract information about possible obstructions.} in the limit $q\ra 0$, that is given by
\begin{multline}
\d\,S(q=0)\,\sim\,\sum_i\tilde{S}_i(\zeta_i)\,\star_{\,i}\,\sum_{\a_i,\b_i}\left\{\Big(2\,\zeta_3\cdot p_{\,i}\Big)^{\a_i}\left[1\,+\,\zeta_i\cdot \zeta_2\,\Big(2\,\zeta_i\cdot\zeta_2\,\zeta_3\cdot p_{\,i}\Big)^{\b_i}\right]\right.\\\left.\star_{\,2,3} \phi_{\,i}(-p_{\,i},\zeta_2)\,\L(-q,\zeta_3)\vphantom{\left[1\,+\,\zeta_i\cdot \zeta_2\,\Big(2\,\zeta_i\cdot\zeta_2\,\zeta_3\cdot p_{\,i}\Big)^\b\right]}\right\}\ ,\label{softlimit3}
\end{multline}
where $\a_i\,\geq\, -1$ and $\b_i\,\geq\, -1$ are some integers\footnote{For $\a_i\,=\,-1$ or $\b_i\,=\,-1$ we simply define the corresponding contribution to be zero.}, we have called $\tilde{S}(\zeta_i)$ the leftover part of the S-matrix together with the propagator numerator, and where the sum over $\a_i$ and $\b_i$ runs over all admissible values that are associated to consistent HS cubic couplings that can be generated from \eqref{softlimit2}. This contribution is \emph{dangerous} since it does not tend to zero when $q\ra 0$, and hence must vanish identically. Restricting the attention to the case in which all external fields but one are scalars and only one HS field $\phi(-q,\xi)$ is present only the scalar exchange contributes to \eqref{softlimit3} and there is only one possible value for $\b$, $\b\,=\,-1$. Thus, in order to set to zero \eqref{softlimit3}, one recovers a non-trivial constraint given by
\be
\sum_i g^i\,p^{\,i}_{\,\m_1}\ldots p^{\,i}_{\,\m_{s-1}}\,=\,0\ ,\label{WeinbergConstr}
\ee
where the $g^i$'s are the corresponding coupling constants. As pointed out by Weinberg, this equation does not admit non-trivial solutions unless in general $s=1$, and eq.~\eqref{WeinbergConstr} reduces to charge conservation
\be
\sum_i g^i\,=\,0\ ,
\ee
or $s=2$, so that eq.~\eqref{WeinbergConstr} reduces to $g^{\,i}\,=\,\kappa$ for any $i$ since, by momentum conservation
\be
\sum_i p^{\,i}_{\,\m}\,\equiv\,0\ .
\ee
We then arrive at a potential inconsistency for HS interactions, since the argument explained so far forces
\be
g^{\,i}\,=\,0
\ee
for spin grater than $2$. Actually, considering a more general HS theory and referring again to \eqref{softlimit3}, we see that as soon as an insertion of a scalar field is present in the amplitude there can be similar obstructions. This happens since one can reiterate this argument, concentrating on the factorized amplitude in which a scalar field is exchanged and in which $\phi_{\,i}$ in eq.~\eqref{softlimit3} is one of the scalar fields participating to the process. This conclusion has actually a deeper meaning, since it forbids the possibility of having an $s-0-0$ coupling whenever $s$ is greater than $2$ within the framework of \emph{local field theories}. This can be understood simply observing that as soon as such cubic couplings are present one generates automatically dangerous contributions to some current exchange amplitude. However this conclusion is true unless, by some mechanism, these dangerous exchanges are eliminated whenever they give rise to this kind of problems. Hence, the only possible way out is related to the fact that we have considered in the $q\ra 0$ limit only the current exchange contribution, so that one is led to a clash with perturbative locality on the Lagrangian side or with commonly accepted S-matrix properties like factorization on the S-matrix side. These anyway are possibly stronger statements than the fundamental unitarity property, and a closer look to them is potentially interesting in view of a better understanding of the tensionless limit of ST.

Let us now turn to see the implications of Weinberg's argument in more general examples. Concentrating on eq.~\eqref{softlimit3}, let us consider an external particle with arbitrary spin $s_i$. The factorization of the amplitude can give rise to problems in the soft limit only if a sufficient number of $\zeta_2$ is contained in \eqref{softlimit3}, otherwise this offending contribution vanishes identically. Hence, we conclude that a dangerous term of the form \eqref{softlimit3} can be generated only if
\be
\b_i\,\geq\,s_i\,-\,1\ .\label{bound}
\ee
In order to analyze the most general case let us restrict the attention to a factorized process in which, referring to \eqref{softlimit3}, $\phi_{\,i}$ is a spin-$s_{\,i}$ particle. If $s_i\,>\,s$ non-vanishing contributions cannot be generated and so, without loss of generality, we can concentrate on the cases in which $s_i\,\leq\,s$. In this case we recover a non-vanishing contribution to \eqref{softlimit3} whenever the bound \eqref{bound} is satisfied, but we also see that as soon as
\be
-1\,\leq\,\b_i\,<\,s_i\,-\,1\ ,\label{bound2}
\ee
no contribution can be generated, so that no inconsistency follows by this argument. Moreover, since for $\b_i\,\geq\,s_i$ one gets simply zero in \eqref{softlimit3}, there is only one dangerous contribution given by $\b_i\,=\,s_i -1$ that is associated to an exchanged particle with spin
\be
s_{\,\text{exchanged}}\,=\,s_i\ .
\ee
We clearly recover the simplest case of before when $s_i\,=\,0$, since in this case there is no solution for $\b_i$, and whenever $s_i\,\geq\,1$ one begins to recover non-trivial solutions to \eqref{bound2}.

Summarizing, one can convince oneself that the only possibly dangerous contributions come in this limit whenever one considers a current exchange built from a coupling of the form $s_i-s_i-s$ with $s$ derivatives when $s\,\geq s_i$ and the exchanged particle with spin $s_i$. As concluded by Weinberg, this argument poses strict restrictions on the long distance behavior of HS fields, that hence cannot interact at zero frequency. In particular all long-range couplings given by the \emph{minimal} ones, can be ruled out in a \emph{local field theory} while other multipolar couplings are not yet forbidden. The former actually entail exactly the leading contribution related to long distance physics on which Weinberg concentrated in \cite{WeinbergNoGo}. Moreover, since for $s\,<\,s_i$ the number of derivatives for these couplings is given by
\be
2s_i\,-\,s\,>\,s\ ,
\ee
we have explicitly shown that the content of Weinberg's argument together with the classification of all consistent cubic couplings completely forbids the minimal coupling for HS particles within the framework of \emph{local field theories}. It is now interesting to compare this result with the scattering amplitudes constructed in the previous sections. As we have remarked a possible solution to the problem can arise resorting to non-local quartic couplings whose job is to cancel the dangerous exchanges contributing in principle to the amplitudes\footnote{It is interesting to comment that as for what concerns the long-distance behavior of HS interactions Weinberg conclusions are still valid also if non-localities are introduced and one can easily check that the general solution to the Noether procedure that we have exhibited satisfies this property.} but without setting to zero all non-abelian cubic couplings. This circumstances of course would possibly violate unitarity but at present we are not able to come up with a definite conclusion on this issue and we believe that more effort is needed in order to clarify the situation. Let us end this section with a simple observation about the consequences of our results. If we concentrate on String Theory we see that all these minimal couplings are indeed generated in the tension-less limit starting from the simplest one that concerns two scalars \cite{cubicstring}. Hence, we see here a very severe obstruction if we insist to use the framework of \emph{local} field theories or the usual factorization properties at the level of the S-matrix in order to describe a would be \emph{tensionless string}. Similar considerations apply to the leading contribution of Fradkin-Vasiliev vertices, as observed in \cite{Seed}. Hence, one can argue that if a background independent underlying theory exists it has to include non-local couplings or, possibly, non-local degrees of freedom, which motivates a closer look at unitarity and its general implications.

%%%%%%%%%%%%%%%%%%%%%%%%%%%%%%%%%%%%%%%%%%%%%%%%

\scs{Non-local field theories?}

%%%%%%%%%%%%%%%%%%%%%%%%%%%%%%%%%%%%%%%%%%%%%%%%

In this section we are going to combine and summarize the results obtained so far on the explicit construction of four-point amplitudes and Weinberg's argument analyzing in detail the form of the non-localities in a particular example. We shall concentrate on the simplest toy-model of \emph{open-string}-like amplitudes built solely from $\cG_{1234}^{(0)}$ in eq.~\eqref{G0off} just to describe the generic setup, even though this example is perhaps a bit too simple. At any rate, starting from the cubic vertex
\be
\cK_{123}\,=\,\exp\left(\xi_1\cdot p_{\,23}\,+\,\xi_2\cdot p_{\,31}\,+\,\xi_3\cdot p_{\,12}\,-\,\theta_1\partial_{\theta^0_{23}}\,-\,\theta_2\partial_{\theta^0_{31}}\,-\,\theta_3\partial_{\theta^0_{12}}\right)\ ,
\ee
we can extract the Lagrangian quartic coupling generating function, or ternary product kernel, from
\begin{multline}
\cK_{1234}\,=\,-\,\left(\int d\theta^0_a\,\cK_{(12|a}\left(\hat{\Xi}_1,\hat{\Xi}_2,\hat{\Xi}_a\right)\,\theta^0_a\right)\frac{\tilde{\star}_a}{p_{\,a}^{\,2}}\left(\int d\theta^0_{\tilde{a}}\,\cK_{\tilde{a}|34)}\left(\hat{\Xi}_{\tilde{a}},\hat{\Xi}_3,\hat{\Xi}_4\right)\, \theta^0_{\tilde{a}}\right)\\+\,\tilde{\cK}_{1234}\left(\hat{\Xi}_1,\hat{\Xi}_2,\hat{\Xi}_3,\hat{\Xi}_4\right)\ .\label{qlag}
\end{multline}
where
\be
\tilde{\cK}_{1234}\,=\,-\,\frac{1}{su}\sum_{n=0}^\infty a_{\,n}(s,t,u)\ \left[-su\,\cG^{(0)\,\text{off-shell}}_{1234}\right]^n
\ee
and where the relative functions $a_n(s,t,u)$ of each single term in $\tilde{\cK}_{1234}$ are to be chosen in such a way that the non-local contributions present in $\tilde{\cK}_{1234}$ are canceled and factorization of the corresponding amplitude holds, albeit on a (infinite) \emph{subset} of all spins. We can see explicitly in this example how this physical requirement allows one to fix, up to local and gauge-invariant quartic couplings under the linearized gauge symmetry, all relative functions in $\tilde{\cK}_{1234}$. For example, the coefficient of the linear term in $\cG_{1234}^{(0)}$ of eq.~\eqref{G0off} must be
\be
a_{\,1}(s,t,u)\,=\,e^{-2t}\,+\,\ldots\ ,
\ee
while going to the term quadratic in $\cG_{1234}^{(0)}$ one should choose
\be
a_{\,2}(s,t,u)\,=\,-\,\frac{1}{t}\ e^{-2t}\,+\,\ldots\ ,
\ee
and so on, where the ellipsis stand for terms proportional to $s\,u$ that kill the pole contribution and give rise to local quartic couplings that are gauge-invariant under the linearized gauge symmetry and that reflect the freedom left by Noether procedure in building a consistent theory. However, we can see that if we want to cancel the non-local contribution given by the scalar exchange we need to introduce an additional pole term proportional to
\be
\frac{1}{t}
\ee
in the four-point functions. Interestingly, had one chosen functions $a_{\,i}(s,t,u)$ with no poles in the Mandelstam variables, like for instance
\be
a_{\,2}(s,t,u)\,=\,-\,\frac{1}{t}\ \left(e^{-2t}\,-\,1\right)\,+\,\ldots\ ,
\ee
and so on for the other $a_{\,i}(s,t,u)$'s, not all non-local contributions in \eqref{qlag} would have been canceled, starting from the scalar exchange given by\footnote{To be precise only part of this non-local coupling is required by the minimal scheme since for instance amplitudes built from abelian and non-deforming cubic couplings can be made gauge invariant without any quartic coupling up to field redefinitions.}
\begin{multline}
\cK_{1234}\,\sim\,\frac{1}{s}\ \exp\left(\xi_{1}\cdot p_{\,2a}\,+\,\xi_2\cdot p_{\,a1}\,+\,\xi_3\cdot p_{\,4\tilde{a}}\,+\,\xi_4\cdot p_{\,\tilde{a}3}\right)\\+\,\frac{1}{u}\ \exp\left(\xi_{1}\cdot p_{\,b4}\,+\,\xi_2\cdot p_{\,3\tilde{b}}\,+\,\xi_3\cdot p_{\,\tilde{b}2}\,+\,\xi_4\cdot p_{\,1b}\right)\ ,
\end{multline}
and so on, considering higher powers of $\cG_{1234}^{(0)}$. Both possibilities reflect the fact that lower-spin exchanges with respect to the external particles participating in the process can be problematic, as was already discussed in the analysis in Section~\eqref{sec:Weinberg}. This is manifest here, since one is either forced to introduce poles in the Mandelstam variables at the level of the four-point amplitude, ruining causality, or alternatively one needs to consider quartic Lagrangian couplings containing \emph{explicit} non-localities thus affecting the factorization property of the corresponding S-matrix amplitude. Since we want to preserve causality, we choose to avoid seemingly unphysical higher order poles. Hence, we are led to believe that the amplitude still \emph{factorizes} but on a subset of all available spins, such that the current exchanges admit local completions that make them gauge-invariant. Non-localities of this kind at the Lagrangian level are actually a mechanism to avoid the dangerous contributions that have been pointed out in Weinberg's argument, and represent, together with our minimal scheme, the mildest form of non-locality that one can choose in order to recover a gauge-invariant result also at the four-point level. Of course we stress that one has to exercise some care when talking about non-local field theories whose properties and systematics are still completely unknown.
\begin{figure}[htbp]
\begin{center}
\resizebox{15cm}{!}{\psfig{figure=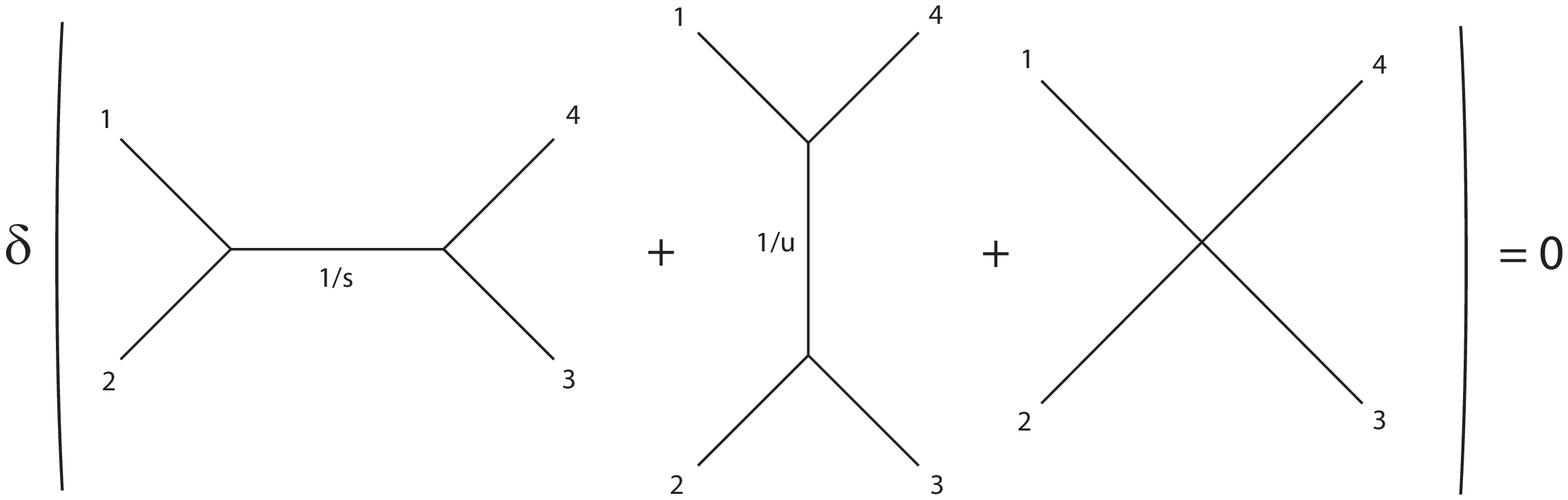,width=15cm}}
\caption{Decoupling condition for unphysical polarizations at the level of the amplitude.} \label{fig1}
\end{center}
\end{figure}
Turning briefly to the case involving also $\cG_{1234}^{\,(1)}$ and $\cG_{1234}^{\,(2)}$, one can give a lower bound for the lowest spin propagating into a process described by a single monomial\footnote{Other options related to the possible derived kernels introduced in the previous sections should be considered in the most general case.}
\be
\tilde{\cK}_{1234}\,\sim\,-\frac{1}{su}\left[-\,su\,\cG^{(0)}_{1234}\right]^\a\left[-\,su\,\cG^{(1)}_{1234}\right]^\b\left[-\,su\,\cG^{(2)}_{1234}\right]^\g\ ,
\ee
where the external particles have spin $s_1$, $s_2$, $s_3$ and $s_4$. Indeed it contains current exchanges among which the lowest propagating spin present is
\be
s_{\text{min}}\,=\,\a\,+\,\b\,+\,2\g\,-\,1\ ,
\ee
so that considering a four-point amplitude with external spin-$s_1$, $s_2$, $s_3$ and $s_4$ particles one has
\be
4\,\g\,+\,2\,\b\,+\,\a\,=\,s_1\,+\,s_2\,+\,s_3\,+\,s_4\ ,
\ee
and hence, generically,
\be
s_{\text{min}}\,=\,\frac{1}{2}\ \left(s_1\,+\,s_2\,+\,s_3\,+\,s_4\right)\,+\,\frac{\a}{2}\,-\,\kappa\,\geq\,\frac{1}{2}\ \left(s_1\,+\,s_2\,+\,s_3\,+\,s_4\right)\,-\,\kappa\ ,
\ee
where $\kappa\,=\,1$ for open-string-like amplitudes while $\kappa\,=\,2$ for closed-string-like amplitudes. Then, as expected since four-point functions encode the fusion rules of the HS algebra, one can prove that an infinite tower of HS fields is required, if the theory involves non-trivial HS-excitations. In fact, considering a four-point amplitude with four external spin-$s$ particles we recover exchanged particles of spin higher than $s$ unless $s\,=\,1$ for \emph{open-string}-like amplitudes, or $s\,=\,2$ for \emph{closed-string}-like ones. To reiterate, the key difference between \emph{local} and \emph{non-local} couplings is that the latter modify in a dramatic way the residue of the current exchange restricting the sum over \emph{all} propagating spins to a given (infinite) subset. This means that the kind of \emph{non-local} couplings discussed here can be somehow eliminated translating the information that they carry into a choice of an (infinite) subset of the exchanges in which scattering amplitudes factorize.
Finally and most importantly, potential clashes with unitarity need to be analyzed taking into account that an infinite number of degrees of freedom ought to contribute to the same residue. We leave further discussions on this important feature, that is intimately related to the still unexplored realm of non-local QFT, for the future awaiting for a better understanding of these problems and of the deformations to (A)dS backgrounds, although we can expect that the corresponding flat limit presents difficulties similar to those encountered here.
\begin{figure}[htbp]
\begin{center}
\resizebox{12cm}{!}{\psfig{figure=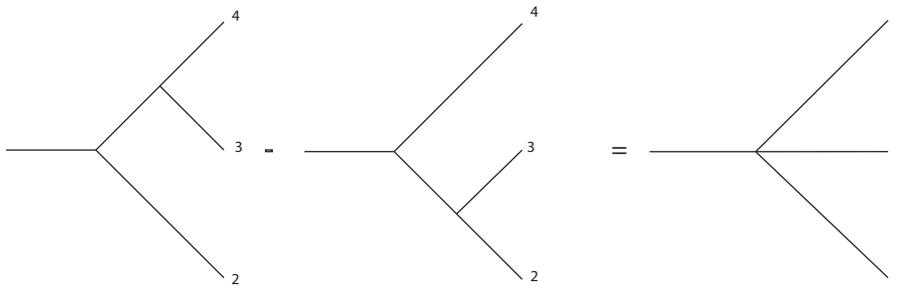,width=12cm}}
\caption{The decoupling condition can be read as an associator equation in which the cubic coupling is considered as a product and the quartic coupling as a trilinear product.}
\label{fig2}
\end{center}
\end{figure}

\begin{figure}[htbp]
\begin{center}
\resizebox{12cm}{!}{\psfig{figure=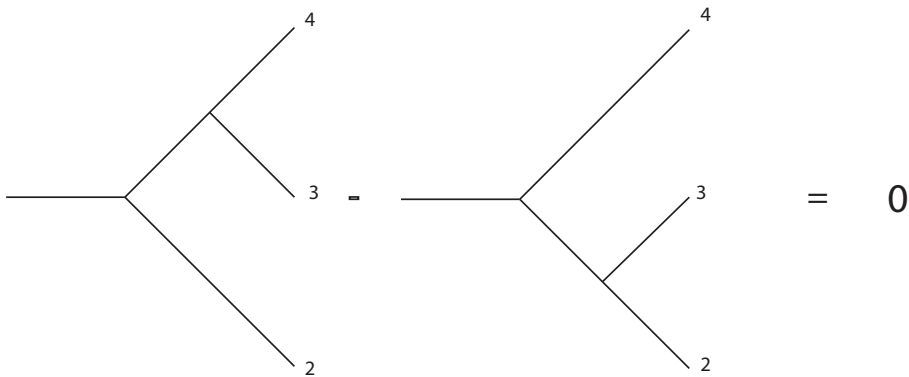,width=12cm}}
\caption{Associative bi-product.}
\label{fig3}
\end{center}
\end{figure}

In light of these observations, let us analyze diagrammatically the requirement of decoupling of the unphysical states at the level of the four-point amplitude. The cubic coupling generating function can be interpreted in full generality as an \emph{on-shell bi-product} and the decoupling conditions of unphysical states translate into the diagrammatic equation in fig.~\ref{fig1}. Evaluating the gauge variation of the amplitude in a diagrammatic form, fig.~\ref{fig1} becomes equivalent to the diagrammatic equation in fig.~\ref{fig2}, that has a key physical meaning since it relates amplitudes in different channels and computes the \emph{associator} related to the on-shell bi-product encoded by the cubic coupling.
If we now consider the Britto-Cachazo-Feng-Witten (BCFW) recursion relations \cite{BCFW}, relating the n-point S-matrix to three-point functions, we can reinterpret, in view of the present discussion, the constructibility criterion
\be
\cA^{(1,2)}(0)\,=\,\cA^{(1,4)}(0)
\ee
where the label in the amplitudes identifies what momenta and particles are deformed into the complex plane. Indeed, it is very tempting to interpret  again this crossing equality as an associator equation pointing out this time an \emph{associative} nature of the on-shell bi-product encoded into the on-shell cubic coupling (fig.\ref{fig3}).
It is suggestive to interpret the non-localities found here as \emph{violations} of this criterion that translate into an intrinsic non-associative nature of the bi-product naturally associated to the fact that HS cubic couplings contain higher derivatives and can be cured, somehow, selecting a subset of the available spins that can be exchanged in the four-point amplitude, with the end result that one recovers in this way again a \emph{constructible} theory but in a different sense. Moreover, it is also tempting to say that \emph{local} four-point couplings can be reabsorbed, somehow, into the current exchange part, and taken care by the BCFW recursion relations, while the non-local contributions have, in some sense, a different meaning. This reasoning actually justifies the special role of \emph{gravity} and \emph{Yang Mills theory} as constructible \emph{local} theories, while HS theories, as we have seen, are constructible only in a generalized sense.
This puts on totally different grounds local and non-local quartic couplings, although we stress that also the latter turn out to be related to the current exchange amplitudes and hence to the cubic couplings, since their role is to compensate some of them. In conclusion, the graphic interpretation that we have presented may give us a view of non localities in terms of associative or non-associative bi-products. Along this line we can wonder wether the non-associative behavior that one can observe for HS external particles may be related to a possible pole at infinity in the recursion relations recently studied in \cite{Benincasa}.

%%%%%%%%%%%%%%%%%%%%%%%%%%%%%%%%%%%%%%%%%%%%%%%%%%%%%%%%

\scs{Conclusions}

%%%%%%%%%%%%%%%%%%%%%%%%%%%%%%%%%%%%%%%%%%%%%%%%%%%%%%%%

In this paper we have studied the Noether procedure drawing some inspiration from the structure encoded into the theory of FDA and proposing to relax the locality hypothesis usually considered in all standard settings in favor of the minimal scheme defined in Section~\ref{sec:four-point} along lines that are actually in the spirit of the previous work \cite{NonLocalNoether}. Reversing the usual perspective of focusing on four-point Lagrangian couplings, we have recovered directly a class of $4$-point functions\footnote{See the Appendix~\ref{app:A} for the extension of the classification to $n$-points, fermions and mixed symmetry fields.} involving \emph{massless} HS fields as well as lower-spin fields from the \emph{linearized} gauge invariance of the free system. Our key result is perhaps the construction of an infinite class of HS $4$-point functions that are related in a relatively simple way to powers of the standard $4$-point functions in a theory with a scalar, a spin-$1/2$ fermion and a gauge boson. This generalizes the construction of \cite{cubicstring}, making it possible to define similar color-ordered kernels in the general case that include, as a particular example, the simpler $\cG_{123}$ kernel from which all cubic couplings originate. One is then able to extract, subtracting the current exchange parts, four-point and in general $n$-point couplings. These, together with the fully non-linear deformation of the linearized gauge symmetries, contain as a special case the familiar low-spin examples as well as an infinite set of local couplings, but manifest also a \emph{non-local} nature as soon as one considers more exotic cases, as for instance a colored spin-$2$, or more generally HS fields. The non-local nature so far observed has an interesting and peculiar structure of the form pointed out in the Appendix of \cite{cubicstring}, although the meaning of non-localities is here to restrict the spins propagating within the amplitude to those whose violation of gauge invariance \emph{can} be compensated by \emph{local} counterterms. This fact entails the key obstruction that has been recognized long ago by Weinberg in \cite{WeinbergNoGo}, as well as other inconsistencies at the level of Jacobi identity and so on \cite{BoulangerCubic}, that disappear as soon higher-derivative and explicitly non-local couplings are considered, as already noticed in \cite{NonLocalNoether}. Of course a non-local solution to the problem, even if explicit, cannot be satisfactory without a full understanding of its implications and in particular of the status of the minimal scheme proposed here. In this respect the only thing we can say is that it is conceivable that potential clashes with the standard form of \emph{tree-level} unitarity, that can come together with \emph{non-localities}, do not materialize as soon as an infinite number of degrees of freedom are present. Even considering the case in which the latter option does not hold, it can be interesting to extract from the classification just presented its deformation to constant curvature background and we leave this for the near future. Nonetheless, let us stress that any residue of the set of amplitudes so far recovered can be related to lower-point couplings via exchange amplitudes if the minimal scheme is enforced in place of the stronger locality constraint. In this respect it can be interesting to ask what plays the role of locality in constant curvature backgrounds and whether the solution will still contain similar non-localties even if controlled by some expansion parameter, thinking to push forward our analysis in order to understand more clearly the possible need of resorting to this kind of picture. Finally, a deeper understanding of the peculiar features involved by HS interactions that seem to imply a clash with commonly accepted ideas about the pole-structure of the S-matrix, can be hopefully related to the difficulties that are encountered in the definition of an S-matrix, and we believe that they deserve a better understanding motivated at least by their appearance within ST in its tension-less limit or within the Vasiliev system in its flat limit \cite{Seed}.

Although the subject of non-local field theories is still a completely unexplored arena, the aforementioned properties of the amplitudes may open the way to a deeper understanding of Field Theory. In this respect, ST appears to contain the seed for interesting generalizations, and hides, in our opinion, some key field theory properties that have surfaced in this work. To wit, the remarkable construction of Closed String Field Theory in \cite{CSFT} is very general in its starting point, but the mechanical model definition of the interactions hides somehow their non-local nature that has long been felt to be related to a broken phase of the HS symmetry. The mechanical model may hide somehow the non-local features that we have presented here linking them to the string tension\footnote{See e.g. \cite{desc,bsb,nonlin} where other examples in which the mechanical model appears to provide an incomplete description are discussed.}. A similar kind of situation may arise in the Vasiliev system, whose spectrum contains a massless spin-$2$ field whose interpretation in terms of gravity has been debated, since it can be dressed with Chan-Paton factors in a much closer analogy with the massive spin-$2$ excitation present in the open string. Indeed, the intrinsic non-local form of the couplings of a colored spin-$2$ exhibited here may shed some light on the \emph{non-local} nature of Vasiliev's system, that seems to be obscured by the presence of the cosmological constant $\L$, whose role is similar to that of the string tension in ST and provides an expansion of local terms in which operators like $\frac{1}{\square}$ could split in terms of $\L$.

Other questions then arise in order to attain a meaningful quantization of systems of this kind, that at any rate can be naturally formulated in terms of the Batalin-Vilkoviski formalism \cite{BV} or in terms of a usual loop expansion. Those can be recovered from the Feynman rules here considered or, alternatively, from recursion relations techniques. Other issues regard the freedom in building a theory of massless HS that we have recursively related to the freedom of choosing a consistent cubic coupling functions with what we have called minimal scheme. We leave this as well as other questions, like the extension of these results to constant curvature backgrounds, that can be naturally addressed starting from the results presented here, for future works.

\vskip 35pt

%%%%%%%%%%%%%%%%%%%%%%%%%%%%%%%%%%%%%%%%%%%%%%%%%%%%%%%%%%%%%%

\section*{Acknowledgments}

%%%%%%%%%%%%%%%%%%%%%%%%%%%%%%%%%%%%%%%%%%%%%%%%%%%%%%%%%%%%%%%%%%%%%

I am grateful to E.~Joung, L.~Lopez, R.~Rahman for several stimulating discussions and to D.~Francia for the reading of the manuscript and for discussions in the final stage of the project. My deep gratitude goes to my advisor Prof. A.~Sagnotti for his invaluable comments and key suggestions and for having encouraged me to publish these results alone. The present research was supported in part by Scuola Normale Superiore, by INFN and by the MIUR-PRIN contract 2009-KHZKRX.

%%%%%%%%%%%%%%%%%%%%%%%%%%%%%%%%%%%%%%%%%%%%%%%%%%%%%%%%%%%%%%%%%%%%%

\begin{appendix}

%%%%%%%%%%%%%%%%%%%%%%%%%%%%%%%%%%%%%%%%%%%%%%%%

\scs{Comments on N-point functions, Fermions and Mixed Symmetry fields}\label{app:A}

%%%%%%%%%%%%%%%%%%%%%%%%%%%%%%%%%%%%%%%%%%%%%%%%

In this appendix we would like to extend the analysis carried out in previous sections to the $n$-point functions and to theories with fermions and mixed-symmetry fields. We shall see that the construction done so far at the level of four-point functions suffice to understand at least in principle how the same problem can be addressed to all orders. The basic idea is again to underline the key role of gauge-invariant amplitudes together with the technical trick of identifying simple objects that are capable of encoding the physical information about the interactions. We first describe how to generalize the previous discussion to $n$-point functions and then we turn to consider the most general theory containing fermions and mixed-symmetry fields, leaving for the future a more detailed discussion of the coupling functions and spectra that are consistent with the minimal scheme.

%%%%%%%%%%%%%%%%%%%%%%%%%%%%%%%%%%%%%%%%%%%%%%%%

\scss{$n$-point functions}

%%%%%%%%%%%%%%%%%%%%%%%%%%%%%%%%%%%%%%%%%%%%%%%%

In order to construct $n$-point functions a key observation concerns the form of both three- and four-point amplitudes for HS fields here proposed. Indeed both $\tilde{\cK}_{123}$ and $\tilde{\cK}_{1234}$ can be related to the color-ordered cubic and quartic amplitudes in a theory with a scalar and a gauge boson.
\begin{figure}[htbp]
\begin{center}
\resizebox{12cm}{!}{\psfig{figure=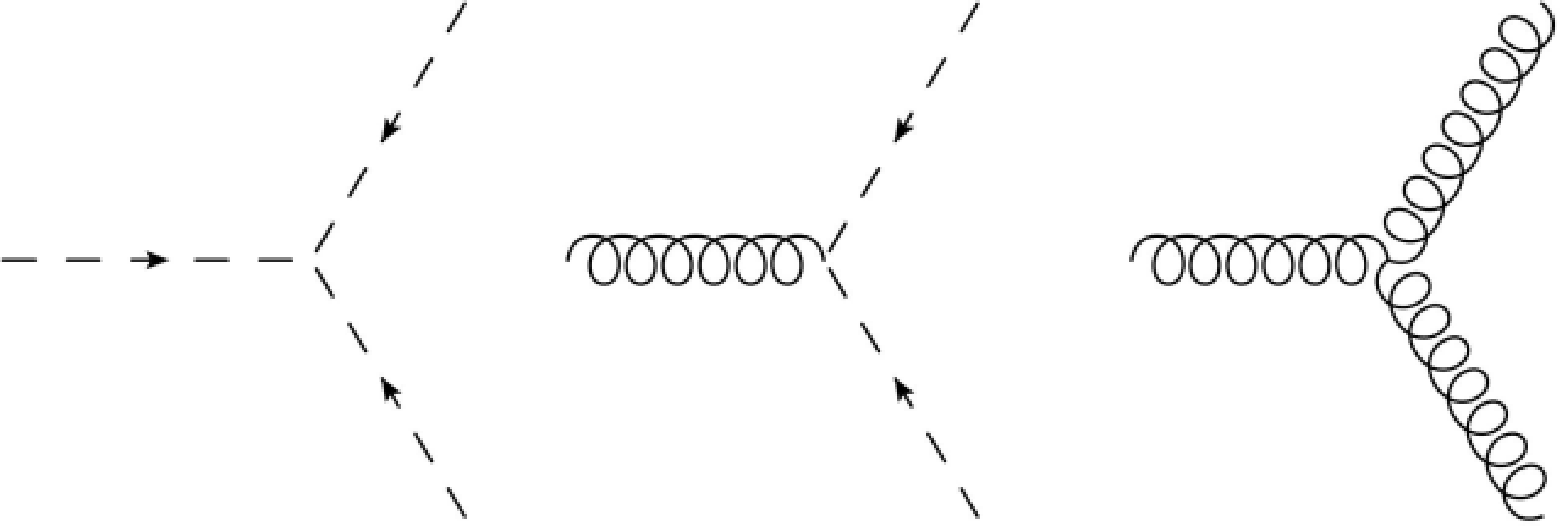,width=12cm}}
\caption{Cubic building blocks for $\cG_{12\ldots n}$.}
\label{fig:cubic}
\end{center}
\end{figure}
This simple fact suggests an interesting generalization to $n$-points, described here by $n$-ary products associated to color-ordered kernels $\cK_{12\ldots n}$. Indeed, one can classify for any $n$ all possible color-ordered $n$-point functions that can be constructed from the cubic couplings in fig.~\ref{fig:cubic}, where the first diagram is just a trivial constant coupling and needs to be considered since it gives non-trivial contributions starting from the quartic order. As we have noticed in Section~\ref{sec:YMfour} one can identify $\cG_{1234}^{\,(0)}$, $\cG_{1234}^{\,(1)}$ and $\cG_{1234}^{\,(2)}$ by the S-matrix diagrams in fig.~\ref{fig:quartic}, where the polarization tensor of the gauge boson has to be replaced by corresponding symbols $\xi_i$ and where one is able to recover automatically the quartic contact terms by requiring gauge invariance of the corresponding color-ordered current exchanges.
For instance, one can extend this result to the fifth order, considering analogous objects that we call $\cG^{\,(0)}_{12345}$, $\cG^{\,(1)}_{12345}$, $\cG^{\,(2)}_{12345}$ and $\cG^{\,(3)}_{12345}$, defined to be respectively the scattering amplitudes of fig.~\ref{fig:quintic}, where the polarization tensor of the gauge bosons have been again replaced by the corresponding symbols $\xi_i$'s so that one recovers again a kernel generating function exponentiating the $\cG_{12345}^{\,(i)}$'s as
\be
\tilde{\cK}_{12345}\,=\,-\,\frac{1}{s_{12}\,s_{23}\,s_{34}\,s_{45}\,s_{51}}\ \exp\left[-\,s_{12}\,s_{23}\,s_{34}\,s_{45}\,s_{51}\left( \cG^{\,(0)}_{12345}\,+\,\cG^{\,(1)}_{12345}\,+\,\cG^{\,(2)}_{12345}\,+ \,\cG^{\,(3)}_{12345}\right)\right]\ .
\ee
\begin{figure}[htbp]
\begin{center}
\resizebox{12cm}{!}{\psfig{figure=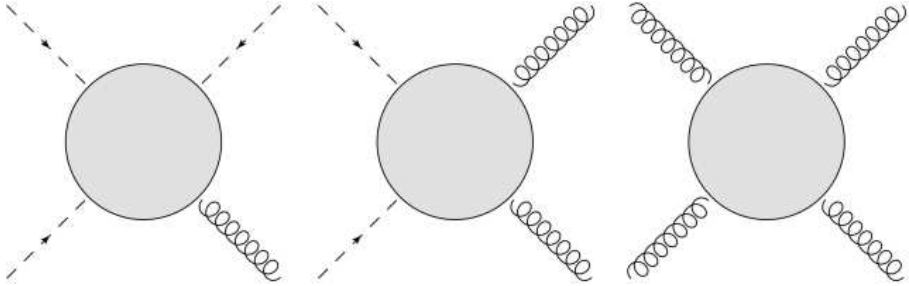,width=12cm}}
\caption{Quartic tree-level diagrams associated to the $\cG^{\,(i)}_{1234}$.}
\label{fig:quartic}
\end{center}
\end{figure}
\!\!Here we have used five independent Mandelstam-like invariants that enter as poles into each $\cG^{(i)}_{12345}$ in order to recover at the exponential a pole-less object. In this form, in analogy with the four-point case, we have still to multiply each totally cyclic combination of the $\xi_i$'s with functions of all Mandelstam-like invariants that \emph{do not} introduce any unphysical pole and that can be constrained, in our minimal scheme, in order to recover factorization on an infinite subset of all the exchanges matching those exchanges that are built from lower-point amplitudes. This fact implies, in general, that even at this order non-local quintic interactions can come into play since, as soon as one considers spins higher than one, in the case of \emph{open-string}-like amplitudes, or higher than two, in the case of \emph{closed-string}-like amplitudes and possible generalizations at higher points, higher powers of the Mandelstam-like invariants arise, which accompany higher and higher propagating spins with respect to the ones propagating at the quartic order. In general, one is thus forced to select different infinite subsets in which the amplitudes factorize at each order.
\begin{figure}[htbp]
\begin{center}
\resizebox{12cm}{!}{\psfig{figure=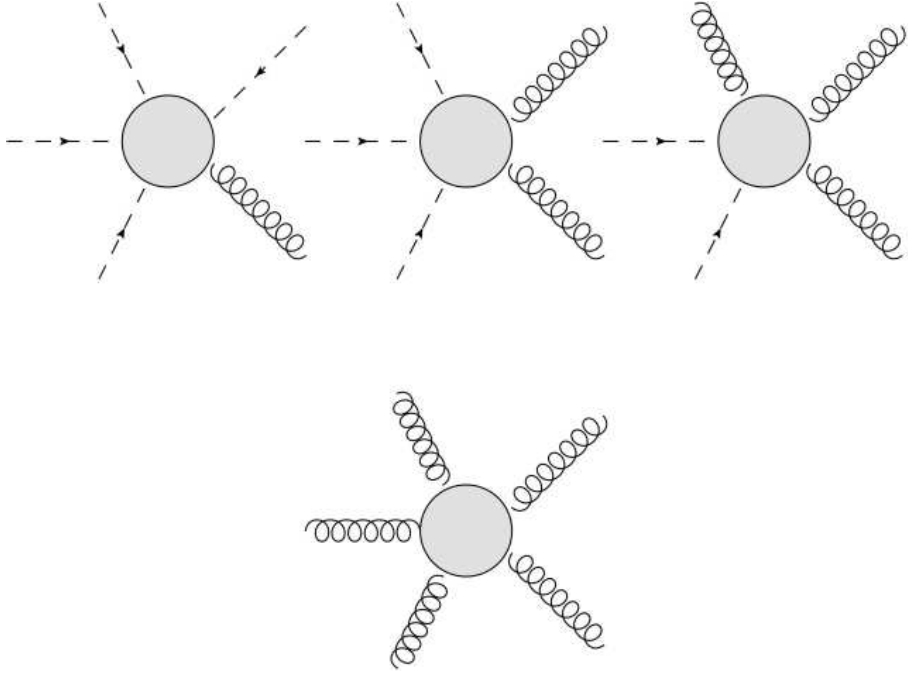,width=12cm}}
\caption{Quintic tree-level diagrams associated to the $\cG^{\,(i)}_{12345}$.}
\label{fig:quintic}
\end{center}
\end{figure}
Leaving aside the $n$-scalar exchange, that is just a function of the Mandelstam-like invariants, at $n$-points we have $n-1$ different color-ordered $\cG_{12\ldots n}^{\,(i)}$'s associated respectively to amplitudes with $n$, $n-2$, $n-3$ down to $1$ gauge bosons while the others are scalars\footnote{Admitting further redundancies one can also consider amplitudes built from the higher derivative couplings of two gauge bosons and one scalar and three gauge bosons in eq.~\eqref{higher}.}. The formal expression for the color-ordered kernel generating function $\tilde{\cK}_{12\ldots n}$ is then
\be
\tilde{\cK}_{12\ldots n}\,=\,-\,\frac{1}{\r(P_{ij})}\ \exp\left[-\,\r(P_{ij})\left(\cG^{(0)}_{12\ldots n}\,+\,\cG^{(1)}_{12\ldots n}\,+\,\ldots\,+\,\cG^{(n-2)}_{12\ldots n}\right)\right]\ ,\label{npointgen}
\ee
where the general form of an exchanged momentum is
\be
P_{ij}\,=\,(p_{\,i}\,+\,\ldots\,+\,p_{\,j})^2\ ,\qquad i\,<\,j\ ,
\ee
while $\r(P_{ij})$ is defined to be a product of all possible exchanged momenta, in which a color-ordered $\cG_{12\ldots n}$ factorizes, containing the minimum number of them that eliminates all poles from the $\cG^{(i)}_{12\ldots n}$'s. Within this setting, the only freedom left is a choice of relative functions of the Mandelstam-like invariants that do not introduce unphysical poles playing here the role of some arbitrary generalized relative coefficients. For instance, from eq.~\eqref{npointgen} one recovers objects of the form
\be
\tilde{\cK}_{12\ldots n}\,=\,-\,\frac{1}{\r(P_{ij})}\ a_{\a_1\ldots\a_{n-2}}\left(\{s_{ij}\}\,,\{s_{ijk}\}\,,\ldots\right)\prod_{l\,=\,0}^{n-2}\frac{1}{\a_l!}\ \left[-\,\r(P_{ij})\ \cG^{(l)}_{12\ldots n}\right]^{\a_l}\ ,
\ee
where we have explicitly considered a function $a_{\a_1\ldots\a_{n-2}}\left(\{s_{ij}\}\,,\{s_{ijk}\}\,,\ldots\right)$ of all Mandelstam-like invariants for any available choice of the $\a_l$'s.
We stress again that, apart from local $n$-point couplings that are gauge invariant under the linearized gauge transformations and that arise whenever the relative functions of the Mandelstam-like invariants cancel the pole factor, the minimal scheme continue to impose constraints on the propagating spins and on the consistent coupling-functions that have to be chosen accordingly. Indeed, at higher points all the functions of the Mandelstam-like invariants that weight the various gauge invariant contributions can be recursively related to the exchanges involving lower-point functions, enforcing factorization on the available exchanges and associating the missing residues, that cannot belong to a $\tilde{\cK}$, to the presence of non-local $n$-point couplings. Before concluding this section we want to point out that further interesting options, that can be considered as a kind of \emph{generalized} closed-string-like kernels, show up from the fifth order, since more than two permutations with respect to the external legs of the kernels $\cG_{12\ldots n}$ are independent. The result has, in general, the following structure
\be
\tilde{\cK}(\xi^1_i\,,\ldots\,,\xi^m_i)\,=\, \left(\sum_{\text{perm}}\ \prod_{\a\in\cI}\tilde{\cK}_{\s_\a(1)\s_\a(2)\s_\a(3)\ldots\s_\a(n)}(p_{\,i},\xi_{\,i}^\a)\right)\ ,\label{Shs5}
\ee
where the first sum over the permutations of the external legs is the usual sum that give rise to a non-planar kernel (see e.g. the sum over $\s$ in eqs.~\eqref{Shs2} and \eqref{Shs3}) while the product over $\a$ encodes the possible independent $\cG_{12\ldots n}$'s associated to some non-cyclic permutations of $\{1,2\ldots,n\}$. We leave a more detailed analysis of these kind of potentially interesting options for the future.

%%%%%%%%%%%%%%%%%%%%%%%%%%%%%%%%%%%%%%%%%%%%%%%%

\scss{Fermions}\label{sec:fermionic}

%%%%%%%%%%%%%%%%%%%%%%%%%%%%%%%%%%%%%%%%%%%%%%%%

In order to address the fermionic case it is very instructive to analyze the structure of the three-point couplings of two fermions and one boson recovered in \cite{cubicstring}. One can naturally generalize the kernels $\cK_{12\ldots n}$ adding $\g$-matrix contributions in order to deal also with fermions without changing the formal structure discussed in this paper. In the following we shall concentrate mostly on the on-shell gauge-fixed results, involving irreducible $\g$-traceless tensor-spinors, since the off-shell extension works exactly as in the bosonic case and all physical information is encoded already at this simpler level. Actually, the fermionic cubic couplings recovered in \cite{cubicstring} can be translated in the generating function
\be
\cK_{123}\,=\,\left(1\,+\,\slashed{\xi}_1\,+\,\slashed{\xi}_2\,+\,\slashed{\xi}_3\right)\,e^{\,\cG_{123}}\ ,
\ee
where the fermionic indices are left implicit, $\cG_{123}$ is defined in eq.~\eqref{G} and where one is free, at this level, to choose arbitrary relative coefficients between the various totally cyclic contributions\footnote{See Appendix~\ref{app:mostgeneral} for more details on the most general form for this result.}. The form of these on-shell gauge-fixed couplings is again relatively simple, and one recovers two different kinds of structures. One is directly related to the bosonic couplings discussed previously, and is obtained simply contracting together the fermionic indices with a Kronecker $\d$, so that one ends simply with a generalization of the Pauli coupling, that on-shell takes the form
\be
A_{\,3}\cdot p_{\,12}\,\bar{\psi}_{\,1}\,\psi_{\,2}\ .
\ee
The second type, on the contrary, generalizes the Yang-Mills minimal coupling
\be
\bar{\psi}_{\,1}\,\slashed{A}_{\,3}\,\psi_{\,2}\ ,
\ee
and is related to the term proportional to
\be
\slashed{\xi}_1\,+\,\slashed{\xi}_2\,+\,\slashed{\xi}_3\ .
\ee
\begin{figure}[htbp]
\begin{center}
\resizebox{7cm}{!}{\psfig{figure=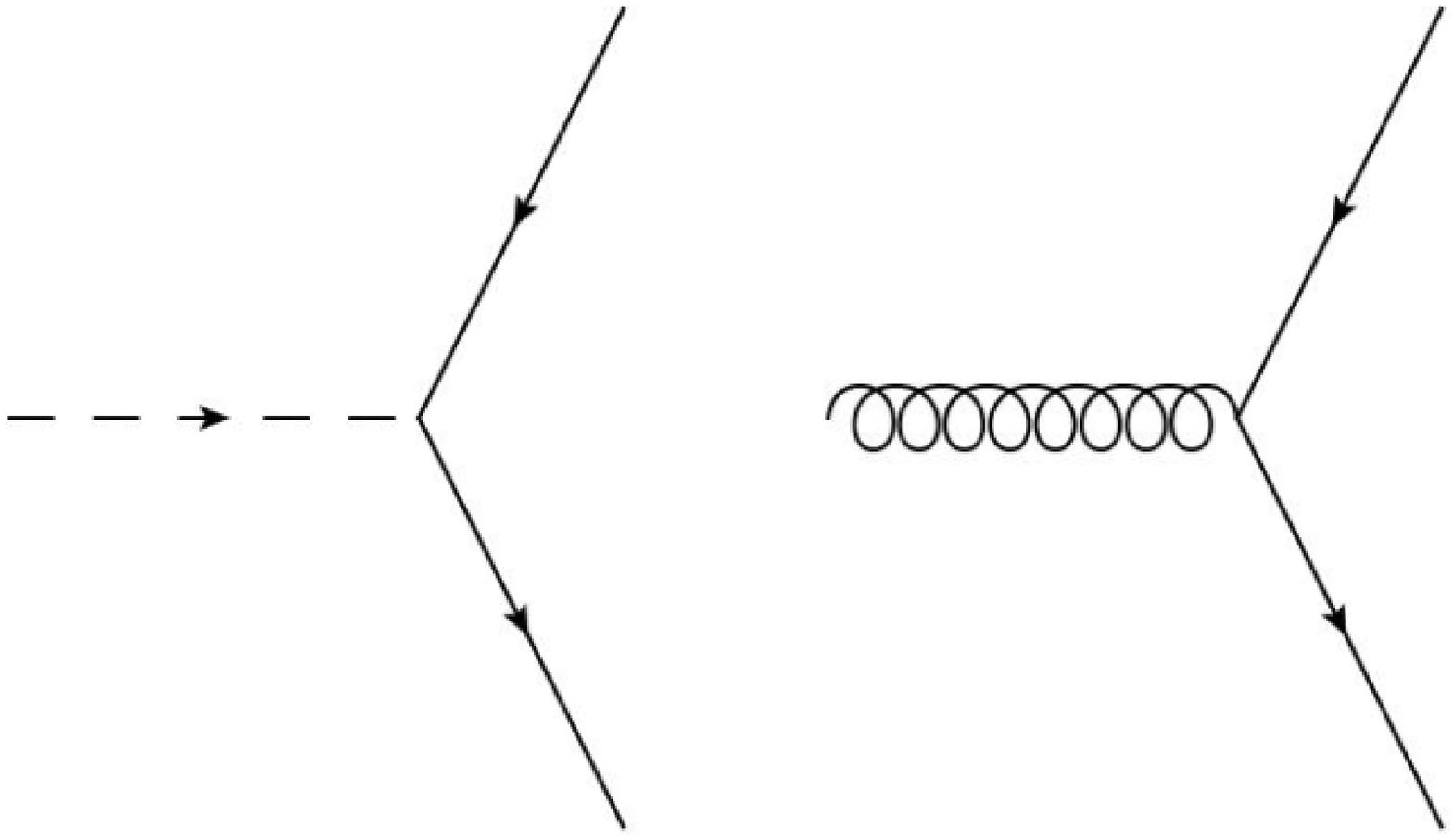,width=12cm}}
\caption{Cubic couplings respectively related to $\cG^{\,(0)\,f}_{123}$ and $\cG^{\,(1)\,f}_{123}$.}
\label{fig:CubicFermion}
\end{center}
\end{figure}
\!\!It is actually possible to reduce on-shell to this form any other combination of $\g$-matrices, so that there are no other options at the cubic order. The crucial observation is now to insist with the same idea of the bosonic case noticing that the cubic coupling kernel is nothing but
\be
\cK_{123}\,=\,\left(\cG^{\,(0)\,f}_{123}\,+\,\cG^{\,(1)\,f}_{123}\right)\,e^{\,\cG^{\,b}_{123}}\,=\,\left(\cG^{\,(0)\,f}_{123}\,+\, \cG^{\,(1)\,f}_{123}\right)\,\cK_{123}^b \ ,\label{FermiCubic}
\ee
where we have defined $\cG^{\,(0)\,f}_{123}$ and $\cG^{\,(1)\,f}_{123}$ to be respectively the cubic couplings of one scalar field with two fermions and of one gauge boson with two fermions drawn in fig.~\ref{fig:CubicFermion}, while labeling with a ``b'' the bosonic ones, that are respectively drawn in fig.~\ref{fig:cubicboson} and are encoded in $\cG^{\,b}_{123}$. We observe here a peculiar feature of fermionic amplitudes: they cannot be exponentiated, since the non-commutative nature of $\g$-matrices would lead to violations of gauge invariance, and hence eq.~\eqref{FermiCubic} exhausts all the possibilities. Starting from eq.~\eqref{FermiCubic} one can now generalize the cubic results just presented to the most general tree-level $n$-point functions containing also massless HS fermionic fields. The result can be expressed again, in analogy with the bosonic case, in terms of some $\cG_{12\ldots n}^{\,(i)\,f}$'s that by definition are linked with the color-ordered scattering amplitudes of a theory with scalars, fermions and gauge bosons and that can be constructed combining the cubic couplings in fig.~\ref{fig:cubicboson} and fig.~\ref{fig:CubicFermion} and completing the result with possible quartic terms. For instance, eliminating redundant pieces, for four-point functions we recover the following result
\be
\tilde{\cK}_{1234}^{\,f}\,=\,\left(\cG_{1234}^{\,(0)\,f}\,+\,\cG_{1234}^{\,(1)\,f}\right)\,\tilde{\cK}_{1234}^{\,b}\ .
\ee
Here $\tilde{\cK}_{1234}^{\,b}$ is exactly the bosonic generating function given in eq.~\eqref{Opengenfunc}, while $\cG_{1234}^{\,(0)\,f}$ and $\cG_{1234}^{\,(1)\,f}$ can be related to the four-point amplitudes in fig.~\ref{fig:fermifour},
\begin{figure}[htbp]
\begin{center}
\resizebox{12cm}{!}{\psfig{figure=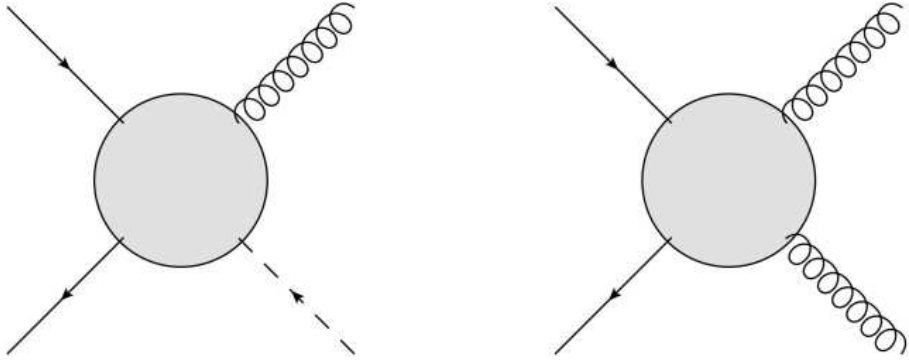,width=12cm}}
\caption{Quartic amplitudes respectively related to $\cG^{\,(0)\,f}_{1234}$ and $\cG^{\,(1)\,f}_{1234}$. We have not displayed here the amplitudes without a gauge boson insertion whose tensorial structure contains only $\d$-functions contracting the fermionic indices.}
\label{fig:fermifour}
\end{center}
\end{figure}
and can be computed obtaining\footnote{The contribution of the amplitude with two fermions, one gauge boson and one scalar turns out to be automatically produced by the combination of $\cG^{\,(0)\,f}_{1234}$ together with $\cG^{\,(0)\,b}_{1234}$ coming from the bosonic part of the kernel.}
\be
\left(\cG_{1234}^{\,(0)\,f}\right)\,=\,-\,\frac{1}{s\,u}\,\left(\d_{12}\,+\,\d_{23}\,+\,\d_{34}\,+\,\d_{41}\,+\,\d_{12}\,\d_{34}\,+ \,\d_{13}\,\d_{24}\,+\,\d_{14}\,\d_{23}\right)\ ,
\ee
where the $\d_{ij}$ contracts the fermionic indices between the fields $\psi_{\,i}$ and $\psi_{\,j}\,$, and
\begin{multline}
\left(\cG^{\,(1)\,f}_{1234}\right)(p_{\,i},\xi_{i})\,=\,-\,\left[\frac{1}{s}\,\left( \cG^{\,(1)\,f}_{12a}\,\star_{a}\cG^{\,(1)\,b}_{a34}\,+\, \cG^{\,(1)\,b}_{12a}\,\star_{a}\cG^{\,(1)\,f}_{a34}\,+\,\cG^{(1)\,f}_{12a}\,\cG^{\,(1)\,f}_{a34}\Big|_{\xi_a\,=\,0}\right)\right.\\\left.
+\,\frac{1}{u}\,\left( \cG^{\,(1)\,f}_{41a}\,\star_{a}\cG^{\,(1)\,b}_{23a}\,+\, \cG^{\,(1)\,b}_{41a}\,\star_{a}\cG^{\,(1)\,f}_{23a}\,+\,\cG^{\,(1)\,f}_{12a}\,\cG^{\,(1)\,f}_{a34}\Big|_{\xi_a\,=\,0}\right)\right]\ ,
\end{multline}
with the bosonic $\cG_{123}^{\,b}$ given in Section~\eqref{sec:binary}, the fermionic ones $\cG_{123}^{\,f}$ defined in this section and where, by convention, a contraction between the left-over fermionic indices in the four-fermion case is understood. To conclude, in the general case of $n$-point functions one is led to the Bose-Fermi kernel
\begin{multline}
\tilde{\cK}_{12\ldots n}\left(\Xi_i\right)\,=\,-\,\frac{1}{\r(P_{ij})}\ \left[1\,-\,\r(P_{ij})\sum_{c\in\cF}\cG^{(c)\,\text{fermi}}_{12\ldots n}\left(\Xi_i\right)\right]\,\exp\left[-\,\r(P_{ij})\sum_{d\in\cB}\cG^{(d)\,\text{bose}}_{12\ldots n}\left(\Xi_i\right)\right]\ ,\label{npointfermi}
\end{multline}
where the sums run over all possible structures related to some $n$-point function in a theory containing a scalar a spin-$1/2$ and a gauge boson, the terms proportional to $1$ within the first parenthesis entail the purely bosonic case already discussed in the previous section, while all the other ingredients have been defined in previous sections. Again, we consider this object as a generalized generating function whose relative coefficients are arbitrary functions of the Mandelstam-like invariants that do not introduce additional physical poles. For instance, a generic portion that one can recover from eq.~\eqref{npointfermi} is of the form
\begin{multline}
\tilde{\cK}_{12\ldots n}\,=\,-\,\frac{1}{\r(P_{ij})}\ a_{c;\a_1\ldots\a_{n-2}}\left(\{s_{ij}\}\,,\{s_{ijk}\}\,,\ldots\right)\\\times\, \left[-\,\r(P_{ij})\ \cG^{(c)\,\text{fermi}}_{12\ldots n}\left(\Xi_i\right)\right] \prod_{l\,=\,0}^{n-2}\frac{1}{\a_l!}\ \left[-\,\r(P_{ij})\ \cG^{(l)\,\text{bose}}_{12\ldots n}\left(\Xi_i\right)\right]^{\a_l}\ ,
\end{multline}
The discussion then extends to (generalized) \emph{closed-string}-like amplitudes along similar lines as in the bosonic case.

%%%%%%%%%%%%%%%%%%%%%%%%%%%%%%%%%%%%%%%%%%%%%%%%%%%%%%%%

\scss{Mixed-symmetry fields}

%%%%%%%%%%%%%%%%%%%%%%%%%%%%%%%%%%%%%%%%%%%%%%%%%%%%%%%%

Let us conclude this discussion mentioning the generalization of these results to the case of mixed-symmetry fields. A mixed-symmetry field is a Lorentz tensor of the form
\be
\phi_{\,\m_1\ldots\m_{s_1};\n_1\ldots\n_{s_2};\ldots}\ ,
\ee
where each index family (say $(\m_1\ldots\m_s)$) is totally symmetrized and no definite symmetrization is enforced between different families in the \emph{reducible} case, or a projection into some Young Tableaux is enforced in the \emph{irreducible} case. Here, for simplicity, we restrict the attention to reducible mixed-symmetry fields leaving open the option of choosing later a Young projection. The symbol calculus that we have developed so far extends naturally to the mixed-symmetry case simply defining more sets of variables
\be
\Xi_i^{\,m}\,=\,(p_{\,i}\,,\xi_i^m\,,\partial_{\theta^0_i}\,,\theta^m_i\,,\bar{\theta}^m_i)\ ,
\ee
as discussed in \cite{BRSTrev,mixedBRST}. Here the label $m$ identifies the index family to whom the symbol is related. We then define analogous generating functions, whose purely bosonic part is given by
\be
\phi_{\,i}(p_{\,i}\,,\xi_i^m)\,=\,\sum_{s_1\,s_2\,\ldots}\,\frac{1}{s_1!s_2!\ldots}\ \phi_{i\,\m_1\ldots\m_{s_1};\n_1\ldots\n_{s_2};\ldots}\,\xi_i^{1\,\m_1}\ldots\,\xi_i^{1\,\m_{s_1}}\,\xi_i^{2\,\n_1}\ldots\,\xi_i^{2\,\n_{s_2}}\ \ldots\ ,
\ee
together with related ghost parts, in the FDA framework. The corresponding BRST charge extends to
\be
Q\,=\,\theta^{\,0}p^{\,2}\,+\,\theta_{\,n}\,p\cdot\partial_{\xi_n}\,+\,\partial_{\bar{\theta}_{\,n}}\,p\cdot\xi_n\, -\,\theta_n\,\partial_{\bar{\theta}_{\,n}}\,\partial_{\theta^{\,0}}\ ,
\ee
with an implicit sum over $n$, while the various $\star$-contractions become
\begin{multline}
\star\,:\ \Big(\Phi_{\,1}(p_{\,1}\,,\xi^m_1)\,,\,\Phi_{\,2}(p_{\,2}\,,\xi^m_2)\Big) \ \ra \\ \Phi_{\,1}\star\,\Phi_{\,2}\,=\,\exp\Big(\sum_m\,\partial_{\xi^m_1}\cdot\,\partial_{\xi^m_2}\Big)\,\Phi_{\,1}(p_1\,,\xi^m_1)\ \Phi_{\,2}(p_2\,,\xi^m_2)\Big|_{\xi^m_i\,=\,0}\ ,\label{contraM}
\end{multline}
where $\xi^m_1$ and $\xi^m_2$ are only the commuting symbols, or, in the FDA framework,
\begin{multline}
\tilde{\star}\,:\ \Big(\Phi_{\,1}(p_{\,1},\theta^0_1,\xi^m_1,\theta^m_1,\bar{\theta}^m_1)\,,\,\Phi_{\,2}(p_{\,2},\theta^0_2,\xi^m_2,\theta^m_2, \bar{\theta}^m_2)\Big) \ \ra \\ \Phi_{\,1}\,\tilde{\star}\ \Phi_{\,2}\,=\,\exp\Big(\partial_{\xi^m_1}\cdot\partial_{\xi^m_2}\,-\,\partial_{\theta^m_1}\,\partial_{\bar{\theta}^m_2}\,+ \,\partial_{\bar{\theta}^m_1}\,\partial_{{\theta}^m_2}\Big)\\\times\,\Phi_{\,1}(p_1,\theta^0,\xi^m_1,\theta^m_1,\bar{\theta}^m_1)\ \Phi_{\,2}(p_2,\theta^0,\xi^m_2,\theta^m_2,\bar{\theta}^m_2)\,\Big|_{\xi^m_i\,,\theta^m_i\,,\bar{\theta}^m_i\,=\,0}\ ,\label{supercontra2M}
\end{multline}
and
\be
\star\,:\ \Big(\Phi_{\,1}(p_{\,1},\theta^0_1,\xi^m_1,\theta^m_1,\bar{\theta}^m_1)\,,\,\Phi_{\,2}(p_{\,2},\theta^0_2,\xi^m_2,\theta^m_2, \bar{\theta}^m_2)\Big) \ \ra \ \Phi_{\,1}\star\,\Phi_{\,2}\,=\,\int d\theta^0\,\left[\Phi_{\,1}\,\tilde{\star}\ \Phi_{\,2}\right]\ \theta^0\ ,\label{supercontraM}
\ee
where by convention
\be
p_{\,1}\,+\,p_{\,2}\,=\,0\ .
\ee
The two latter contractions over the super phase-space $\cH$ for mixed-symmetry fields reduce then to eq.~\eqref{contraM} if one restricts the attention to the bosonic coordinates and ghost-number zero super-fields. Remarkably, it turns out that with a trick it is possible to extract mixed-symmetry color-ordered kernels $\tilde{\cK}_{12\ldots n}$ from the totally symmetric ones. In the case of mixed-symmetry (spinor-)tensors one is then able to obtain the result
\begin{multline}
\tilde{\cK}_{1 2\ldots n}\left(\Xi^{k}_i\right)\,=\,-\,\frac{1}{\r(P_{ij})}\ \left[1\,-\,\r(P_{ij})\sum_{k_i}\sum_{c\in\cF}\cG^{(c)\,\text{fermi}}_{1^{k_1} 2^{k_2}\ldots n^{k_n}}\left(\Xi^{k_i}_i\right)\right]\\\times\,\exp\left[-\,\r(P_{ij})\sum_{k_j}\sum_{d\in\cB}\cG^{(d)\,\text{bose}}_{1^{k_1} 2^{k_2}\ldots n^{k_n}}\left(\Xi^{k_j}_j\right)\right]\ ,\label{mix}
\end{multline}
where the following generalized building blocks
\be
\cG_{12\ldots n}^{ij\ldots z}\,\equiv\,\cG_{12\ldots n}(\Xi_1^{i},\,\Xi_2^j,\,\ldots,\Xi_n^z)\ ,
\ee
have been used\footnote{I want to thank Euihun Joung and Luca Lopez for discussions on this point.}.
In this way we avoid a manifest symmetrization that is intrinsic in the definition of the kernel for totally symmetric fields, recovering the possibility to extract mixed-symmetry components from the generating function. Moreover, as before, the generating function needs to be interpreted in a generalized fashion modulo arbitrary relative functions of the Mandelstam-like invariants. The generic portion of $\tilde{\cK}_{12\ldots n}$ reads
\begin{multline}
\tilde{\cK}_{1 2\ldots n}\left(\Xi^{k}_i\right)\,\sim\,-\,\frac{1}{\r(P_{ij})}\ a\left(\{s_{ij}\,,s_{ijk}\,,\ldots\}\right)\\\times\, \left[-\,\r(P_{ij})\ \cG^{(c)\,\text{fermi}}_{1^{k_1} 2^{k_2}\ldots n^{k_n}}\left(\Xi^{k_i}_i\right)\right]\ \prod_{l_j}\prod_{d\,=\,0}^{n-2}\ \frac{1}{\a_{d,l}!}\,\left[-\,\r(P_{ij})\ \cG^{(d)\,\text{bose}}_{1^{l_1} 2^{l_2}\ldots n^{l_n}}\left(\Xi^{l_j}_j\right)\right]^{\,\a_{d,l}}\ ,\label{mostgenexp}
\end{multline}
where the $a\left(\{s_{ij}\,,s_{ijk}\,,\ldots\}\right)$ are functions of all Mandelstam-like invariants that do not introduce further physical poles and encodes both local $n$-point couplings that are gauge invariant under the linearized gauge transformations and the exchanges built from lower-point amplitudes so that, in our minimal scheme, are constrained once a consistent choice for the cubic coupling function has been found accordingly.

%%%%%%%%%%%%%%%%%%%%%%%%%%%%%%%%%%%%%%%%%%%%%%%%%%%%%%%%%%%%%%%%%%%%%%%%%%%%%%%%%%%%%%%%%%%%%%%%%%%%%%%%%%%%%%%%%%%%%%%%

\scs{Cubic couplings for bosonic and fermionic totally symmetric fields}\label{app:mostgeneral}

%%%%%%%%%%%%%%%%%%%%%%%%%%%%%%%%%%%%%%%%%%%%%%%%%%%%%%%%%%%%%%%%%%%%%%%%%%%%%%%%%%%%%%%%%%%%%%%%%%%%%%%%%%%%%%%%%%%%%%%%

In this Appendix we describe briefly the results of \cite{cubicstring} from a field theory perspective\footnote{I wish to thank Euihun Joung for very useful discussions and comments on the content of the first part of this appendix.}. From this point of view we analyze the constraints that Lorentz invariance, together with gauge invariance, impose on the generic on-shell and gauge-fixed generating function of bosonic and fermionic cubic couplings for totally symmetric (spinor-)tensors. We exploit the simplification of the on-shell gauge-fixed system that we describe in more detail in the following appendices in order to recognize simple objects that are actually the building blocks of any HS three-point function and contain the main physical information about cubic interactions. Considering irreducible HS bosonic fields and leaving aside the possibility of using the totally anti-symmetric tensor $\e_{\,\m_1\ldots\m_D}$ that gives non-trivial options only for lower dimensions, at three points the most general kernel that one can write down requiring Lorentz symmetry is given by
\be
\cK_{123}^{\,b}\,=\,A\left(\xi_1\cdot p_{\,23}\,,\xi_2\cdot p_{\,31}\,,\xi_3\cdot p_{\,12}\,,\xi_1\cdot\xi_2\,,\xi_2\cdot\xi_3\,,\xi_3\cdot\xi_1\right)\ .
\ee
Here we have enforced the transversality and traceless constraints by
\be
\xi_i\cdot\xi_i\,=\,0\ ,\qquad \xi_i\cdot p_{\,i}\,=\,0\ ,
\ee
while
\be
A(z_1,z_2,z_3;y_1,y_2,y_3)\ ,
\ee
is a function such that $\cK_{123}^{\,b}$ is totally cyclic under permutations of the labels $\{1,2,3\}$. Gauge invariance then puts constraints on the dependence on $y_1$, $y_2$ and $y_3$ since the on-shell equations
\be
p_{\,i}\cdot\partial_{\xi_i}\,\cK_{123}^{\,b}\,\approx\,0\ ,
\ee
are nothing but the differential equations
\be
p_{\,i}\cdot\partial_{\xi_i}\,A\left(\xi_1\cdot p_{\,23}\,,\xi_2\cdot p_{\,31}\,,\xi_3\cdot p_{\,12}\,,\xi_1\cdot\xi_2\,,\xi_2\cdot\xi_3\,,\xi_3\cdot\xi_1\right)\,\approx\,0\ ,\qquad i\,=\,1,\ 2,\ 3\ ,
\ee
that actually, by cyclic symmetry, reduce to $2$ independent conditions on $A(z_1,z_2,z_3;y_1,y_2,y_3)$. Hence, considering the following useful change of variable
\begin{multline}
A\left(\xi_1\cdot p_{\,23}\,,\xi_2\cdot p_{\,31}\,,\xi_3\cdot p_{\,12}\,,\xi_1\cdot\xi_2\,,\xi_2\cdot\xi_3\,,\xi_3\cdot\xi_1\right)\\=\,\tilde{A}\left(\xi_1\cdot p_{\,23}\,,\xi_2\cdot p_{\,31}\,,\xi_3\cdot p_{\,12}\,,\xi_1\cdot\xi_2\,\xi_3\cdot p_{\,12}\,,\xi_2\cdot\xi_3\,\xi_1\cdot p_{\,23}\,,\xi_3\cdot\xi_1\,\xi_2\cdot p_{\,31}\right)\ ,\label{change}
\end{multline}
one ends up with the following differential equations for $\tilde{A}(z_1,z_2,z_3;y_1,y_2,y_3)$:
\be
\partial_{\,y_{\,ij}}\,\tilde{A}(z_1,z_2,z_3;y_1,y_2,y_3)\,=\,0\ ,
\ee
where, as usual
\be
\partial_{\,y_{\,ij}}\,=\,\partial_{\,y_{\,i}}\,-\,\partial_{\,y_{\,j}}\ .
\ee
The solution is then shown to be
\be
\tilde{A}(z_1,z_2,z_3;y_1,y_2,y_3)\,=\,a(z_1,z_2,z_3; y_1\,+\,y_2\,+\,y_3)\ ,
\ee
where $a(z_1,z_2,z_3; w)$ is the coupling function that was introduced in Section~\ref{sec:binary}, and where the term $y_1\,+\,y_2\,+\,y_3$ gives rise to the object that we call $\cG_{123}^{(1)}$ in eq.~\eqref{G2}.

Turning now to the fermionic case we need to study the most general gauge-invariant object containing at least one $\g$-matrix. The basic additional ingredients are then
\be
\slashed{p}_{\,i}\ ,\quad \slashed{\xi}_i\ ,
\ee
so that labeling the two tensor-spinors with indices $1$ and $2$ we can use momentum conservation in order to reduce the number of building blocks to
\be
\slashed{p}_{\,1}\ ,\quad \slashed{p}_{\,2}\ ,\quad \slashed{\xi}_{\,i}\ .
\ee
Exploiting the standard relations
\be
\slashed{a}\slashed{b}\,+\,\slashed{b}\slashed{a}\,=\,2\,a\cdot b\ ,
\ee
one can easily convince that any monomial of the form
\be
\bar{\psi}_{\,1}\,\left(\ldots\, \slashed{p}_{\,1}\,\ldots\right)\,\psi_{\,2}\ ,
\ee
will generate terms proportional to $p_{\,12}\cdot\xi_3$ or $p_{\,31}\cdot\xi_2$ while all other being zero by mass-shell condition of the three external states. Iteratively we are then able to eliminate all $\slashed{p}_{\,i}$'s leaving out only the $\slashed{\xi}_i$'s. Among those, we are again able to eliminate all terms proportional to $\slashed{\xi}_1$ and $\slashed{\xi}_2$ since we can anticommute them till they act on the polarization spinor-tensors giving zero by the $\g$-traceless constraint but leaving out terms of the form $\xi_i\cdot\xi_j$ that we have dealt with previously, in the bosonic case. To reiterate, we are actually left only with a single possible object given by
\be
\left(1\,+\,\l\ \slashed{\xi}_3\right)^{12}\ ,
\ee
with $\l$ a constant, that exhausts all the possibilities for cubic fermionic couplings described in Section~\ref{sec:fermionic} once combined with the terms of the form $\xi_{i}\cdot p_{\,jk}$ and $\xi_i\cdot\xi_j$ that have to satisfy the same equations discussed for bosons before. Hence,
\be
\left(\cK_{123}^{\,f}\right)\,=\,\left(1\,+\,\slashed{\xi}_1\,+\,\slashed{\xi}_2\,+\,\slashed{\xi}_3\right)\,\cK_{123}^{\,b}\ ,
\ee
where we have written the result in a totally cyclic form. We have thus recovered that the general gauge-invariant cubic HS coupling is built from those of a gauge boson, a spin $1/2$ fermion and a scalar. Similar arguments can be used to construct the results for the higher correlation functions that we have found in this work.

%%%%%%%%%%%%%%%%%%%%%%%%%%%%%%%%%%%%%%%%%%%%%%%%%%%%%%%%%%%%%%%%%%%%%%%%%%%%%%%%%%%%%

\scs{``Off-shell'' Theory vs. ``On-shell'' Theory: quadratic level}\label{app:quadratic}

%%%%%%%%%%%%%%%%%%%%%%%%%%%%%%%%%%%%%%%%%%%%%%%%%%%%%%%%%%%%%%%%%%%%%%%%%%%%%%%%%%%%%

This Appendix is devoted to clarifying the relation between the on-shell gauge-fixed theory and the various off-shell completions one can in principle consider, revisiting the constructions so far known \cite{Fronsdal,FranciaSagnotti,minimal,francia10} as well as other possible options and putting them on the same footing. For brevity, we consider only the bosonic case while we expect similar considerations to be possible for fermions and mixed-symmetry fields. In order to try and be general and to clarify the basic viewpoint of the present paper, our starting point is \emph{Lorentz invariance}. It is indeed well known that any massive bosonic totally symmetric and unitary representation of the Poincarè group can be \emph{uniquely} specified by the Fierz system
\be
\begin{split}
\left(p^{\,2}\,+\,m^{\,2}\right)\,\phi(p\,,\xi)\,&=\,0\ ,\\
p\cdot\partial_{\xi}\ \phi(p\,,\xi)\,&=\,0\ ,\label{massiveFP}
\end{split}
\ee
where $\phi(p\,,\xi)$ is a generating function, the first equation selects the value of the Casimir operator $p^{\,2}$ and the second enforces the cancelation of ghosts projecting the various polarization tensors into their positive definite components.
We stress that these two equations encode the physical requirements needed to select a generic massive bosonic representation of the Poincarè group. To this system one usually adds, starting from the works of Fierz and others \cite{Majorana:1932rj}, the third \emph{algebraic} equation
\be
\partial_{\xi}\cdot\partial_{\xi}\ \phi(p\,,\xi)\,=\,0\ ,\label{traceconstr}
\ee
that is meant to select an \emph{irreducible} representation, and hence has a slightly different origin than the first two.
If one insists in using a covariant formalism also in the massless case, subtleties will soon arise since a Lorentz transformation ${\L_{\,\m}}^{\n}$ for massless fields will generate transverse components to the polarization tensors
\be
\phi(p\,,\xi)\,\ra\,e^{\left(\xi\cdot\L\cdot\partial_{\xi}\right)}\,\phi(p\,,\xi)\,-\,p\cdot\xi\,\L(p\,,\xi)\ .
\ee
so that in general the covariance, that one would like to preserve when using a tensorial representation, is lost unless one requires on-shell gauge invariance. The on-shell description of a massless HS particle can then be carried out considering the analogous Fierz system
\be
\begin{split}
p^{\,2}\,\phi(p\,,\xi)\,&=\,0\ ,\\
p\cdot\partial_{\xi}\ \phi(p\,,\xi)\,&=\,0\ ,\label{masslessFP}
\end{split}
\ee
where the field has to be considered as an equivalence class defined by the relation
\be
\phi(p\,,\xi)\,\sim\,\phi(p\,,\xi)\,+\,p\cdot\xi\,\L(p\,,\xi)\ .
\ee
In this fashion one is able to recover manifest covariance for the transformation properties of tensors in terms of their equivalence classes. By consistency with gauge invariance one then recovers an analogous Fierz system for the gauge parameter $\L(p\,,\xi)$ given by
\be
\begin{split}
p^{\,2}\,\L(p\,,\xi)\,&=\,0\ ,\\
p\cdot\partial_{\xi}\,\L(p\,,\xi)\,&=\,0\ .\label{GmasslessFP}
\end{split}
\ee
Moreover, any constraint added to the system, selecting some particular representation of the Poincarè group, will propagate to the gauge parameter as well selecting the needed degrees of freedom that serve in order to maintain covariance of the representation. For instance, if we choose to impose the traceless constraint \eqref{traceconstr} this automatically propagates over the gauge parameter that is to be chosen in a way that preserves such constraint. Hence, one is led to consider a constrained gauge parameter
\be
\partial_{\xi}\cdot\partial_{\xi}\,\L(p\,,\xi)\,=\,0\ .
\ee
We want to stress that, starting from an on-shell system together with a choice of the propagating degrees of freedom gives a completely meaningful setting to address the problem of \emph{consistent deformations}, that acquires in this way its \emph{simplest} covariant form. One should then address the tedious issue of off-shell completion, that here we refer to as the procedure of relaxing the transversality constraint on $\phi(p\,,\xi)$ as well as other possible constraints depending on the particular representation that has been considered and also on the kind of off-shell completion. Hence, we decide to start from the Lagrangian
\be
\cL\,=\,\frac{1}{2}\,\phi(-p\,,\xi_{\,1})\,\star \, p^{\,2} \,\phi(p\,,\xi_{\,2})\ ,\label{totallygauge}
\ee
written in terms of the bosonic $\star$ of eq.~\eqref{contra} and that is gauge invariant if the transversality constraint
\be
p\cdot\partial_{\xi}\,\phi(p\,,\xi)\,=\,0\ ,
\ee
is enforced. Additional constraints like
\be
\partial_{\xi}\cdot\partial_{\xi}\,\phi(p\,,\xi)\,=\,0\ ,\label{traceless}
\ee
that select irreducible representations are not required here by gauge invariance and will be considered in detail in the following depending on the off-shell completion. We begin by analyzing the standard cases of spin-$1$ and spin-$2$ fields before going on to consider the most general case of arbitrary spin.

In the spin-$1$ case the quadratic action reads
\be
\cL\,=\,\frac{1}{2}\ A_{\,\m}\,(p^{\,2})\,A^{\,\m}\ ,\label{spin1}
\ee
where now we want to relax the constraint $p\cdot A\,=\,0$. In order to complete this action to a fully gauge-invariant one we can start computing its gauge variation under $\delta A_{\,\m}\,=\,p_{\,\m}\ \L$ recovering
\be
\d\,\cL\,=\,A_{\,\m}\,p^{\,2}\,p^{\,\m}\,\L\ .\label{unwanted}
\ee
This term can nonetheless be compensated adding a divergence that now is non-vanishing and whose gauge variation produces
\be
\delta (p\cdot A)\,=\,p^{\,2}\,\L\ .\label{Useful}
\ee
To reiterate, a counter-term proportional to the divergence of $A_{\,\m}$ cancels the unwanted non--zero gauge variation \eqref{unwanted} and leads off-shell to
\be
\cL\,=\,\frac{1}{2}\ \Big[A_{\,\m}\,(p^{\,2})\,A^{\,\m}\,+\,(p\cdot A)_{\,\m}\,(p\cdot A)^{\,\m}\Big]\ .
\ee
In a similar fashion for spin-2 the starting point is
\be
\cL\,=\,\frac{1}{2}\ h_{\,\m\n}\,(p^{\,2})\,h^{\,\m\n}\ ,\label{spin2}
\ee
where $h_{\,\m\n}$ is a transverse polarization tensor, and we would like to relax this constraint. Computing as before the gauge variation of \eqref{spin2}, under the gauge transformation
\be
\d\,h_{\,\m\n}\,=\,p_{\,\m}\,\L_{\,\n}\,+
\,p_{\n}\,\L_{\,\m}\ ,\label{gauge2}
\ee
one obtains the result
\be
\d\,\cL\,=\,2\,h_{\,\m\n}\,(p^{\,2})\,p^{\,\m} \,\L^{\,\n}\ ,
\ee
In order to cancel again the unwanted gauge variation, one can proceed as in the spin-$1$ case, constructing a tensor proportional in general to traces and divergences that maintains the useful gauge transformation \eqref{Useful}, but now for a spin two field. Hence, it is convenient to define a \emph{generalized} de Donder tensor $\cD$ by the condition
\be
\d\cD\,=\,p^{\,2}\,\L(p)\ ,\label{dedeq}
\ee
with the ansatz
\be
\cD_{\,\m}\,=\,p\cdot h_{\,\m}\,+\,p_{\,\m}\,\cA(h)\ ,
\ee
so that eq.~\eqref{dedeq} translates into the equivalent condition
\be
\d\cA_{\,\m}(h)\,=\,-\,p\cdot \L(p)\ ,\label{aeq}
\ee
and hence the full gauge-invariant Lagrangian is given in a manifestly self-adjoint form by
\be
\cL\,=\,\frac{1}{2}\ h_{\,\m\n}\,(p^{\,2})\,h^{\,\m\n}\,+\,\cD_{\,\m}\,\cD^{\,\m}\,-\,\frac{1}{2}\,\cA(h)\,p^{\,2}\,\cA(h)\ .\label{generallag}
\ee
This is a very general form for the Lagrangian and $\cA(h)$ can in principle depend \emph{also} on additional auxiliary fields. Solving now equation \eqref{aeq} for spin-$2$ one has various options. If one chooses to describe a \emph{reducible} spin-$2$ field, the $h_{\,\m\n}$ would not be constrained on-shell by
\be
h^{\,\prime}\,=\,0\ ,
\ee
and this means that one is not allowed to introduce traces in order to solve for $\cA$. If one insists not adding auxiliary fields, one can easily convince oneself that the solution is \emph{non-local} and is given by
\be
\begin{split}
\cD^{(1)}_{\,\m}\,&=\,(p\cdot h)_{\,\m}\,-\,\frac{1}{2}\,\frac{p_{\,\m}}{p^{\,2}}\ (p\cdot p\cdot h)\ ,\\
\cA^{(1)}\,&=\,-\,\frac{1}{2}\,\frac{1}{p^{\,2}}\ (p\cdot p\cdot h)\ .
\end{split}
\ee
On the contrary, if one chooses to describe an irreducible spin-$2$ field the previous solution is still admissible, since one can excite only the spin-$2$ components with a further projection, but since terms proportional to the trace can now be considered in the solution for $\cD_{\,\m}$, other solutions show up like, for instance
\be
\begin{split}
\cD_{\,\m}^{(2)}\,&=\,(p\cdot h)_{\,\m}\,-\,\frac{1}{2}\, p_{\,\m}\,h^{\,\prime}\ ,\\
\cA^{(2)}\,&=\,-\frac{1}{2}\,h^{\,\prime}\ .
\end{split}
\ee
The latter, if substituted into \eqref{generallag} give rise to the standard linearized Einstein-Hilbert Lagrangian whose equations of motion can be expressed in the usual form
\be
\cR_{\,\m\n}\,-\,\frac{1}{2}\,\eta_{\m\n}\,\cR^{\,\prime}\,=\,0\ ,
\ee
where $\cR_{\,\m\n}$ is the linearized Ricci tensor.

The spin-$2$ discussion generalizes directly to the HS case, so that starting from the constrained system identified by the Lagrangian \eqref{totallygauge} one recovers the full off-shell Lagrangian
\begin{multline}
\cL\,=\,\frac{1}{2}\,\left[\phi_{\,1}(p_{\,1}\,,\xi_{\,1})\,\star \, p^{\,2}_{\,2} \,\phi_{\,2}(p_{\,2}\,,\xi_{\,2})\,+\,\cD_1(p_{\,1}\,,\xi_1)\,\star\,\cD_2(p_{\,2}\,,\xi_2)\vphantom{\frac{1}{2}}\right.\\\left.\vphantom{\frac{1}{2}}- \,\cA_{\,1}(p_{\,1}\,,\xi_1) \,\star\, p_{\,2}^{\,2}\,\cA_{\,2}(p_{\,2}\,,\xi_2)\right]\ ,\label{quadHSlag}
\end{multline}
where by convention
\be
p_{\,1}\,+\,p_{\,2}\,=\,0\ ,
\ee
and that is gauge invariant without requiring any transversality constraint on the field or on the gauge parameter if $\cD(p\,,\xi)$ and $\cA(p\,,\xi)$ satisfy
\be
\cD(p\,,\xi)\,=\,(p\cdot\partial_{\xi})\,\phi(p\,,\xi)\,+\,p\cdot\xi\,\cA(p\,,\xi)\ ,\qquad\d\cA(p\,,\xi)\,=\,-\,p\cdot\partial_{\xi}\,\L(p\,,\xi)\ ,\label{Ded}
\ee
and where $\cA$ is needed in order to relax the on-shell constraint
\be
p\cdot\partial_{\xi}\,\L(p\,,\xi)\,=\,0\ .
\ee
At this point we have recovered a general \emph{form} for a quadratic manifestly self-adjoint and gauge-invariant Lagrangian in terms of two objects $\cD$ and $\cA$ that can depend on the field $\phi(p\,,\xi)$ as well as other auxiliary fields. The next step is to solve eqs.~\eqref{Ded}, and in order to do this one needs to specify as before whether one is restricting the attention to a single irreducible HS tensor enforcing the traceless constraint on-shell or, alternatively, one is describing a reducible representation containing, for instance, spin-$s$ down to spin $1$ or $0$ polarizations respectively for $s$ odd or even, or also other intermediate possibilities. The second option, as for the spin-$2$ case, brings about non-localities since one is not allowed to add to the completely gauge-fixed Lagrangian trace terms being them non-zero already at the gauge-fixed level. Then, a solution to eqs.~\eqref{Ded} can be given as
\be
\begin{split}
\cD(p\,,\xi)\,&=\ :\exp\left[\,\frac{1}{p^{\,2}}\,(p\cdot\xi)\,(p\cdot\partial_{\,\xi})\,\right]\!:\,(p\cdot\partial_{\,\xi})\,\phi(p\,,\xi)\ ,\\
\cA(p\,,\xi)\,&=\ \left(:\exp\left[\,\frac{1}{p^{\,2}}\,(p\cdot\xi)\,(p\cdot\partial_{\,\xi})\,\right]\!:\,-\,1\right)\,(p\cdot\partial_{\,\xi})\,\phi(p\,,\xi)\ ,
\end{split}
\ee
that actually, when substituted into the Lagrangian \eqref{quadHSlag}, produces the result obtained by Francia in \cite{francia10}
\be
\cL\,=\,\cR_1(p_{\,1},\xi_1)\,\star\,\left(\frac{1}{p^{\,2}}\right)^{s-1}\cR_2(p_{\,2},\xi_2)\ .
\ee
The other option is to consider solutions for $\cD(p\,,\xi)$ and $\cA(p\,,\xi)$ containing \emph{single} traces together with divergences trying to address the case in which a \emph{single} irreducible spin-$s$ polarization is propagated. A solution is in this case the usual de Donder term
\be
\cD(p\,,\xi)\,=\,\left[(p\cdot\partial_{\,\xi})\,-\,\frac{1}{2}\,(p\cdot\xi)\,\partial_{\,\xi}\cdot\partial_{\,\xi}\right]\,\phi(p\,,\xi)\ , \label{DeDonderFronsdal}
\ee
together with
\be
\cA(p\,,\xi)\,=\,-\,\frac{1}{2}\,\partial_{\,\xi}\cdot\partial_{\,\xi}\,\phi(p\,,\xi)\ .
\ee
Their gauge variation turns out to be
\be
\begin{split}
\d\,\cD(p\,,\xi)\,&=\,-\,\frac{1}{2}\,(p\cdot\xi)^{\,2}\,\partial_{\,\xi}\cdot\partial_{\,\xi}\,\L(p\,,\xi)\ ,\\
\d\,\cA(p\,,\xi)\,&=\,-\,p\cdot\partial_{\xi}\,\L(p\,,\xi)\,-\,\frac{1}{2}\,p\cdot\xi\,\partial_{\,\xi}\cdot\partial_{\,\xi}\,\L(p\,,\xi)\ ,
\end{split}
\ee
so that eq.~\eqref{DeDonderFronsdal} is a solution only for traceless gauge parameters. This reflects the traceless constraint on the gauge parameter that we have discussed at the beginning of this Appendix, in eq.~\eqref{GmasslessFP}. Therefore, while the transversality constraints can be relaxed, this is not true for the traceless constraint
\be
\partial_{\xi}\cdot\partial_{\xi}\,\L(p\,,\xi)\,=\,0\ ,
\ee
that has to be enforced also off-shell in order to maintain the same number of gauge symmetries, while an additional number of them would spoil the original Fierz system if no auxiliary field is added.
Then, having recovered a constrained gauge parameter, one can also prove that the double trace of the field is gauge invariant, so that it is no more possible in general\footnote{More explicitly when the double trace of the field is not a scalar.} to choose a gauge of the form
\be
\cD(p\,,\xi)\,=\,0\ ,
\ee
to return to the on-shell system we started from. This actually reflects the fact that no gauge-invariant massless field can carry a covariant tensorial representation of the Poincarè group. Hence, to avoid inconsistencies, one is also forced to put in general the additional constraint
\be
(\partial_{\xi}\cdot\partial_{\xi})^{\,2}\,\phi(p\,,\xi)\,=\,0\ ,
\ee
that, from a more general point of view, serves here to recover an off-shell system that is equivalent to the one we started from on-shell.
We have seen here that the constraints on the gauge parameters and on the fields originate already at the level of the Fierz system \eqref{masslessFP}. Hence, it is tempting to consider the on-shell system \eqref{totallygauge} as more fundamental so that if one wants to recover an unconstrained gauge symmetry one can think to address the problem starting from the traceless constraint in the Fierz equations. One option is to introduce compensators via a Stueckelberg shift so that the traceless constraint becomes
\be
\partial_{\xi}\cdot\partial_{\xi}\,\phi(p\,,\xi)\,=\,0\,\ra\,\partial_{\xi}\cdot\partial_{\xi}\,\phi(p\,,\xi)\,-\,p\cdot\xi\,\a(p\,,\xi)\,=\,0\ .
\ee
with
\be
\d\a(p\,,\xi)\,=\,\partial_{\xi}\cdot\partial_{\xi}\,\L(p\,,\xi)\ .
\ee
With this choice one can eliminate the traceless constraint on the gauge parameter, recovering a situation analogous to the first example. One can then eliminate the transversality constraints with the Lagrangian \eqref{quadHSlag} and we can solve for $\cD(p\,,\xi)$ and $\cA(p\,,\xi)$ recovering now the solution
\be
\begin{split}
\cD(p\,,\xi)\,&=\,\Big[(p\cdot\partial_{\,\xi})\,-\,\frac{1}{2}\,(p\cdot\xi)\,\partial_{\,\xi}
\cdot\partial_{\,\xi}\Big]\,\phi(p\,,\xi)\,+\,\frac{1}{2}\,(p\cdot\xi)^{\,2}\,\a(p\,,\xi)\ ,\\
\cA(p\,,\xi)\,&=\,-\,\frac{1}{2}\Big(\partial_{\,\xi}
\cdot\partial_{\,\xi}\,\phi(p\,,\xi)\,-\,(p\cdot\xi)\,\a(p\,,\xi)\Big)\ ,
\end{split}
\ee
that satisfy \eqref{Ded} without any constraint on the gauge parameter. Still there is a subtlety that is related to the fact that there exists a completion of the double trace of the field that is now gauge invariant so that in order to have the possibility of fixing the gauge recovering the on-shell system we started from one still needs to add the constraint
\be
\partial_{\,\xi}\cdot\partial_{\,\xi}\,\Big[\partial_{\,\xi}\cdot\partial_{\,\xi}\,\phi(p\,,\xi)\,- \,(p\cdot\xi)\,\a(p\,,\xi)\Big]\,=\,0\ ,
\ee
that can be enforced with the addition of a Lagrange multiplier as in \cite{FranciaSagnotti,minimal} and reduces to the previous double-traceless constraint once the gauge symmetry is fixed in order to eliminate the compensator $\a(p\,,\xi)$. To conclude this Appendix, let us stress again the key role of the on-shell system that actually is the common origin of any off-shell completion since even the BRST charge $Q$ defined in \eqref{BRST} is a gauge-invariant completion of $p^{\,2}$ whose Lagrangian can be obtained along the same lines starting from the corresponding Fierz system enlarged in order to contain the auxiliary fields $C$ and $D$. We expect that changing the constraints on the field $\phi$ one is able to recover, in principle, any kind of theory propagating intermediate subset of the fields contained into a polarization tensor of spin-$s$. The same logic that we have discussed at length in the relatively simple case of the free Lagrangian is expected to apply also at the cubic and quartic level, so that a consistent deformation of the on-shell system should be extendable off-shell. In this work we have restricted the attention to the formalism associated to the super phase-space $\cH$ that has been, somehow, a source of inspiration, but any other off-shell completion should be equally meaningful modulo the fact that some of them can be more convenient to recover a covariant representation of HS gauge symmetry and for this reason deserve a closer study. Analogous observations are expected to work for fermionic fields, as well as for mixed-symmetry fields, of course, with different solutions for $\cD$ and $\cA$, together with different options depending on the kind of representation of the Lorentz group that is being considered.

%%%%%%%%%%%%%%%%%%%%%%%%%%%%%%%%%%%%%%%%%%%%%%%%%%%%%%%%%%%%%%%%%%%%%%%%%%%%%%

\scs{``Off-shell'' Theory vs. ``On-shell'' Theory: interactions}\label{app:cubic}

%%%%%%%%%%%%%%%%%%%%%%%%%%%%%%%%%%%%%%%%%%%%%%%%%%%%%%%%%%%%%%%%%%%%%%%%%%%%%%

Considerations similar to those present in the previous Appendix are expected to hold at each order. Here, for brevity, we consider the general result for the cubic bosonic case in the irreducible Fronsdal setting, while higher-order results as well as the extension to the fermionic case can be recovered in a similar way starting from the deformations of the on-shell system \eqref{on-shell} that we have discussed or from the corresponding deformation of the fermionic system \eqref{on-shellf}. The \emph{on-shell} $n$-point functions are solutions of eq.~\eqref{decoupling} that we can rewrite in the following form\footnote{In what follows for convenience we make the substitution $\xi_i\ra\partial_{\xi_i}$ in all the $\cG^{(i)}$'s.}
\be
\left[\tilde{\cK}_{12\ldots n}(\xi_i\ra\partial_{\xi_i})\,,p_{\,i}\cdot\xi_i\right]\,\approx\,0\ ,\label{decoupling2}
\ee
where the symbol $\approx$ means that the equation is satisfied in the on-shell system \eqref{on-shell}. In order to go off-shell one is to relax the transversality constraint so that eq.~\eqref{decoupling2} holds modulo the full \emph{off-shell} EoM's. The procedure is tedious but straightforward and rests on finding the needed counterterms proportional to traces and divergences that compensate the traces and divergences coming from the original on-shell result. Moreover, since any $\tilde{\cK}$ is recursively related to $\cG_{123}$ in eq.~\eqref{G}, one can recognize the off-shell completion of the former exploiting the off-shell completion of the latter paying attention to the fact that the analog of the $\star$-contraction is off-shell the \emph{propagator numerator}.

Restricting the attention to the off-shell completion of
\be
\tilde{\cK}_{123}\,=\,\exp\left(\cG_{123}\right)\ ,
\ee
we can exploit the general setting of the previous Appendix considering the following identity
\be
p^{\,2}\phi(p\,,\xi)\,=\,-\,\cF(p\,,\xi)\,+\,p\cdot\xi\,\cD(p\,,\xi)\ ,
\ee
where we may call $\cF(p\,,\xi)$ \emph{generalized} Fronsdal tensor and where the EoM's coming from the off-shell system are exactly
\be
\cF(p\,,\xi)\,\approx\,0\ .\label{onshelleq}
\ee
Hence, evaluating the linearized gauge variation of the vertex on-shell in the sense of eq.~\eqref{onshelleq} the following commutators show up
\be
\begin{split}
[\cG_{123}\,,\,p_{\,1}\cdot\xi_{\,1}]\,\phi_{\,1}\,\phi_{\,2}\,\L_{\,3}&\approx-\,p_{\,2}\cdot\xi_{\,2} \Big[(\partial_{\xi_{\,3}}\cdot\partial_{\xi_{\,2}}\,+\,1) \,\cD_{\,2}\,+\,p_{\,2}\cdot\partial_{\xi_{\,3}}\,\cA_2\Big]\phi_{\,1}\,\L_{\,3}
\\ \vphantom{\left(\pm\sqrt{\frac{\a^{\,\prime}\!\!}{2}}\right)^{\,3}}
&+ \d_3\Big[(\partial_{\xi_{\,2}}\cdot\partial_{\xi_{\,3}}\,+\,1)\ \cD_{\,3}\,+\,p_{\,3}\cdot\partial_{\xi_{\,2}}\,\cA_3\Big]\, \phi_{\,1}\,\phi_{\,2}\ ,\\
[\cG_{123}\,,\,p_{\,2}\cdot\xi_{\,2}]\,\phi_{\,1}\,\phi_{\,2}\,\L_{\,3} & \approx\,p_{\,1}\cdot\xi_{\,1} \Big[(\partial_{\xi_{\,3}}\cdot\partial_{\xi_{\,1}}\,+\,1) \,\cD_{\,1}\,+\,p_{\,1}\cdot\partial_{\xi_{\,3}}\,\cA_1\Big]\phi_{\,2}\,\L_{\,3}
\\ \vphantom{\left(\pm\sqrt{\frac{\a^{\,\prime}\!\!}{2}}\right)^{\,3}}
&- \d_3\Big[(\partial_{\xi_{\,1}}\cdot\partial_{\xi_{\,3}}\,+\,1)\ \cD_{\,3}\,+\,p_{\,3}\cdot\partial_{\xi_{\,1}}\,\cA_3\Big]\, \phi_{\,1}\,\phi_{\,2}\ ,\\
[\cG_{123}\,,\,p_{\,3}\cdot\xi_{\,3}]\,\phi_{\,1}\,\phi_{\,2}\,\L_{\,3} & \approx\,p_{\,2}\cdot\xi_{\,2} \Big[(\partial_{\xi_{\,1}}\cdot\partial_{\xi_{\,2}}\,+\,1) \,\cD_{\,2}\,+\,p_{\,2}\cdot\partial_{\xi_{\,1}}\,\cA_2\Big]\phi_{\,1}\,\L_{\,3}
\\ \vphantom{\left(\pm\sqrt{\frac{\a^{\,\prime}\!\!}{2}}\right)^{\,3}}
& -\,p_{\,1}\cdot\xi_{\,1} \Big[(\partial_{\xi_{\,2}}\cdot\partial_{\xi_{\,1}}\,+\,1) \,\cD_{\,1}\,+\,p_{\,1}\cdot\partial_{\xi_{\,2}}\,\cA_1\Big]\phi_{\,2}\,\L_{\,3}\ ,
\label{commutators}
\end{split}
\ee
so that defining the tensor structure
\be
\cH_{ij}\,=\,(\partial_{\xi_{\,i}}\cdot\partial_{\xi_{\,j}}\,+\,1) \,\cD_{\,j}\,+\,p_{\,j}\cdot\partial_{\xi_{\,i}}\,\cA_j\ ,
\ee
one can determine recursively all totally cyclic counterterms whose gauge variations cancel the contributions that are not proportional to the EoM's. In the Fronsdal setting one then ends up with the full \emph{off-shell} result
\be
\begin{split}
\tilde{\cK}_{123}^{\,\text{off-shell}}\,&=\,\left(1\,+\,\slashed{\partial}^{\,23}_{\xi_1}\,+\,\slashed{\partial}^{\,31}_{\xi_2} \,+\,\slashed{\partial}^{\,12}_{\xi_3}\right) \exp\Big(\cG_{123}\Big)\\&\times\,\left[1\,+\,\frac{\a^{\,\prime}\!\!}{2}\ \hat{\cH}_{\,12}\,\hat{\cH}_{\,13} +\,\left(\frac{\a^{\,\prime}\!\!}{2}\,\right)^{\frac{3}{2}}:\hat{\cH}_{21}\,\hat{\cH}_{\,32}\,\hat{\cH}_{\,13}:\,+\ \text{cyclic}\right]\\
&+\,\exp\Big(\cG_{123}\Big)\ \left\{\slashed{\partial}_{\xi_1}^{\,12}\ \left[\sqrt{\frac{\a^{\,\prime}\!\!}{2}}\ \hat{\cH}_{\,23}\,+\,\frac{\a^{\,\prime}\!\!}{2}\ \hat{\cH}_{\,32}\,\hat{\cH}_{\,13}\right]\right.\\&\left.\,\qquad\quad\quad \qquad-\,\slashed{\partial}_{\xi_2}^{\,12}\ \left[\sqrt{\frac{\a^{\,\prime}\!\!}{2}}\ \hat{\cH}_{\,13}\,-\,\frac{\a^{\,\prime}\!\!}{2}\ \hat{\cH}_{\,31}\,\hat{\cH}_{\,23}\right]\,+\,\text{cyclic}\right\}\ ,\label{genoffshell}
\end{split}
\ee
where we have considered also fermionic labels that give zero contribution in the purely bosonic case, so that for instance $\slashed{\partial}^{ij}$ contracts the fermionic indices between the field $\psi_{\,i}$ and $\psi_{\,j}$ while the $1$ simply contracts the two fermionic indices together, whenever they are present. Moreover, we have defined $\hat{\cH}_{\,ij}$ by
\be
\cH_{\,ij}\,=\,\hat{\cH}_{\,ij}\,\phi_{\,j}\ ,
\ee
while, regarding the terms of order $\tilde{\cH}^{\,3}$, an ordering prescription simplifies the expressions in the Appendix of \cite{cubicstring}. Explicitly, in each term the de Donder operators are to be put to the right, making them act directly on the generating function of the fields, so that for instance,
\be
:\cD_i\,\partial_{\xi_i}:\,=\,\partial_{\xi_i}\,\cD_i\ .
\ee
Similar results supplemented by eventual further terms should hold in all off-shell extensions reviewed in the Appendix~\ref{app:quadratic}.
In the $n$-point case analogous completions should be recovered by a recursive procedure starting from the results so far presented in the main body of the paper and exploiting eq.~\eqref{genoffshell} and its generalizations.

\end{appendix}

\end{document}